\documentclass{aa}
\usepackage{graphics,epsfig,txfonts}

\usepackage{natbib}
\bibpunct{(}{)}{;}{a}{}{,}

\newcommand{\ergs}[1]{$\times 10^{#1}$ erg s$^{-1}$}
\newcommand{\oergs}[1]{$10^{#1}$ erg s$^{-1}$}
\newcommand{\hcm}[1]{$\times 10^{#1}$ cm$^{-2}$}
\newcommand{\ohcm}[1]{$10^{#1}$ cm$^{-2}$}
\newcommand{\expo}[1]{$\times 10^{#1}$}
\newcommand{\oexpo}[1]{$10^{#1}$}

\newcommand{\nh}{N$_{\rm H}$}
\newcommand{\nhi}{N$_{\rm HI}$}

\newcommand{\ct}[1]{$\times 10^{#1}$ cts s$^{-1}$}
\newcommand{\oct}{cts s$^{-1}$}

\newcommand{\Hone}{\ion{H}{I}}

\newcommand{\Halp}{H${\alpha}$}
\newcommand{\ltsima}{$\buildrel < \over \sim$}
\newcommand{\lsim}{\lower.5ex\hbox{\ltsima}}
\newcommand{\gtsima}{$\buildrel > \over \sim$}
\newcommand{\gsim}{\lower.5ex\hbox{\gtsima}}
\newcommand{\xmm}{XMM-Newton}

\begin{document}
 
\title{XMM-Newton observations of the Small Magellanic Cloud:\\ 
       Be/X-ray binary pulsars active between October 2006 and June 2007
       \thanks{Based on observations with 
               XMM-Newton, an ESA Science Mission with instruments and contributions 
               directly funded by ESA Member states and the USA (NASA).}}
\author{F.~Haberl \and P.~Eger \and W.~Pietsch}

\titlerunning{New Be/X-ray binary pulsars in the SMC}
\authorrunning{Haberl et al.}
 
\institute{Max-Planck-Institut f\"ur extraterrestrische Physik,
           Giessenbachstra{\ss}e, 85748 Garching, Germany\\
	   \email{fwh@mpe.mpg.de}}
 
\date{Received 30 April 2008 / Accepted 25 June 2008}
 
\abstract{}
         {We analysed eight XMM-Newton observations toward the Small Magellanic Cloud (SMC), 
	  performed between October 2006 and June 2007, to investigate high mass X-ray binary 
	  systems.}
         {We produced images from the European Photon Imaging Cameras (EPIC) and 
	  extracted X-ray spectra and light curves in different energy bands 
	  from sources which yielded a 
	  sufficiently high number of counts for a detailed temporal and spectral 
	  analysis. To search for periodicity we applied Fourier transformations and
	  folding techniques and determined pulse periods using a Bayesian approach.
	  To identify optical counterparts we produced X-ray source lists
	  for each observation using maximum likelihood source detection techniques 
	  and correlated them with optical catalogues. The 
	  correlations were also used for astrometric boresight corrections of the X-ray
	  source positions.}
	 {We found new X-ray binary pulsars with periods of 202~s (XMMU\,J005929.0-723703), 
	  342~s (XMMU\,J005403.8-722632), 645~s (XMMU\,J005535.2-722906)
	  and 325~s (XMMU\,J005252.1-721715), in the latter case confirming the independent
	  discovery in Chandra data.
          In addition we detected sixteen known Be/X-ray binary pulsars and 
	  six ROSAT-classified candidate high mass X-ray binaries. From one of the 
	  candidates, RX\,J0058.2-7231, we discovered X-ray pulsations with a period 
	  of 291~s which makes it the likely counterpart of XTE\,J0051-727.
	  From the known pulsars, we revise the pulse period of CXOU\,J010206.6-714115 
	  to 967 s, and we detected the 18.37~s pulsar XTE\,J0055-727 
	  (=~XMM\,J004911.4-724939) in outburst, which allowed us to localise the source.
	  The pulse profiles of the X-ray pulsars show a large variety of shapes 
	  from smooth to highly structured patterns and differing energy dependence.
	  For all the candidate high mass X-ray binaries optical counterparts can be 
	  identified with magnitudes and colours consistent with Be stars.	  
	  Twenty of the Be/X-ray binaries were detected with X-ray luminosities
	  in the range 1.5\ergs{35} - 5.5\ergs{36}.
	  The majority of the spectra is well represented by an absorbed power-law
	  with an average power-law index of 0.93.
          The absorption (in addition to the Galactic foreground value)
	  varies over a wide range between a few \ohcm{20} and several \ohcm{22}. 
	  An overall correlation of the absorption with the total SMC \Hone\
	  column density suggests that the absorption seen in the X-ray spectra
	  is often largely caused by interstellar gas.}
	  {}

 
\maketitle
 
\section{Introduction}

Together with the Milky Way, the Small Magellanic Cloud (SMC) is the galaxy with 
the highest number of known high mass X-ray binaries (HMXBs). 
In these systems a compact object is accreting mass from an early 
type companion star. With only one exception (the supergiant system SMC\,X-1), 
the vast majority of HMXBs in the SMC comprises systems with a Be star as mass donor and 
a neutron star as compact object that is usually recognised as an X-ray pulsar. 
Depending on the eccentricity of the
binary orbit, the mass accretion rate can be enhanced around the time of periastron passage 
when the neutron star approaches the circumstellar disc of the Be star 
\citep[see, e.g.,][]{2001A&A...377..161O}. These outbursts usually last a few days.
Longer outbursts with durations of weeks (type II) can be caused by a drastic expansion of the 
disc around the Be star.

Many of the Be/X-ray binaries in the SMC were discovered during X-ray outburst. 
Then pulsations could be found as periodic modulation of the X-ray flux. During 
quiescence the sources were often not detected at all. 
It is not clear how many of these X-ray transients we are still missing, but present 
day X-ray observations with Chandra, the Rossi X-ray Timing Explorer (RXTE) and 
\xmm\ keep finding new Be/X-ray binaries in the SMC with a rate of a few per year. 
Currently, in the SMC nearly 50 Be/X-ray binary pulsars are known 
\citep{2004A&A...414..667H,2005MNRAS.356..502C}. In addition, more than two dozen 
Be/X-ray binary candidates are discussed in the literature which are either
classified as such from their X-ray properties and/or are optically identified 
with a Be star counterpart. In these cases, however, uncertain X-ray positions and
missing X-ray pulsations make the cases less clear.

\begin{table*}
\caption[]{XMM-Newton EPIC observations of SMC fields in 2006/2007.}
\begin{tabular}{lllllrrc}
\hline\hline\noalign{\smallskip}
\multicolumn{1}{c}{Observation} &
\multicolumn{2}{c}{Pointing direction} &
\multicolumn{1}{c}{Sat.} &
\multicolumn{1}{c}{EPIC$^{(a)}$} &
\multicolumn{1}{c}{Start time (UT)} &
\multicolumn{1}{c}{End time (UT)}\\

\multicolumn{1}{c}{ID} &
\multicolumn{1}{c}{R.A.} &
\multicolumn{1}{c}{Dec.} &
\multicolumn{1}{c}{Rev.} &
\multicolumn{1}{c}{Instrument} &
\multicolumn{1}{c}{} &
\multicolumn{1}{c}{} \\

\multicolumn{1}{c}{} &
\multicolumn{2}{c}{(J2000.0)} &
\multicolumn{1}{c}{} &
\multicolumn{1}{c}{configuration} &
\multicolumn{1}{c}{} &
\multicolumn{1}{c}{} \\
\noalign{\smallskip}\hline\noalign{\smallskip}
 0404680101 & 00 47 36.0 & -73 08 24 & 1249 & PN FF thin      & 2006-10-05 00:44:47 & 2006-10-05 06:51:09 \\
            &            &           &      & M1/M2 FF medium &            00:22:05 &            06:50:54 \\
 0404680201 & 00 52 26.4 & -72 52 12 & 1263 & PN FF thin      & 2006-11-01 01:18:41 & 2006-11-01 09:58:46 \\
            &            &           &      & M1/M2 FF medium &            00:55:59 &            09:58:31 \\
 0403970301 & 00 47 39.4 & -72 59 31 & 1329 & PN EFF thin     & 2007-03-12 21:02:47 & 2007-03-13 06:53:05 \\
            &            &           &      & M1/M2 FF thin   &            20:01:50 &            06:52:50 \\
 0404680301 & 00 51 00.7 & -73 24 17 & 1344 & PN FF thin      & 2007-04-11 20:00:39 & 2007-04-12 02:15:32 \\
            &            &           &      & M1/M2 FF medium &            19:37:57 &            02:15:17 \\
 0404680501 & 01 07 42.3 & -72 30 11 & 1344 & PN FF thin      & 2007-04-12 03:29:37 & 2007-04-12 09:44:42 \\
            &            &           &      & M1/M2 FF medium &            03:06:55 &            09:44:27 \\
 0501470101 & 00 59 41.8 & -71:38:15 & 1371 & PN FF thin      & 2007-06-04 09:22:05 & 2007-06-04 18:20:14 \\
            &            &           &      & M1/M2 FF thin   &            08:59:23 &            18:19:59 \\
 0500980201 & 01 00 00.0 & -72 27 00 & 1372 & PN FF thin      & 2007-06-06 09:14:26 & 2007-06-06 16:52:51 \\
            &            &           &      & M1/M2 FF medium &            08:51:44 &            16:52:36 \\
 0500980101 & 00 53 02.4 & -72 26 17 & 1380 & PN FF thin      & 2007-06-23 06:13:53 & 2007-06-23 13:03:10 \\
            &            &           &      & M1/M2 FF medium &            05:51:11 &            13:02:55 \\
\noalign{\smallskip}\hline\noalign{\smallskip}
\end{tabular}

$^{(a)}$ FF: full frame CCD readout mode with 73~ms frame time for PN and 2.6~s for MOS; 
         EFF: PN `extended FF' with 200~ms frame time;
         thin and medium optical blocking filters.
\label{tab-obs}
\end{table*}

We used the European Photon Imaging Cameras (EPIC) on \xmm\ to investigate HMXBs in the SMC.
Here we report on results from eight observations - seven from our own dedicated programs to 
investigate candidate HMXBs and supersoft X-ray sources and one from the public archive - 
which cover six candidate Be/X-ray binaries and sixteen known Be/X-ray binary pulsars. 
We present a temporal and spectral analysis of the majority of these objects and 
additional newly discovered Be/X-ray binary pulsars in the SMC.

\section{Observations and data analysis}

\begin{figure*}
\sidecaption
  \includegraphics[width=12cm,clip=]{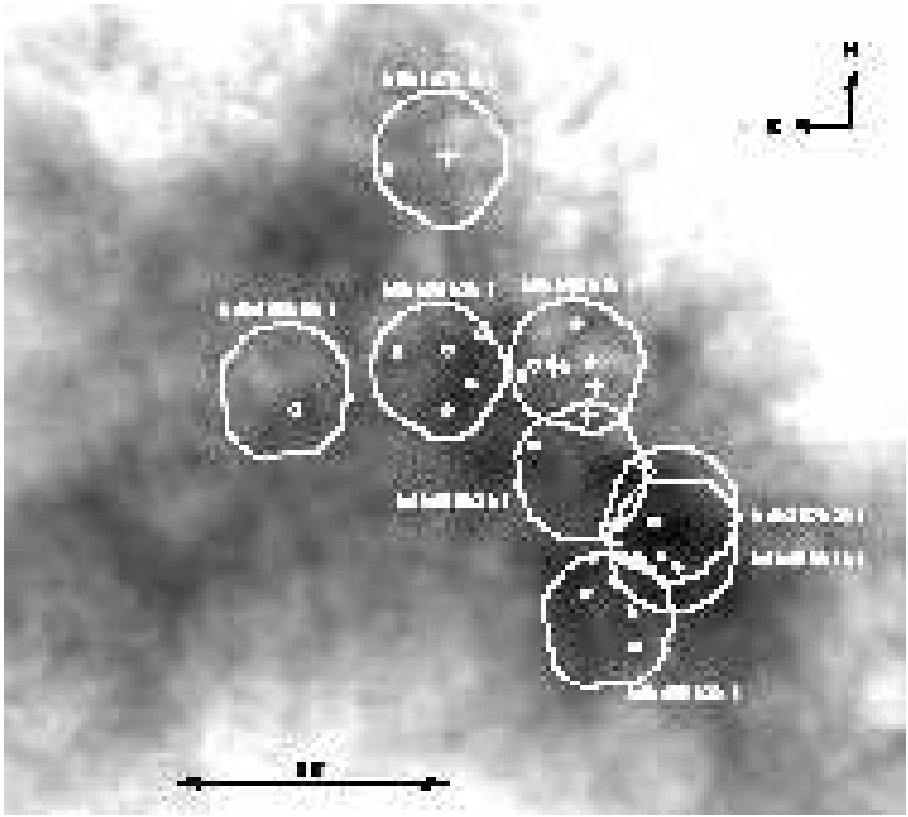}
  \caption{Observed fields (marked by contours which are derived from 
           the combined EPIC exposure maps) superimposed on an \Hone\ map of the SMC 
	   \citep{1999MNRAS.302..417S}. The positions of the investigated Be/X-ray binaries
	   are marked by different symbols:  
	   thin circles are used for Be/X-ray binary pulsars with previously known pulse period, 
	   thick circles indicate new discoveries from this work, boxes show the positions 
	   of Be/X-ray binaries with no detected pulse period and crosses mark the positions of 
	   known Be/X-ray binaries which we did not detect. An X-ray image of field 0500980101 
	   is presented in Fig.~\ref{fig-image}.}
  \label{fig-obs}
\end{figure*}

The eight SMC fields were observed with \xmm\ \citep{2001A&A...365L...1J} between October 2006 and 
June 2007. Details of the observations (in chronological order) are summarised in Table~\ref{tab-obs}.
The EPIC-MOS \citep{2001A&A...365L..27T} and EPIC-PN \citep{2001A&A...365L..18S} cameras were 
operated in imaging mode, covering an area with about 28\arcmin\ in diameter. 
The locations of the fields are superimposed on an \Hone\ image of the SMC in Fig.~\ref{fig-obs}. 
Our dedicated observations cover part of the main bar of the SMC where most of the HMXBs are found.
The archival field 0403970301, pointed at the emission nebula N19, overlaps largely with 
field 0404680101 (shifted by about 9\arcmin\ to the north). 
Observation 0501470101 is part of our programme to observe supersoft X-ray sources in the SMC
\citep{2006A&A...452..431K}. The field is located in the north, outside the main bar of the SMC.

For the X-ray analysis we applied the same procedures as outlined in \citet{2008fysx.conf...32H} 
and \citet{2008arXiv0801.4679H} using the \xmm\ Science Analysis System (SAS) version 7.1.0 
together with tools from the FTOOLS package.  
To be most sensitive to faint sources, intervals with increased background flaring activity 
were removed from our source detection analysis (standard SAS maximum likelihood method).
X-ray source lists were produced for each observation and astrometric 
boresight corrections were applied using the optical catalogue of \citet{2002AJ....123..855Z} 
as reference system. In this way we could reduce the systematic 1$\sigma$ uncertainty of the 
positions to 1.1\arcsec.

The X-ray source lists were correlated with optical catalogues. We selected HMXB
candidates using X-ray colours (hardness ratios) and the optical brightness and colours
of likely optical counterparts.
The resulting HMXB candidate list comprises sixteen known Be/X-ray binary pulsars
\citep[see recent compilations in][]{2000A&A...359..573H,2004A&A...414..667H,2005MNRAS.356..502C}, 
six candidate Be/X-ray binaries as classified in previous work 
\citep[see][]{2000A&A...359..573H,2004A&A...414..667H} 
and four new candidates, i.e. in total we selected twenty-six X-ray sources for further 
investigation. The results from the new Be/X-ray binary 25.55~s pulsar were published 
separately \citep{2008arXiv0803.2473H} and we count this source here as ``known'' pulsar.
The X-ray positions (summarised in Table~\ref{tab-sources}) with a typical accuracy 
of 1\arcsec$-$2\arcsec\ (total 1$\sigma$ error inferred from the EPIC data) allowed 
us to securely identify the optical counterparts. 
In Table~\ref{tab-ids} optical brightness and colours from different optical surveys are 
summarised. For completeness and easier comparison we list the values for all our investigated 
objects.
Optical counterparts from the emission line star catalogue of 
\citet[][ hereafter MA93]{1993A&AS..102..451M} are listed in Table~\ref{tab-sources}.
For the newly identified Be/X-ray binaries we extracted light curves of their optical 
counterparts in the I-band from the Optical Gravitational Lensing Experiment (OGLE) 
photometry database \citep[][]{2005AcA....55...43S,1997AcA....47..319U}
and the approximate B- and R-band from the Massive Compact Halo Object project (MACHO).
We applied an FFT analysis to the optical light curves 
\citep{1976Ap&SS..39..447L,1982ApJ...263..835S} to 
search for periodic variations which could indicate orbital periods.
For this purpose we removed the long-term variations from the optical light curves
which was difficult for times of fast brightness changes. In these cases we restricted 
the analysis to time intervals with slow and smooth variations.

\begin{table*}
\caption[]{SMC Be/X-ray binaries detected by \xmm\ between October 2006 and June 2007}
\renewcommand{\tabcolsep}{5pt}
\begin{center}
\begin{tabular}{lcccllcccc}
\hline\hline\noalign{\smallskip}
\multicolumn{1}{l}{Name} &
\multicolumn{1}{c}{X-ray Position} &
\multicolumn{1}{c}{Err.$^{(a)}$} &
\multicolumn{1}{c}{MA} &
\multicolumn{1}{l}{Comment$^{(b)}$} &
\multicolumn{1}{c}{Sect.} &
\multicolumn{1}{c}{Period} &
\multicolumn{1}{c}{PF$^{(d)}$} &
\multicolumn{1}{c}{Orbital} \\
\multicolumn{1}{l}{} &
\multicolumn{1}{c}{R.A. and Dec. (J2000.0)} &
\multicolumn{1}{c}{\arcsec} &
\multicolumn{1}{c}{93} &
\multicolumn{1}{c}{} &
\multicolumn{1}{c}{} &
\multicolumn{1}{c}{s} &
\multicolumn{1}{c}{\%} &
\multicolumn{1}{c}{Phase$^{(e)}$} \\

\noalign{\smallskip}\hline\noalign{\smallskip}
\multicolumn{8}{c}{Observation ID 0404680101} \\
 XMMU\,J004723.7-731226 & 00 47 23.42 --73 12 27.3 & 0.25 &  172 & 263 s XP 	  & 4.1   & 262.23$\pm$0.66   & 22$\pm$9     & --            \\
 XMMU\,J004814.1-731003 & 00 48 14.10 --73 10 04.0 & 0.45 &   -- & 25.5 s XP (1)  & 4.2   & 25.550$\pm$0.002  & 45$\pm$15    & --	     \\
 RX\,J0049.2-7311	& 00 49 13.60 --73 11 36.8 & 0.55 &   -- & (2)  	  & 5.2   & --  	      & --	     & --	     \\
 RX\,J0049.5-7310	& 00 49 29.92 --73 10 59.2 & 0.81 &  300 & (2)  	  & 5.2   & --  	      & --	     & --	     \\
\noalign{\smallskip}\hline\noalign{\smallskip}
\multicolumn{8}{c}{Observation ID 0404680201} \\
 CXOU\,J005455.6-724510 & 00 54 55.91 --72 45 10.6 & 0.49 &  809 & 500 s XP       & 4.3   & 497.5$\pm$1.0     & 39$\pm$16    & 0.00$\pm$0.04 \\
\noalign{\smallskip}\hline\noalign{\smallskip}
\multicolumn{8}{c}{Observation ID 0403970301} \\
 XMMU\,J004723.7-731226 & 00 47 23.43 --73 12 27.2 & 0.35 &  172 & 263 s XP 	  & 4.1   & $^{(c)}$	      & --	     & --            \\
 XMMU\,J004814.1-731003 & 00 48 14.41 --73 10 04.6 & 0.91 &   -- & 25.5 s XP (1)  & 4.2   & --  	      & --	     & --	     \\
 RX\,J0048.5-7302	& 00 48 34.15 --73 02 31.0 & 1.08 &  238 & (2)  	  & 5.1   & --  	      & --	     & --	     \\
 AX\,J0049-729  	& 00 49 03.29 --72 50 51.9 & 0.53 &   -- & 74.7 s XP	  & 4.4   & --  	      & --	     & 0.08$\pm$0.15 \\
 XMMU\,J004911.4-724939 & 00 49 11.54 --72 49 37.3 & 0.08 &   -- & 18.37 s XP	  & 4.5   & 18.3814$\pm$0.0001& 21$\pm$3     & 0.97$\pm$0.11 \\
 RX\,J0049.2-7311	& 00 49 14.12 --73 11 35.8 & 0.49 &   -- & (2)  	  & 5.2   & --  	      & --	     & --	     \\
\noalign{\smallskip}\hline\noalign{\smallskip}
\multicolumn{8}{c}{Observation ID 0404680301} \\
 RX\,J0049.5-7331	& 00 49 30.59 --73 31 09.1 & 0.33 &  302 & (2)  	  & 5.3   & --  	      & --	     & --	     \\
 RX\,J0049.7-7323	& 00 49 42.02 --73 23 14.5 & 0.23 &  315 & 755 s XP 	  & 4.6   & 746.15$\pm$0.79   & 25$\pm$5     & 0.14$\pm$0.10 \\
 RX\,J0050.8-7316	& 00 50 44.70 --73 16 05.1 & 0.18 &  387 & 323 s XP 	  & 4.7   & 317.44$\pm$0.26   & 31$\pm$7     & 0.00$\pm$0.11 \\
 AX\,J0051.6-7311	& 00 51 51.91 --73 10 32.6 & 0.59 &  504 & 172 s XP 	  & 4.8   & --  	      & --	     & --	     \\
 RX\,J0052.1-7319	& 00 52 15.30 --73 19 14.6 & 0.82 &   -- & 15.3 s XP	  & 4.9   & --  	      & --	     & --	     \\
\noalign{\smallskip}\hline\noalign{\smallskip}
\multicolumn{8}{c}{Observation ID 0404680501} \\
 CXOU\,J010712.6-723533 & 01 07 12.81 --72 35 33.1 & 0.22 & 1619 & 65.8 s XP      & 4.10  & 65.95$\pm$0.02    & 31$\pm$9     & --	     \\
\noalign{\smallskip}\hline\noalign{\smallskip}
\multicolumn{8}{c}{Observation ID 0501470101} \\
 CXOU\,J010206.6-714115 & 01 02 06.68 --71 41 16.1 & 0.17 & 1301 & 967 s XP       & 3.1   & 966.97$\pm$0.47   & 32$\pm$17    & --	     \\
\noalign{\smallskip}\hline\noalign{\smallskip}
\multicolumn{8}{c}{Observation ID 0500980201} \\
 CXOU\,J005736.2-721934 & 00 57 35.84 --72 19 34.2 & 1.14 & 1020 & 565 s XP 	  & 4.11  & --  	      & --	     & 0.43$\pm$0.18 \\
 RX\,J0058.2-7231	& 00 58 12.64 --72 30 48.0 & 0.20 &   -- & (2), 291 s XP  & 3.2   & 291.327$\pm$0.057 & 17$\pm$6     & --            \\
 RX\,J0059.3-7223	& 00 59 21.02 --72 23 16.8 & 0.14 &   -- & 202 s XP 	  & 4.12  & 200.50$\pm$0.29   & 19$\pm$6     & --            \\
 XMMU\,J005929.0-723703 & 00 59 29.05 --72 37 03.2 & 0.08 & 1147 & NXT, 202 s XP  & 3.3   & 202.52$\pm$0.02   & 12.4$\pm$0.1 & --            \\
 RX\,J0101.8-7223	& 01 01 52.34 --72 23 32.6 & 0.19 & 1288 & (2)  	  & 5.4   & --                & --           & --            \\
\noalign{\smallskip}\hline\noalign{\smallskip}
\multicolumn{8}{c}{Observation ID 0500980101} \\
 SMC\,X-3		& 00 52 05.68 --72 26 05.3 & 0.14 &  531 & 7.78 s XP	  & 4.13  & 7.7912$\pm$0.0096 & 18$\pm$6     & 0.05$\pm$0.09 \\
 XMMU\,J005252.1-721715 & 00 52 52.12 --72 17 15.7 & 0.21 &   -- & NXT, 325 s XP  & 3.4   & 325.36$\pm$0.59   & 20$\pm$7     & --            \\
 CXOU\,J005323.8-722715 & 00 53 23.97 --72 27 15.3 & 0.16 &  667 & 138 s XP 	  & 4.14  & 139.136$\pm$0.016 & 42$\pm$8     & 0.09$\pm$0.14 \\
 XMMU\,J005403.8-722632 & 00 54 03.88 --72 26 32.8 & 0.27 &   -- & NXT, 342 s XP  & 3.5   & 341.87$\pm$0.15   & 54$\pm$12    & --            \\
 XTE\,J0055-724 	& 00 54 56.20 --72 26 47.1 & 0.36 &  810 & 59 s XP  	  & 4.15  & 58.858$\pm$0.001  & 90$\pm$23    & 0.12$\pm$0.08 \\
 XMMU\,J005535.2-722906 & 00 55 35.24 --72 29 06.2 & 0.19 &   -- & NXT, 645 s XP  & 3.6   & 644.55$\pm$0.72   & 14$\pm$14    & --            \\
\noalign{\smallskip}\hline
\end{tabular}
\end{center}
$^{(a)}$ Statistical 1$\sigma$ uncertainty, the systematic error is 1.1\arcsec.\\
$^{(b)}$ NXT: New Be/X-ray transient; XP: X-ray pulsar
References: (1) published separately in \citet{2008arXiv0803.2473H}, (2) candidate BeX from \citet{2000A&A...359..573H}.\\
$^{(c)}$ Although sufficient number of counts, pulsations not detected (see text).\\
$^{(d)}$ Pulsed fraction obtained from 0.2-10.0 keV EPIC-PN pulse profiles.\\
$^{(e)}$ Orbital phase calculated from the X-ray ephemeris listed in \citet{2008arXiv0802.2118G}.
\label{tab-sources}
\end{table*}

\begin{table*}
\caption[]{SMC high mass X-ray binaries not detected by \xmm.}
\begin{center}
\begin{tabular}{llccl}
\hline\hline\noalign{\smallskip}
\multicolumn{1}{l}{Observation} &
\multicolumn{1}{l}{Name} &
\multicolumn{1}{c}{Upper limit} &
\multicolumn{1}{c}{Orbital} &
\multicolumn{1}{l}{Comment} \\
\multicolumn{1}{l}{ID} &
\multicolumn{1}{l}{} &
\multicolumn{1}{c}{10$^{-3}$\oct / \oergs{34}} &
\multicolumn{1}{c}{Phase} &
\multicolumn{1}{l}{Comment} \\

\noalign{\smallskip}\hline\noalign{\smallskip}
0404680101 & RX\,J0048.5-7302       & 13.1 / 5.2 & --             & no pulsations found yet, see also detection in 0403970301 \\
\noalign{\smallskip}\hline\noalign{\smallskip}
0404680201 & XTE\,J0052-725         & 7.9  / 3.2 & 0.39$\pm$0.07  & 82.4 s pulsar \citep{2002IAUC.7932....2C} \\
\noalign{\smallskip}\hline\noalign{\smallskip}
0403970301 & RX\,J0049.5-7310       & 11.9 / 4.8 & --             & no pulsations found yet, see also detection in 0404680101 \\
\noalign{\smallskip}\hline\noalign{\smallskip}
0501470101 & RX\,J0059.2-7138       & 7.5  / 3.0 & --             & 2.76 s pulsar \citep{1994ApJ...427L..25H} \\
\noalign{\smallskip}\hline\noalign{\smallskip}
0500980101 & RX\,J0051.8-7231       & 6.0  / 2.4 & 0.11$\pm$0.12  & 8.9 s pulsar  \citep{1997ApJ...484L.141I} \\
           & XTE\,J0052-725         & 8.9  / 3.6 & 0.03$\pm$0.08  & 82.4 s pulsar \citep{2002IAUC.7932....2C} \\
           & XTE\,J0052-723         & --         & --             & 4.78 s pulsar \citep{2003MNRAS.339..435L} uncertain position \\
           & XTE\,J0053-724         & 4.2  / 1.7 & 0.42$\pm$0.05  & 46.6 s pulsar \citep{1998IAUC.6803....1C} \\
\noalign{\smallskip}\hline
\end{tabular}
\end{center}
\label{tab-nondet}
\end{table*}

\begin{table*}
\caption[]{Optical identifications (continued next page).}
\begin{center}
\begin{tabular}{llclcccc}
\hline\hline\noalign{\smallskip}
\multicolumn{1}{l}{Source} &
\multicolumn{1}{l}{Catalogue$^{(a)}$} &
\multicolumn{1}{c}{R.A. and Dec. (J2000.0)} &
\multicolumn{1}{c}{Vmag} &
\multicolumn{1}{c}{B$-$V} &
\multicolumn{1}{c}{U$-$B} &
\multicolumn{1}{c}{V$-$R} &
\multicolumn{1}{c}{V$-$I} \\

\noalign{\smallskip}\hline\noalign{\smallskip}
XMMU\,J004723.7-731226  & UBVR & 00 47 23.97 --73 12 25.9 & 14.37$^{(b)}$ & $-$0.06 & $-$0.83 & $+$0.05 &  --	  \\
                        & MCPS & 00 47 23.42 --73 12 27.3 & 16.03 & $+$0.08 & $-$0.97 & --      & $+$0.05 \\
                        & OGLE & 00 47 23.35 --73 12 27.0 & 16.14 & $-$0.04 & --      & --      & $+$0.14 \\
XMMU\,J004814.1-731003  & UBVR & 00 48 14.10 --73 10 04.0 & 15.25 & $+$0.13 & $-$0.64 & $+$0.12 &  --	  \\
                        & MCPS & 00 48 14.18 --73 10 03.9 & 15.30 & $+$0.26 & $-$0.50 & --      & $-$0.21 \\
                        & OGLE & 00 48 14.13 --73 10 03.5 & 15.71 & $+$0.01 & --      & --      & $+$0.09 \\
RX\,J0049.2-7311        & UBVR & 00 49 13.59 --73 11 37.0 & 15.67 & $+$0.25 & $-$0.71 & $+$0.19 &  --	  \\
                        & MCPS & 00 49 13.68 --73 11 37.5 & 16.44 & $+$0.19 & $-$0.92 & --      & $+$0.16 \\
                        & OGLE & 00 49 13.63 --73 11 37.4 & 16.51 & $+$0.10 & --      & --      & $+$0.30 \\
RX\,J0049.5-7310        & UBVR & 00 49 29.78 --73 10 58.7 & 15.66 & $+$0.19 & $-$0.49 & $+$0.28 &  --	  \\
                        & MCPS & 00 49 29.87 --73 10 58.2 & 16.15 & $+$0.20 & $-$0.88 & --      & $+$0.14 \\
                        & OGLE & 00 49 29.81 --73 10 58.0 & 16.30 & $+$0.09 & --      & --      & $+$0.33 \\
\noalign{\smallskip}\hline\noalign{\smallskip}
CXOU\,J005455.6-724510  & UBVR & 00 54 55.88 --72 45 10.5 & 14.78 & $-$0.05 & $-$0.90 & $+$0.05 &  --	  \\
                        & MCPS & 00 54 55.91 --72 45 10.5 & 15.00 & $-$0.03 & $-$0.93 & --      & $+$0.10 \\
                        & OGLE & 00 54 55.87 --72 45 10.7 & 14.99 & $-$0.02 & --      & --      & $+$0.20 \\
\noalign{\smallskip}\hline\noalign{\smallskip}
RX\,J0048.5-7302        & UBVR & 00 48 34.08 --73 02 31.1 & 14.84 & $-$0.05 & $-$0.79 & $+$0.18 & --      \\
                        & MCPS & 00 48 34.17 --73 02 30.8 & 14.78 & $+$0.00 & $-$0.81 & --      & $+$0.02 \\
                        & OGLE & 00 48 34.12 --73 02 30.9 & 15.06 & $-$0.23 & --      & --      & $+$0.31 \\
AX\,J0049-729	        & UBVR & 00 49 03.35 --72 50 52.0 & 16.09 & $+$0.12 & $-$0.59 & $+$0.12 & --      \\
                        & MCPS & 00 49 03.41 --72 50 52.3 & 16.78 & $+$0.08 & $-$0.84 & --      & $+$0.17 \\
                        & OGLE & 00 49 03.34 --72 50 52.1 & 16.92 & $+$0.09 & --      & --      & $+$0.24 \\
XMMU\,J004911.4-724939  & UBVR & 00 49 11.40 --72 49 37.2 & 15.61 & $+$0.19 & $-$0.64 & $+$0.19 & --      \\
			& MCPS & 00 49 11.53 --72 49 37.2 & 15.96 & $+$0.05 & $-$0.79 & --      & $+$0.07 \\
			& OGLE & 00 49 11.45 --72 49 37.1 & 15.71 & $+$0.07 & --      & --      & $+$0.21 \\
\noalign{\smallskip}\hline\noalign{\smallskip}
RX\,J0049.5-7331        & UBVR & 00 49 30.39 --73 31 10.2 & 13.83$^{(b)}$ & $-$0.04 & $-$0.92 & $-$0.89 &  --	  \\
                        & MCPS & 00 49 30.55 --73 31 09.1 & 14.64 & $-$0.04 & $-$0.86 & --      & $+$0.09 \\
                        & OGLE & 00 49 30.53 --73 31 08.8 & 14.66 & $-$0.06 & --      & --      & $+$0.16 \\
RX\,J0049.7-7323        & UBVR & 00 49 41.99 --73 23 14.6 & 14.86 & $+$0.04 & $-$0.86 & $-$0.89 &  --	  \\
                        & MCPS & 00 49 42.03 --73 23 14.4 & 14.86 & $+$0.15 & $-$0.91 & --      & $+$0.16 \\
                        & OGLE & 00 49 42.02 --73 23 14.2 & 14.98 & $+$0.05 & --      & --      & $+$0.25 \\
RX\,J0050.8-7316        & UBVR & 00 50 44.66 --73 16 05.4 & 15.25 & $-$0.03 & $-$0.94 & $+$0.07 &  --	  \\
                        & MCPS & 00 50 44.75 --73 16 05.3 & 15.48 & $-$0.11 & $-$0.91 & --      & $+$0.17 \\
                        & OGLE & 00 50 44.71 --73 16 05.0 & 15.44 & $-$0.04 & --      & --      & $+$0.17 \\
AX\,J0051.6-7311        & UBVR & 00 51 52.01 --73 10 33.8 & 14.33 & $-$0.04 & $-$0.94 & $+$0.09 &  --	  \\
                        & MCPS & 00 51 52.08 --73 10 33.9 & 14.45 & $-$0.07 & $-$0.99 & --      & $+$0.16 \\
                        & OGLE & 00 51 52.02 --73 10 33.6 & 14.45 & $-$0.07 & --      & --      & $+$0.09 \\
RX\,J0052.1-7319        & MCPS & 00 52 15.48 --73 19 15.1 & 15.90 & $-$0.14 & $-$0.95 & --      & $-$0.20 \\
                        & OGLE & 00 52 15.39 --73 19 15.0 & 15.91 & $-$0.10 & --      & --      & $-$0.03 \\
\noalign{\smallskip}\hline\noalign{\smallskip}
CXOU J010712.6-723533   & UBVR & 01 07 12.58 --72 35 33.6 & 15.64 & $-$0.12 & $-$0.75 & $-$0.12 &  --	  \\
                        & MCPS & 01 07 12.58 --72 35 33.3 & 15.64 & $-$0.09 & $-$0.83 & --      & $-$0.09 \\
                        & OGLE & 01 07 12.59 --72 35 33.5 & 15.76 & $-$0.14 & --      & --      & $+$0.00 \\
\noalign{\smallskip}\hline\noalign{\smallskip}
CXOU\,J010206.6-714115  & UBVR & 01 02 06.65 --71 41 15.9 & 14.56 & $-$0.08 & $-$1.12 & $-$1.06 &  --     \\
                        & MCPS & 01 02 06.68 --71 41 16.1 & 14.37 & $+$0.21 & $-$1.04 &         & $-$0.05 \\
\noalign{\smallskip}\hline\noalign{\smallskip}
CXOU\,J005736.2-721934  & UBVR & 00 57 36.06 --72 19 33.8 & 15.84 & $-$0.10 & $-$0.85 & $+$0.05 & --      \\
                        & MCPS & 00 57 36.09 --72 19 33.6 & 15.99 & $+$0.01 & $-$1.06 & --      & $+$0.19 \\
                        & OGLE & 00 57 36.01 --72 19 33.8 & 15.97 & $-$0.02 & --      & --      & $+$0.26 \\
RX\,J0058.2-7231        & UBVR & 00 58 12.65 --72 30 48.5 & 14.78 & $+$0.06 & $-$0.91 & $-$0.95 & --      \\
                        & MCPS & 00 58 12.63 --72 30 48.2 & 14.87 & $+$0.11 & $-$1.07 & --      & $+$0.26 \\
                        & OGLE & 00 58 12.58 --72 30 48.5 & 14.89 & $+$0.06 & --      & --      & $+$0.27 \\
RX\,J0059.3-7223        & UBVR & 00 59 21.10 --72 23 17.2 & 14.65 & $-$0.06 & $-$0.92 & $-$0.88 & --      \\
                        & MCPS & 00 59 21.04 --72 23 16.7 & 14.98 & $-$0.01 & $-$1.16 & --      & $-$0.10 \\
                        & OGLE & 00 59 21.04 --72 23 17.0 & 14.83 & $-$0.07 & --      & --      & $+$0.03 \\
XMMU\,J005929.0-723703  & UBVR & 00 59 28.75 --72 37 04.1 & 15.31 & $+$0.01 & $-$0.77 & $+$0.16 & --      \\
                        & MCPS & 00 59 28.70 --72 37 04.2 & 15.53 & $+$0.05 & $-$1.01 & --      & $+$0.20 \\
                        & OGLE & 00 59 28.67 --72 37 03.9 & 15.60 & $-$0.03 & --      & --      & $+$0.21 \\
RX\,J0101.8-7223        & UBVR & 01 01 52.21 --72 23 34.1 & 14.61 & $-$0.04 & $-$0.90 & $-$0.87 & --      \\
                        & MCPS & 01 01 52.25 --72 23 33.8 & 14.94 & $-$0.01 & $-$1.09 & --      & $+$0.09 \\
                        & OGLE & 01 01 52.28 --72 23 33.7 & 14.98 & $-$0.06 & --      & --      & $+$0.14 \\
\noalign{\smallskip}\hline
\end{tabular}
\end{center}
\label{tab-ids}
\end{table*}
\addtocounter{table}{-1}
\begin{table*}
\caption[]{Continued.}
\begin{center}
\begin{tabular}{llclcccc}
\hline\hline\noalign{\smallskip}
\multicolumn{1}{l}{Source} &
\multicolumn{1}{l}{Catalogue} &
\multicolumn{1}{c}{R.A. and Dec. (J2000.0)} &
\multicolumn{1}{c}{Vmag} &
\multicolumn{1}{c}{B$-$V} &
\multicolumn{1}{c}{U$-$B} &
\multicolumn{1}{c}{V$-$R} &
\multicolumn{1}{c}{V$-$I} \\

\noalign{\smallskip}\hline\noalign{\smallskip}
SMC\,X-3                & UBVR & 00 52 05.67 --72 26 04.0 & 14.94 & $-$0.09 & $-$0.99 & $+$0.03 & --      \\
                        & MCPS & 00 52 05.69 --72 26 03.9 & 14.91 & $-$0.00 & $-$0.97 & --      & $+$0.08 \\
XMMU\,J005252.1-721715  & UBVR & 00 52 52.21 --72 17 15.0 & 16.33 & $-$0.09 & $-$0.35 & $-$0.10 & --      \\
                        & MCPS & 00 52 52.26 --72 17 14.7 & 16.62 & $-$0.21 & $-$0.58 & --      & $-$0.2  \\
CXOU\,J005323.8-722715  & UBVR & 00 53 23.78 --72 27 15.4 & 16.04 & $-$0.06 & $-$0.92 & $+$0.17 & --      \\
                        & MCPS & 00 53 23.90 --72 27 15.4 & 16.19 & $-$0.09 & $-$1.05 & --      & $+$0.05 \\
XMMU\,J005403.8-722632  & UBVR & 00 54 03.81 --72 26 33.0 & 14.81 & $-$0.04 & $-$0.90 & $+$0.01 & --      \\
                        & MCPS & 00 54 03.92 --72 26 32.8 & 14.94 & $+$0.01 & $-$1.13 & --      & $+$0.15 \\
XTE\,J0055-724          & UBVR & 00 54 56.13 --72 26 48.1 & 14.39$^{(b)}$ & $-$0.11 & $-$0.79 & $+$0.00 & --      \\
                        & MCPS & 00 54 56.26 --72 26 47.4 & 15.27 & $-$0.05 & $-$1.07 & --      & $+$0.16 \\
                        & OGLE & 00 54 56.17 --72 26 47.6 & 15.28 & $-$0.04 & --      & --      & $+$0.15 \\
XMMU\,J005535.2-722906  & UBVR & 00 55 35.21 --72 29 06.5 & 14.36 & $-$0.01 & $-$0.87 & $+$0.15 & --      \\
                        & MCPS & 00 55 35.20 --72 29 06.1 & 14.69 & $-$0.04 & $-$1.05 & --      & $+$0.01 \\
                        & OGLE & 00 55 35.13 --72 29 06.4 & 14.69 & $-$0.05 & --      & --      & $+$0.11 \\
\noalign{\smallskip}\hline
\end{tabular}
\end{center}
$^{(a)}$ From the UBVR CCD Survey \citep{2002ApJS..141...81M}, 
         the Magellanic Clouds Photometric Survey \citep[MCPS,][]{2002AJ....123..855Z} and 
         the Optical Gravitational Lensing Experiment (OGLE) BVI photometry catalogue \citep{1998AcA....48..147U}
$^{(b)}$ Blend of stars.

\end{table*}

In Table~\ref{tab-nondet} we list known HMXBs which were covered by our observations, but were not detected. 
Two of them, RX\,J0048.5-7302 and RX\,J0049.5-7310, were covered twice and detected in one observation 
(see Table~\ref{tab-sources}). XTE\,J0052-725 was also covered twice, but not detected in either observation. 
Five other known
Be/X-ray binary pulsars were covered once and not detected. However, the coarse position of 
one of them (XTE\,J0052-723 with a large
uncertainty in the X-ray position derived from RXTE observations) is located near the rim of field 0500980101
and the correct position might also be outside. In Table~\ref{tab-nondet} we provide 3$\sigma$ upper limits
of the EPIC-PN (0.2-4.5 keV) count rate (converted to X-ray luminosity with 4\ergs{33} per 1\ct{-3}, see below)
for the sources with accurately known positions. We mark their location in Fig.~\ref{fig-obs}. 
We calculated the binary orbital phase of our \xmm\ observations whenever an X-ray ephemeris from RXTE 
data is available for the covered pulsars \citep{2008arXiv0802.2118G}. The phases with errors are listed 
in Table~\ref{tab-sources} and Table~\ref{tab-nondet}.
Five more Be/X-ray binary candidates from correlations of ROSAT sources
with emission line stars from MA93 \citep{2000A&A...359..573H} are covered by the fields. They were either 
not detected, or the EPIC sources found in or close to their (relatively large) ROSAT error circles are 
inconsistent with Be/X-ray binaries. Hence, we can not confirm nor reject their proposed Be/X-ray binary 
nature.

Twenty of our sources yielded a sufficient number of counts for a detailed spectral and 
timing analysis. 
For the selection of the sources we used a limit of 250 for the total number of counts 
(summed over the available instruments) in the 0.5$-$4.5 keV band as obtained by the source detection 
analysis. The actual counts used for the analysis differ from this value depending on the spectral shape 
(wider energy band used for spectral analysis) and the off-axis angle of the source.
We extracted pulse-phase averaged EPIC spectra for 
PN (single + double pixel events, PATTERN 0$-$4) and MOS (PATTERN 0$-$12) disregarding bad CCD pixels 
and columns (FLAG 0) and removing intervals of increased background flaring activity. 
We used circular or elliptical source extraction regions depending on the 
off-axis angle of the source. In several cases the source extraction region was cut by bad columns 
or nearby CCD borders and therefore, we report fluxes and luminosities obtained from the instrument
with full source extraction region. We used XSPEC version 11.3.2p for spectral modelling.

We searched for X-ray pulsations in the EPIC data (for each instrument individually and 
for the combined data) of each source. We created power spectra 
and applied folding techniques. To further investigate candidate periods and
derive accurate values and errors for the detected pulse periods we used the Bayesian method
\citep{1996ApJ...473.1059G} as described in \citet{2000ApJ...540L..25Z}. 
Given period errors denote 1$\sigma$ uncertainties.
For the pulsars we folded the X-ray light curves on the best period in the standard EPIC energy bands 
(0.2$-$0.5 keV, 0.5$-$1.0 keV, 1.0$-$2.0 keV, 2.0$-$4.5 keV, 4.5$-$10.0 keV and the
broad band 0.2$-$10.0 keV) to investigate
the shape and energy dependence of the pulse profiles.
We fitted sine functions (with harmonics when required) to the pulse profiles in the 0.2$-$10.0 keV
band as obtained from EPIC-PN to derive pulsed fractions. The determined pulse periods and pulsed fractions 
are summarised in Table~\ref{tab-sources}.
From the pulse profiles we computed hardness ratios  as a function of pulse phase 
with HR1 = (R2-R1)/(R2+R1), HR2 = (R3-R2)/(R3+R2), HR3 = (R4-R3)/(R4+R3) and HR4 = (R5-R4)/(R5+R4) 
with R{\it i} denoting the background-subtracted count rate in the 
standard band {\it i} (with energy boundaries as defined above and starting with band 1 at the 
lowest energies).

We present the results of our spectral analysis below. 
For the results on the timing analysis we refer to Sections~3, 4 and 5,
where we divide the sources into Be/X-ray binaries with newly discovered pulse periods, pulsars
with previously known periods and Be/X-ray binaries without detected pulse periods. Within each 
section the sources are sorted by field (i.e. in time) and within each field by right ascension.

\begin{figure*}
\sidecaption
  \includegraphics[width=12cm,clip=]{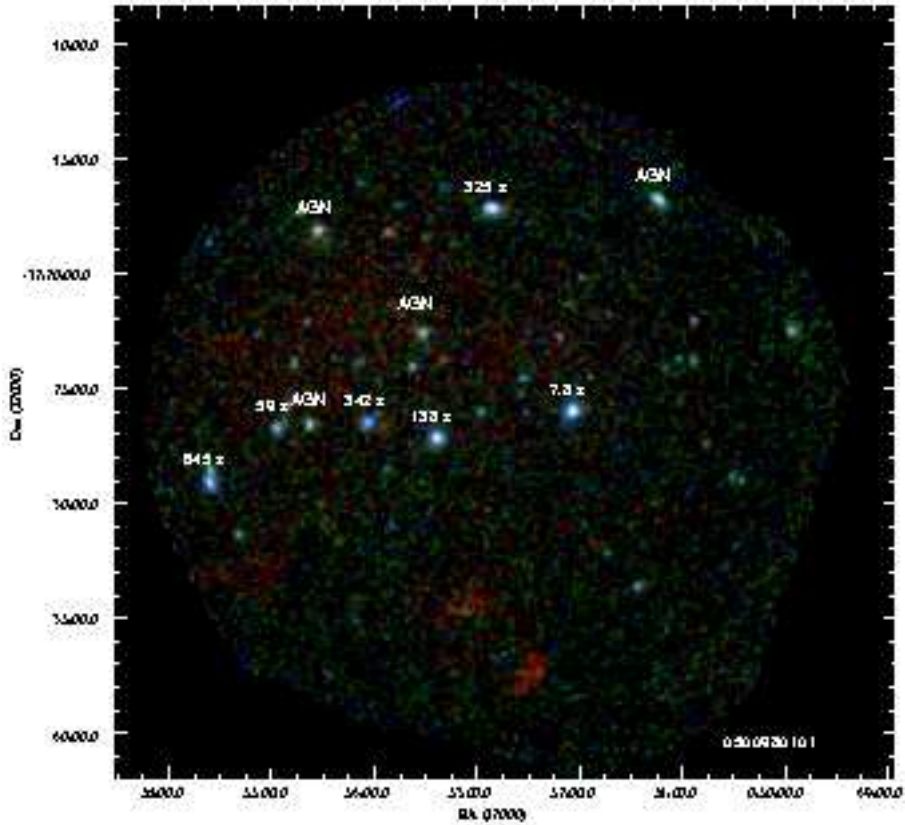}
  \caption{EPIC RGB colour image of the field 0500980101, composed of images from the three 
           energy bands 0.2$-$1.0 keV (red), 1.0$-$2.0 keV (green) and 2.0$-$4.5 keV (blue) 
	   and from all three EPIC instruments.
	   The individual images are exposure corrected and EPIC-PN out-of-time events are subtracted.
	   The ten brightest X-ray sources in the image are marked, the Be/X-ray binary pulsars
	   by their pulse periods. Four likely AGN exhibit power-law X-ray spectra steeper than those of 
	   HMXBs. They appear less blue in the colour representation of the image. The southern of the 
	   two patches of diffuse emission seen in the lower part of the image is related to a supernova 
	   remnant (SNR B0050-72.8) known from radio and optical wavelengths, the other one might also be a 
	   SNR, however, without radio and optical emission \citep{2008A&A...acc.....F}.}
  \label{fig-image}
\end{figure*}

We show the combined EPIC image obtained from observation 0500980101 in Fig.~\ref{fig-image}. 
This field is located in the central bar region, north of field 404680201 with a small overlap.
In the EPIC images of this observation we detected the three known Be/X-ray binary pulsars
SMC\,X-3 \citep[7.79 s;][]{2004ATel..225....1E}, 
CXOU\,J005323.8-722715 \citep[138 s;][]{2004MNRAS.353.1286E}
and XTE\,J0055-724  \citep[=~1SAX\,J0054.9-7226 = 1WGA\,J0054.9-7226 = RX\,J0054.9-7226, 59 s;][]{1998IAUC.6818R...1M}.
During this observation, three additional, previously uncatalogued X-ray sources were detected:
XMMU\,J005252.1-721715, XMMU\,J005403.8-722632 and XMMU\,J005535.2-722906.
All three exhibit X-ray spectra typical for high mass X-ray binaries and their optical counterparts 
are consistent with Be stars.
Moreover, we detect X-ray pulsations from all three, clearly identifying them as Be/X-ray binary pulsars. 
Thus, from the ten brightest sources detected in this field six are Be/X-ray binary pulsars. The other four
are most likely active galactic nuclei (AGN) due to their steeper power-law X-ray spectra and faint 
optical counterparts. One of them is listed as HMXB candidate \citep[SMC 36 in ][]{2005MNRAS.362..879S} from \xmm\ 
data. However, our new observation yielded a factor of four more counts from this source and we obtain 
a spectral index of 1.5$\pm$0.2, probably too steep for a HMXB.

For each source, the EPIC spectra (in some cases MOS spectra are not available) were simultaneously 
fit with a power-law model (which in most cases yields acceptable fits) allowing for a normalization 
factor between the spectra from different cameras. We included two absorption components in
our spectral model, accounting for  the Galactic foreground absorption 
\citep[with a fixed hydrogen column density of 6\hcm{20} and with elemental abundances from][]{2000ApJ...542..914W} 
and the SMC absorption \citep[with column density as free parameter in the fit and with metal abundances 
reduced to 0.2 as typical for the SMC;][]{1992ApJ...384..508R}.

For twenty sources 
\citep[adding the results for XMMU\,J004814.1-731003 from ][]{2008arXiv0803.2473H}
the derived best fit parameters for the 
power-law model together with observed fluxes and source luminosities are summarised in 
Table~\ref{tab-spectra}. The table also lists net exposure times for each EPIC instrument. 
Example spectra together with their best fit model are plotted in  Fig.~\ref{fig-spectra}.
To estimate luminosities for the detected sources which did not yield a sufficient number 
of counts for a spectrum, we converted the count rates, obtained by the source detection 
analysis, assuming an appropriate spectral shape 
(using spectral parameters from the same source from a different observation when 
available or from a nearby Be/X-ray binary to at least get an estimate for the 
absorbing column density; see below and the sections on the individual sources).
The luminosities are listed in Table~\ref{tab-spectra} and more details can be 
found in the description of each individual source in the next sections.

\begin{figure*}
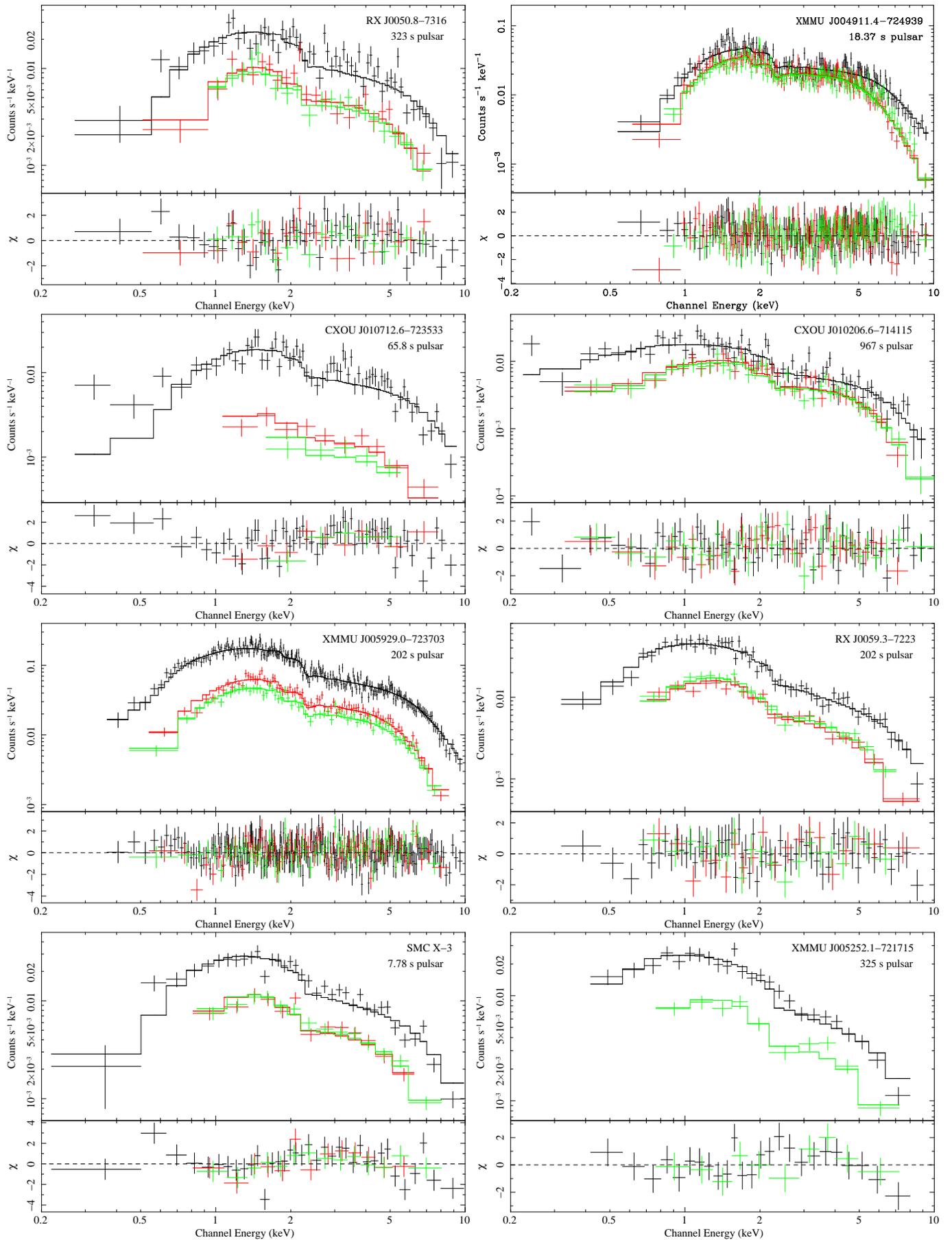

  \hbox{
  \resizebox{0.49\hsize}{!}{\includegraphics[angle=-90,clip=]{spectra_0404680301_1.ps}}
  \resizebox{0.49\hsize}{!}{\includegraphics[angle=-90,clip=]{spectra_0403970301_1.ps}}
  }
  \hbox{
  \resizebox{0.49\hsize}{!}{\includegraphics[angle=-90,clip=]{spectra_0404680501_1.ps}}
  \resizebox{0.49\hsize}{!}{\includegraphics[angle=-90,clip=]{spectra_0501470101_2.ps}}
  }
  \hbox{
  \resizebox{0.49\hsize}{!}{\includegraphics[angle=-90,clip=]{spectra_0500980201_1.ps}}
  \resizebox{0.49\hsize}{!}{\includegraphics[angle=-90,clip=]{spectra_0500980201_2.ps}}
  }
  \hbox{
  \resizebox{0.49\hsize}{!}{\includegraphics[angle=-90,clip=]{spectra_0500980101_1.ps}}
  \resizebox{0.49\hsize}{!}{\includegraphics[angle=-90,clip=]{spectra_0500980101_3.ps}}
  }
  \caption{EPIC spectra of selected Be/X-ray binaries. EPIC-PN is shown in black and 
  EPIC-MOS in red (M1) and green (M2) (both grey in black and white representation). The histograms show
  the best-fit absorbed power-law model.}
  \label{fig-spectra}
\end{figure*}
\addtocounter{figure}{-1}
\begin{figure*}
  \hbox{
  \resizebox{0.49\hsize}{!}{\includegraphics[angle=-90,clip=]{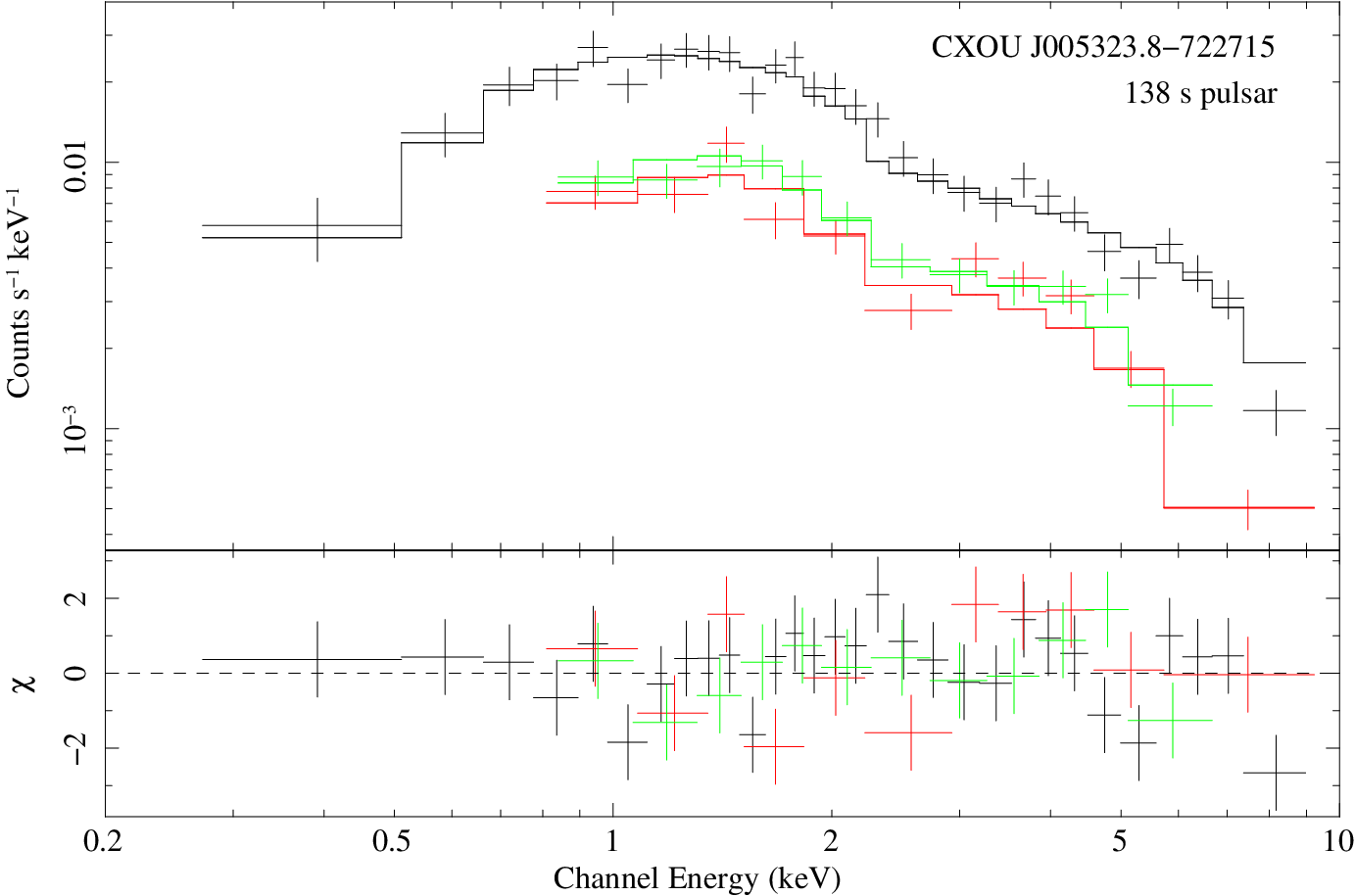}}
  \resizebox{0.49\hsize}{!}{\includegraphics[angle=-90,clip=]{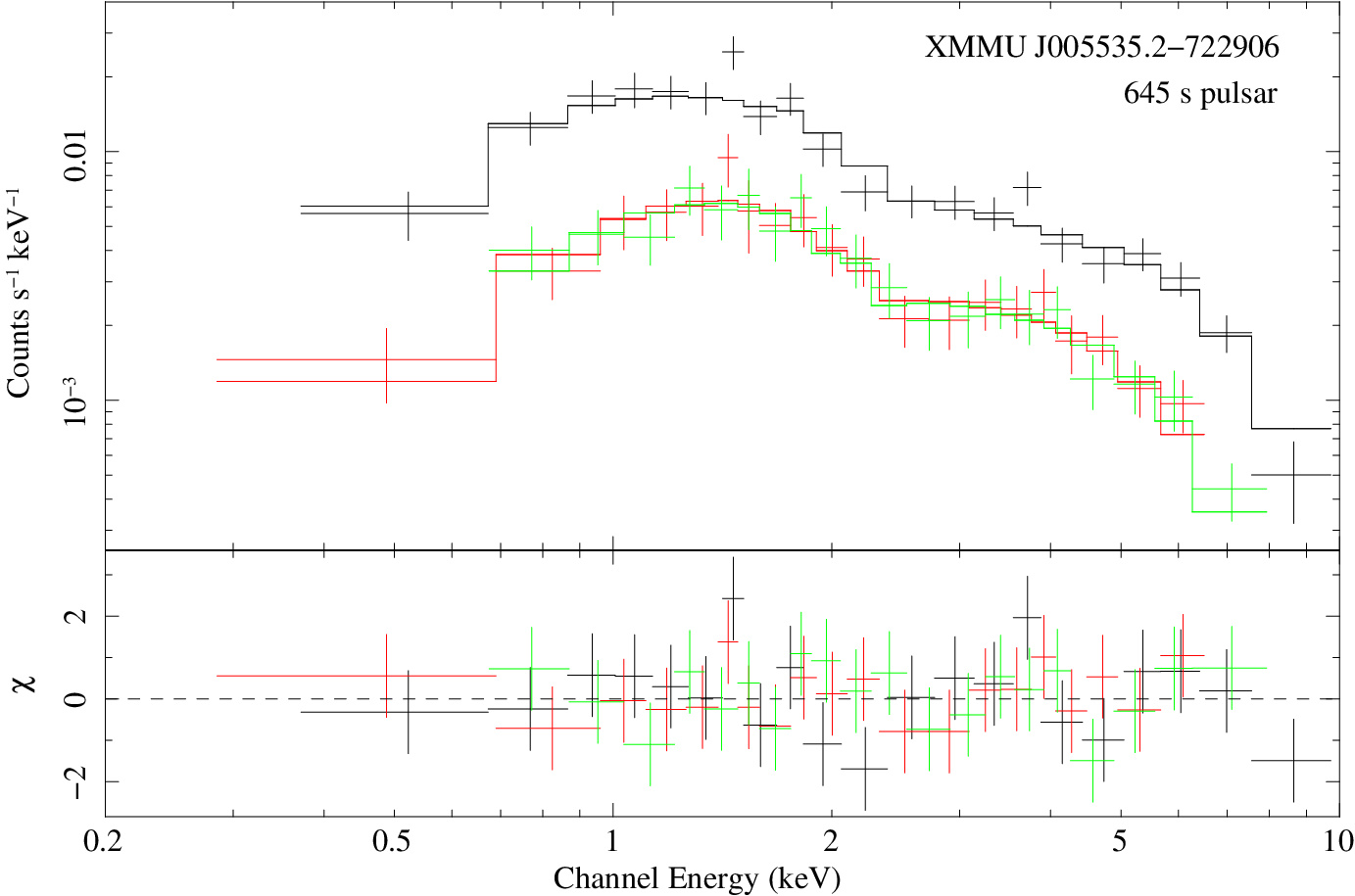}}
  }
  \caption{Continued.}
\end{figure*}

\begin{table*}
\caption[]{Spectral fit results for the power-law model.}
\begin{center}
\begin{tabular}{lcccccrrrc}
\hline\hline\noalign{\smallskip}
\multicolumn{1}{l}{Source} &
\multicolumn{1}{c}{SMC \nh} &
\multicolumn{1}{c}{Photon} &
\multicolumn{1}{c}{Flux$^{(a)}$} &
\multicolumn{1}{c}{L$_{\rm x}^{(b)}$} &
\multicolumn{1}{c}{$\chi^2_{\rm r}$/dof} &
\multicolumn{3}{c}{Net exposure time (s)} &
\multicolumn{1}{c}{SMC \nhi$^{(c)}$}\\
\multicolumn{1}{c}{} &
\multicolumn{1}{c}{[\oexpo{21}cm$^{-2}$]} &
\multicolumn{1}{c}{Index} &
\multicolumn{1}{c}{erg cm$^{-2}$ s$^{-1}$} &
\multicolumn{1}{c}{erg s$^{-1}$} &
\multicolumn{1}{c}{} &
\multicolumn{1}{c}{PN} &
\multicolumn{1}{c}{MOS1} &
\multicolumn{1}{c}{MOS2} &
\multicolumn{1}{c}{[\oexpo{21}cm$^{-2}$]} \\

\noalign{\smallskip}\hline\noalign{\smallskip}
XMMU\,J004723.7-731226         & 7.4$^{+2.4}_{-1.0}$    & 0.97$^{+0.05}_{-0.10}$ & 7.8\expo{-13} &  3.7\expo{35}  &  0.82/110 & 16550 & 18498 & 18508 &  9.6 \\
XMMU\,J004814.1-731003$^{(d)}$ & 52$\pm$20              & 1.33$\pm$0.27          & 3.5\expo{-13} &  2.1\expo{35}  &  0.82/27  & 19044 & 22760 & 22772 & 10.8 \\
RX\,J0049.2-7311               & 13$^{+14}_{-8}$        & 0.96$^{+0.34}_{-0.23}$ & 3.4\expo{-13} &  1.7\expo{35}  &  1.08/41  & 16563 & -     & 18520 & 10.6 \\
RX\,J0049.5-7310               & 9.4$^{+6.9}_{-4.4}$    & 0.83$^{+0.24}_{-0.21}$ & 4.3\expo{-13} &  2.0\expo{35}  &  0.58/23  & 16563 & -     & 18520 & 10.4 \\
\noalign{\smallskip}\hline\noalign{\smallskip}
CXOU\,J005455.6-724510         & 6.2$^{+4.5}_{-1.2}$    & 0.85$^{+0.30}_{-0.14}$ & 2.9\expo{-13}  &  1.3\expo{35} &  1.20/15  & 27174 & 30324 & 30365 &  6.8 \\
\noalign{\smallskip}\hline\noalign{\smallskip}
XMMU\,J004723.7-731226         & 5.6$^{+3.8}_{-2.5}$    & 1.00$\pm$0.17          & 7.8\expo{-13}  &  3.7\expo{35} &  1.13/33  & 17728 & 24785 & 24813 &  9.6 \\
XMMU\,J004814.1-731003         &                        &                        &                &  9.1\expo{33} & & \multicolumn{3}{l}{faint}       & 10.8 \\
RX\,J0048.5-7302               &                        &                        &                &  1.0\expo{34} & & \multicolumn{3}{l}{faint}       & 11.1 \\
AX\,J0049-729	               &                        &                        &                &  1.1\expo{35} & & \multicolumn{3}{l}{faint}       &  9.4 \\
XMMU\,J004911.4-724939         & 19.4$\pm$1.8           & 0.65$\pm$0.05          & 1.2\expo{-11}  &  5.5\expo{36} &  1.02/364 & 17682 & 24804 & 24820 &  9.3 \\
RX\,J0049.2-7311               &                        &                        &                &  2.2\expo{34} & & \multicolumn{3}{l}{faint}       & 10.6 \\
\noalign{\smallskip}\hline\noalign{\smallskip}
RX\,J0049.5-7331               & 0.2$^{+1.6}_{-0.2}$    & 0.70$^{+0.16}_{-0.13}$ & 4.0\expo{-13}  &  1.8\expo{35} &  0.98/22  & 14231 & 17408 & 17458 &  6.6 \\
RX\,J0049.7-7323               & 10.1$^{+3.0}_{-2.4}$   & 0.92$\pm$0.11          & 9.6\expo{-13}  &  4.5\expo{35} &  1.17/59  & 14235 & -     & -     &  7.5 \\
RX\,J0050.8-7316               & 6.5$^{+2.3}_{-0.9}$    & 0.81$^{+0.10}_{-0.09}$ & 1.1\expo{-12}  &  5.3\expo{35} &  1.22/96  & 14228 & 17414 & 17451 &  8.0 \\
AX\,J0051.6-7311               &                        &                        &                &     & & \multicolumn{3}{l}{at edge of FOV, faint} &  7.3 \\
RX\,J0052.1-7319               &                        &                        &                &  4.0\expo{34} & & \multicolumn{3}{l}{faint}       &  8.1 \\
\noalign{\smallskip}\hline\noalign{\smallskip}
CXOU\,J010712.6-723533         & 7.1$^{+3.9}_{-0.3}$    & 0.74$^{+0.12}_{-0.11}$ & 1.6\expo{-12}  &  7.3\expo{35} &  1.69/73  & 19121 & 22871 & 22916 &  6.4 \\
\noalign{\smallskip}\hline\noalign{\smallskip}
CXOU\,J010206.6-714115         & 0.15$^{+0.47}_{-0.15}$ & 0.71$^{+0.07}_{-0.06}$ & 2.1\expo{-12}  &  9.3\expo{35} &  0.93/135 & 20222 & 24276 & 24320 &  3.5 \\
\noalign{\smallskip}\hline\noalign{\smallskip}
CXOU\,J005736.2-721934         &                        &                        &                &     & & \multicolumn{3}{l}{at edge of FOV, faint} &  6.3 \\
RX\,J0058.2-7231               & 13.5$\pm$3.2           & 1.27$\pm$0.12          & 8.3\expo{-13}  &  4.4\expo{35} &  0.97/48  & 23810 & 28132 & 28198 &  8.3 \\
RX\,J0059.3-7223               & 3.4$\pm$0.7            & 1.20$\pm$0.06          & 8.1\expo{-13}  &  3.9\expo{35} &  0.78/95  & 23816 & 28203 & 28205 &  7.5 \\
XMMU\,J005929.0-723703         & 5.4$\pm$0.5            & 0.90$\pm$0.03          & 9.6\expo{-12}  &  4.5\expo{36} &  1.05/364 & 23777 & 28212 & 28256 &  7.6 \\
RX\,J0101.8-7223               & 4.6$\pm$1.6            & 1.01$\pm$0.09          & 8.4\expo{-13}  &  3.9\expo{35} &  0.94/49  & 23798 & 28163 & 28189 &  5.4 \\
\noalign{\smallskip}\hline\noalign{\smallskip}
SMC\,X-3                       & 5.0$^{+1.7}_{-1.4}$    & 0.92$\pm$0.07          & 9.3\expo{-13}  &  4.3\expo{35} &  1.49/58  & 20107 & 24016 & 24031 &  2.9 \\
XMMU\,J005252.1-721715         & 1.4$^{+1.0}_{-0.8}$    & 1.05$\pm$0.09          & 7.2\expo{-13}  &  3.3\expo{35} &  1.10/31  & 20109 & -     & 24034 &  4.0 \\
CXOU\,J005323.8-722715         & 2.9$^{+1.2}_{-0.9}$    & 0.97$\pm$0.07          & 6.6\expo{-13}  &  3.0\expo{35} &  1.26/48  & 20101 & 24016 & 24031 &  4.1 \\
XMMU\,J005403.8-722632         & 0.6$^{+5.5}_{-0.6}$    & 0.37$^{+0.19}_{-0.12}$ & 3.5\expo{-13}  &  1.5\expo{35} &  1.37/26  & 20095 & 24016 & 24031 &  3.7 \\
XTE\,J0055-724                 &                        &                        &                &  9.2\expo{34} & & \multicolumn{3}{l}{faint}       &  5.1 \\
XMMU\,J005535.2-722906         & 2.7$^{+1.3}_{-1.0}$    & 0.85$^{+0.10}_{-0.09}$ & 1.1\expo{-12}  &  4.9\expo{35} &  0.69/56  & 20090 & 24004 & 24050 &  5.3 \\
\noalign{\smallskip}\hline\noalign{\smallskip}
\end{tabular}
\end{center}
$^{(a)}$ Observed 0.2-10.0 keV flux.
$^{(b)}$ Source intrinsic X-ray luminosity in the 0.2-10.0 keV band (corrected for absorption)
for a distance to the SMC of 60 kpc \citep{2005MNRAS.357..304H}.
$^{(c)}$ Total SMC \Hone\ column density from \citet{1999MNRAS.302..417S}.
$^{(d)}$ From \citet{2008arXiv0803.2473H}.
\label{tab-spectra}
\end{table*}

\section{New Be/X-ray binary pulsars}
\label{sect-newpuls}

In this section we present the results of our analysis of the new pulsars.
This includes CXOU\,J010206.6-714115 for which we revise the period; RX\,J0058.2-7231, a 
former candidate Be/X-ray binary known from ROSAT; the new transient XMMU\,J005929.0-723703 
which is the second 202~s 
pulsar in the SMC (both are only 13.8\arcmin\ apart); and the three previously unknown sources
XMMU\,J005252.1-721715, XMMU\,J005403.8-722632 and XMMU\,J005535.2-722906 which were discovered in
field 0500980101.

\subsection{A revised period of 967~s for the Be/X-ray binary pulsar CXOU\,J010206.6-714115}

We included observation 0501470101 into our analysis to investigate the X-ray properties of 
CXOU\,J010206.6-714115 for which a pulse period of 700$\pm$34~s was claimed \citep{2007MNRAS.376..759M}. 
The error on the period is large and the value is close to the Chandra dithering period of 707 s.

We detected a source in the EPIC images (the second brightest source after the supersoft 
source RX J0058.6-7136) within 0.31\arcsec\ of the Chandra source. 
From the EPIC data we find a strong peak in the power spectrum at 1.037\expo{-3} Hz (Fig.~\ref{xmmp-pspec}). 
Bayesian ana\-ly\-sis yields 966.97$\pm$0.47~s for the pulse period. 
Pulse profiles in the standard energy bands are presented in Fig.~\ref{xmmp-pulse} which show a modulation
with a broad peak at all energies. The X-ray spectrum shows relatively low absorption and the
power-law index of 0.7 indicates a somewhat harder spectrum than the average SMC Be/X-ray binaries
(Table~\ref{tab-spectra}, Fig.~\ref{fig-spectra}). 
The X-ray position obtained from the EPIC data coincides exactly with that of the optical 
star from the MCPS (Table~\ref{tab-ids}) which was suggested as the optical counterpart of 
CXOU\,J010206.6-714115 \citep{2007MNRAS.376..759M}. 

Although we can not confirm the 700~s period, reported for CXOU\,J010206.6-714115, 
it is extremely unlikely that the EPIC (observation date 2007 June 4) and Chandra (2006 Feb. 6)
detections do not arise from the same source. 
We therefore re-analysed the Chandra data of CXOU\,J010206.6-714115. Using the standard level 2 
event file, which is cleaned for CCD columns with higher background than adjacent columns, we 
produced images in detector coordinates. The images show missing columns which are crossed
by the source during the dithering motion. Timing analysis based on these data reveals a 
period of $\sim$700~s. We then created a new level 2 event file that did not reject these 
columns (keeping all other standard screening criteria). Timing analysis of the new data 
does not show the periodicity at 700~s, suggesting that the period found in the standard 
level 2 data is caused  by the satellite dithering. 
In the new dataset, we searched without success for the periodic signal at 967~s, found 
in the \xmm\ data. This may be caused by the exposure time of 9.4 ks, which covers only 
less than 10 periods.

\subsection{Discovery of 291~s pulsations from RX\,J0058.2-7231}

The power spectrum of the combined EPIC data revealed peaks at the fundamental frequency 
of 3.44\expo{-3} Hz, at the first harmonic, which dominates in power, and at the 
third harmonic (Fig.~\ref{xmmp-pspec}).
The resulting pulse period from the Bayesian analysis is 291.327$\pm$0.057 s.
This period is close to the values found for XTE\,J0051-727, which was discovered with periods 
of 293.9$\pm$0.4 s and 292.7$\pm$0.4~s on 2004, April 19 and April 22, respectively 
\citep{2004ATel..273....1C}.
The \xmm\ position is 32\arcmin\ away from the published RXTE position, outside the formal error box,
but still inside the field of view of the PCA. XTE\,J0051-727 and RX\,J0058.2-7231 are very likely
the same source. The pulsar was detected by RXTE several times in 2003 and 2004 with periods around 
293~s. After the last outburst which lasted longer than one week, the period had decreased to 
$\sim$291~s \citep{2008arXiv0802.2118G}, similar to the value measured in our \xmm\ observation.
The pulse profiles obtained from the EPIC data (Fig.~\ref{xmmp-pulse}) show a broad peak below 2.0 keV 
which changes to a depression at higher energies.
\citet{1999AJ....117..927S} suggested a V=14.9 mag Be star as optical counterpart for this ROSAT 
HRI source \citep[source 76 in][]{2000A&AS..147...75S}. Our improved \xmm\ position confirms
the identification with this star and its optical magnitudes and colours are listed in Table~\ref{tab-ids}.
The spectral type was determined to B0IIIe by \citet{2006A&A...457..949M}.

\subsection{XMMU\,J005929.0-723703, a new transient 202~s Be/X-ray binary pulsar}

The \xmm\ observation 0500980201 revealed a previously unreported source for which we 
assign the name XMMU\,J005929.0-723703. It was the brightest source in the field and yielded
the highest number of counts from all sources in our sample.
During the observation XMMU\,J005929.0-723703 showed strong flaring activity on timescales of an hour 
with intensity increases up to a factor of three. The broad band EPIC-PN light curve is shown 
in Fig.~\ref{xmmp-lcurve}. 
A Fourier timing analysis of the EPIC-PN data revealed the presence of strong peaks at four harmonic 
frequencies with the fundamental frequency corresponding to a period of 202.52$\pm$0.02~s (Fig.~\ref{xmmp-pspec}).

Pulse profiles and the corresponding hardness ratio variations versus pulse phase are 
shown in Figs.~\ref{xmmp-pulse} and \ref{xmmp-hr}. 
The pulse profile of XMMU\,J005929.0-723703 is dominated by two broad main pulses which show additional
sub-structure. The shape and the relative intensity of the main pulses are energy dependent.
While at energies below $\sim$2 keV the two main pulses are different in shape and amplitude, they become
more and more similar at higher energies. This causes the highest power in the 
power spectrum in the first harmonic at 101 s. Significant differences in HR3 between the main 
pulses indicate a change in the intrinsic source spectrum during the pulse period.

From the X-ray position we identify a V~14.9 Be star as optical counterpart (Table~\ref{tab-ids}).
This star ([MA93]\,1147) is listed in the 2dF survey of the SMC with a spectral type 
of B0-5(III)e \citep{2004MNRAS.353..601E}. The light curves obtained from the 
MACHO database show a long-term brightness
increase with $\sim$0.05 mag variations on top of it (Fig.~\ref{fig-macho}). We applied an 
FFT analysis to the MACHO R-band data \citep{1976Ap&SS..39..447L,1982ApJ...263..835S} 
to look for periodicity. The power spectrum shows the strongest peaks  
at 334 days and 220 days (with power of 26 and 16, respectively). 
Although similarly long orbital periods were suggested for SMC Be/X-ray 
binaries \citep{2008arXiv0802.2118G}, the broad peaks indicate more quasi-periodic variations 
as they might be expected from changes in the Be disk.

\subsection{XMMU\,J005252.1-721715, a new transient 325~s Be/X-ray binary pulsar}
\label{sect-newtra}

Three previously unreported sources were detected in the EPIC data of observation 0500980101,
the first is designated XMMU\,J005252.1-721715. 
The power spectrum (Fig.~\ref{xmmp-pspec}) shows a single peak and the
Bayesian analysis yields 325.36$\pm$0.59~s for the period.
The pulse profiles (Fig.~\ref{xmmp-pulse}) reveal significant modulation mainly below 2.0 keV.
The period of this pulsar is close to that of AX\,J0051-733 (=~RX\,J0050.7-7316) with 
323.2$\pm$0.4~s detected in ASCA data in November 1997 \citep{1998IAUC.6853....2Y}. The 
angular separation of the two pulsars is 59.6\arcmin.
During the preparation of this paper \citet{2008arXiv0803.3941C} reported an optical and
RXTE X-ray study of a new X-ray binary pulsar (period $\sim$327~s) based on a detection in 
Chandra data. The Chandra position is consistent with that of XMMU\,J005252.1-721715 and the 
similar periods confirm that the same new Be/X-ray binary pulsar was independently detected in 
Chandra and \xmm\ data. For information on the optical counterpart we refer to 
\citet{2008arXiv0803.3941C}.

\subsection{XMMU\,J005403.8-722632, a new transient 342~s Be/X-ray binary pulsar}

The second transient source from observation 0500980101 also shows a single peak in the 
power spectrum (Fig.~\ref{xmmp-pspec}). 
The pulse period was determined to 341.87$\pm$0.15 s.
The pulse profile (Fig.~\ref{xmmp-pulse}) is characterised by a broad, nearly triangular-shaped
peak at all energies.
The pulse period of XMMU\,J005403.8-722632 is close to the value of SAX\,J0103.2-7209 
\citep[345.2$\pm$0.1~s detected in BeppoSAX data of July 1998;][]{1998IAUC.6999....1I}.
The two pulsars are separated by 45.2\arcmin.
\citet{2008arXiv0802.2118G} investigated the spin history of the pulsar SXP348 obtained from 
the RXTE SMC monitoring program. They identify this pulsar with SAX\,J0103.2-7209.
The large scatter seen in the pulse periods may, however, be caused by the RXTE detection of two different 
pulsars, SAX\,J0103.2-7209 and our new pulsar XMMU\,J005403.8-722632.

As optical counterpart of XMMU\,J005403.8-722632 we identify a star with V$\sim$14.9 
(Table~\ref{tab-ids}) which shows large (0.7 mag, 0.5 mag) variations in the MACHO 
(R-band, B-band) light curves (Fig.~\ref{fig-macho}). 
This star (entry 28942 in \citet{2002ApJS..141...81M}) was also proposed as possible 
counterpart of an INTEGRAL source with a potential periodicity of 6.8~s \citep{2007MNRAS.382..743M}. 
If this periodicity is confirmed, the star can not be the optical counterpart of 
the INTEGRAL source. On the other hand, if the periodicity is spurious, 
XMMU\,J005403.8-722632 might be the X-ray counterpart of the INTEGRAL source. 
XMMU\,J005403.8-722632 shows the hardest power-law X-ray spectrum of all our 
investigated Be/X-ray binaries in the SMC, which could explain a detection by 
INTEGRAL.

\subsection{XMMU\,J005535.2-722906, a new transient 645~s Be/X-ray binary pulsar}

\begin{figure*}
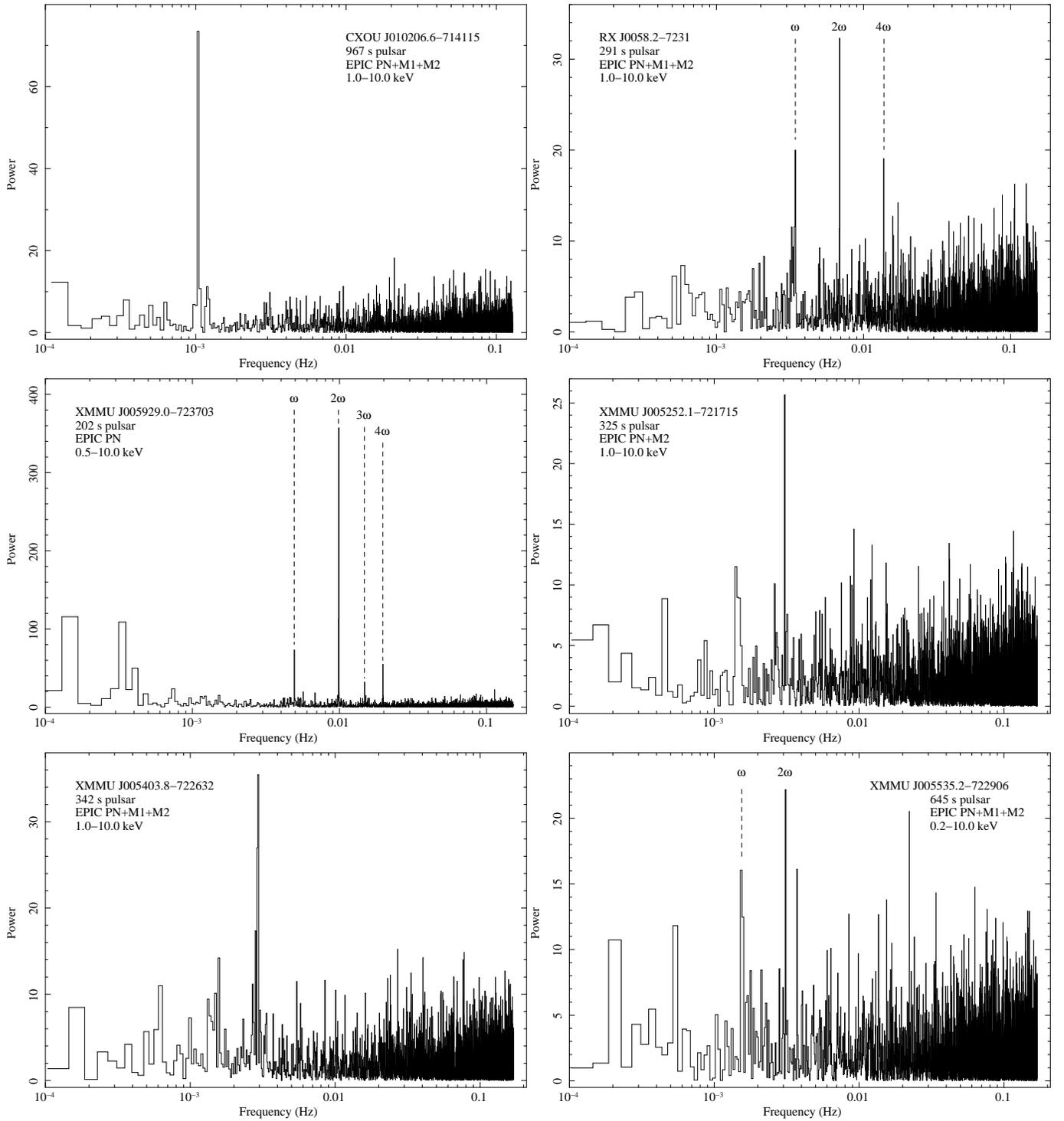

  \hbox{
  \resizebox{0.49\hsize}{!}{\includegraphics[angle=-90,clip=]{P0501470101_PN_M1_M2_IEVLI0000_sc2_1000_10000_powspec.ps}}
  \resizebox{0.49\hsize}{!}{\includegraphics[angle=-90,clip=]{P0500980201_PN_M1_M2_IEVLI0000_sc4_1000_10000_powspec.ps}}
  }
  \vspace{1mm}
  \hbox{
  \resizebox{0.49\hsize}{!}{\includegraphics[angle=-90,clip=]{P0500980201PNS003PIEVLI0000_sc1_500_10000_powspec.ps}}
  \resizebox{0.49\hsize}{!}{\includegraphics[angle=-90,clip=]{P0500980101_PN_M2_IEVLI0000_sc3_1000_10000_powspec.ps}}
  }
  \vspace{1mm}
  \hbox{
  \resizebox{0.49\hsize}{!}{\includegraphics[angle=-90,clip=]{P0500980101_PN_M1_M2_IEVLI0000_sc5_1000_10000_powspec.ps}}
  \resizebox{0.49\hsize}{!}{\includegraphics[angle=-90,clip=]{P0500980101_PN_M1_M2_sc4_200_10000_powspec.ps}}
  }
  \caption{Power spectra of newly discovered pulsars obtained from the EPIC data in broad
           energy bands. To improve statistics in most cases the three instruments PN, MOS1 and MOS2,
	   when available, were combined for the timing analysis.}
  \label{xmmp-pspec}
\end{figure*}

\begin{figure*}
  \hbox{
  \resizebox{0.33\hsize}{!}{\includegraphics[clip=]{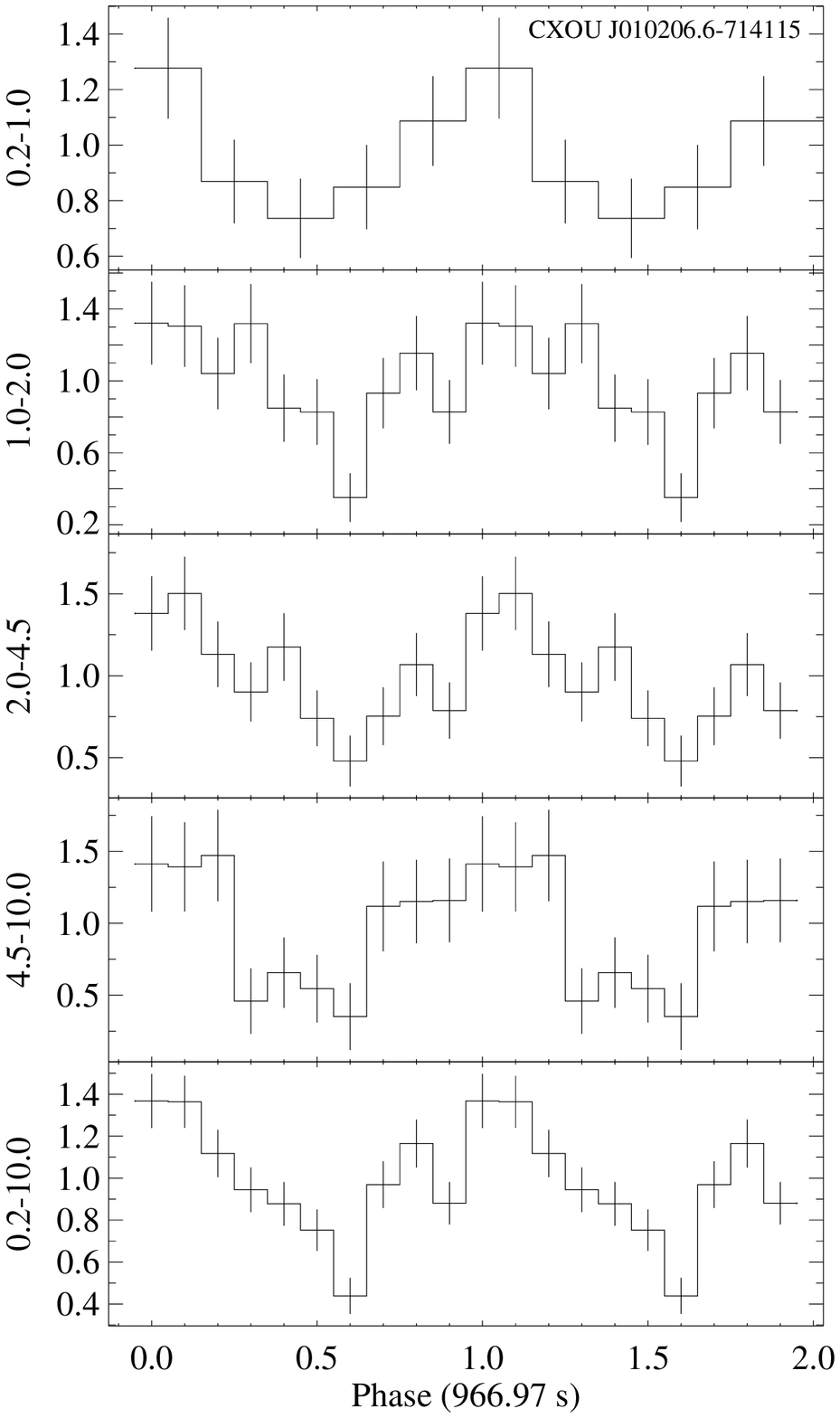}}
  \resizebox{0.33\hsize}{!}{\includegraphics[clip=]{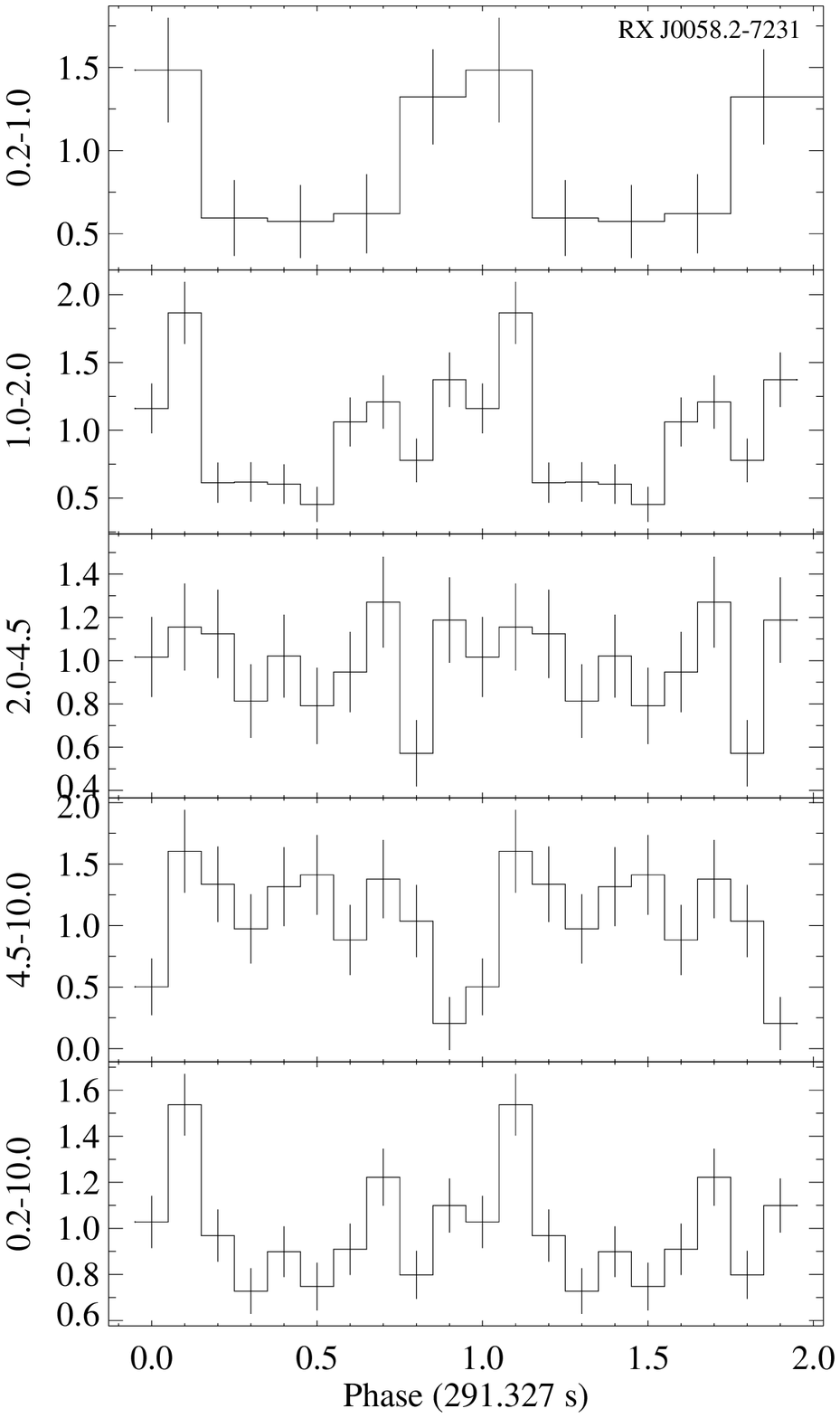}}
  \resizebox{0.33\hsize}{!}{\includegraphics[clip=]{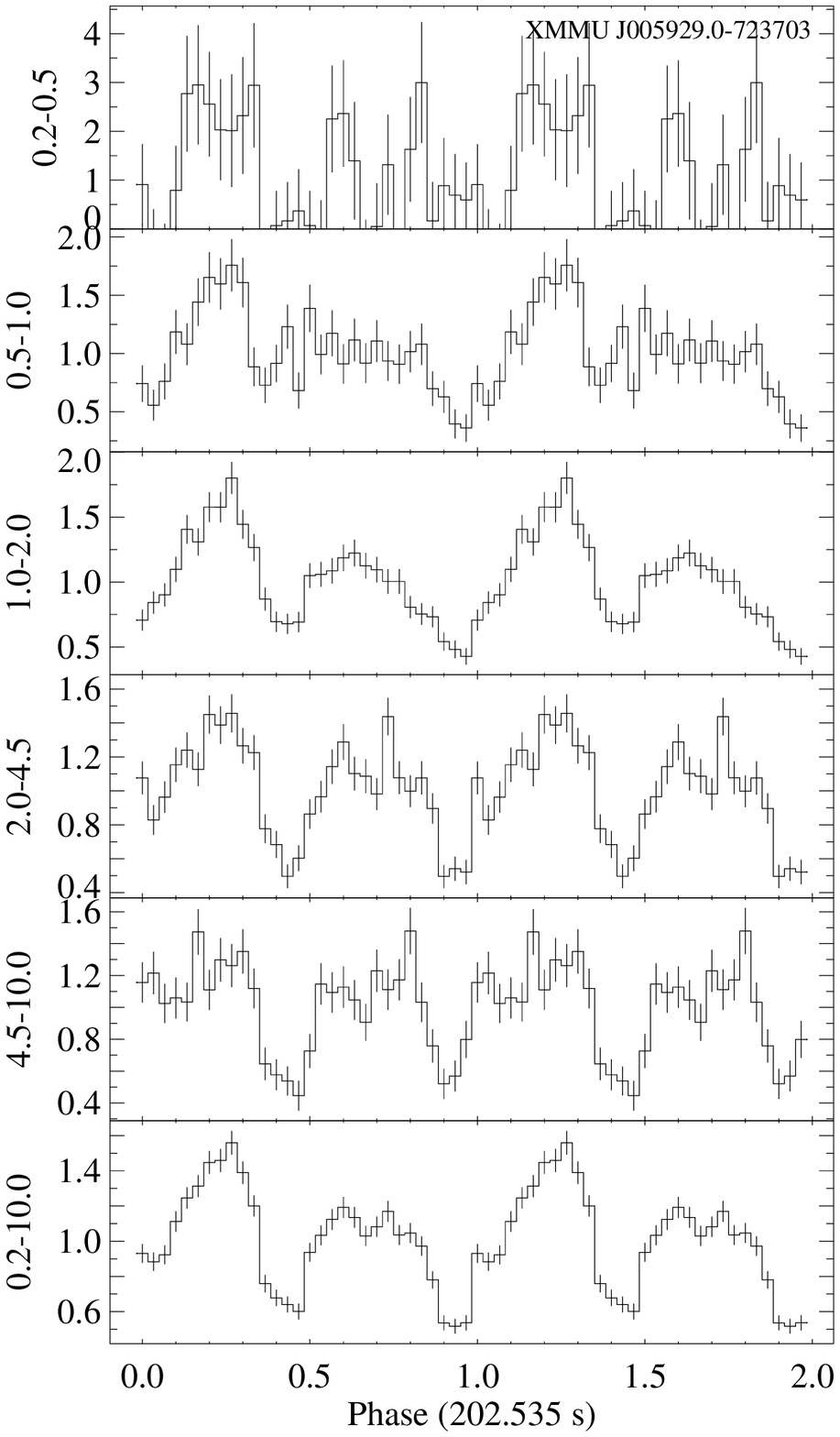}}
  }
  \hbox{
  \resizebox{0.33\hsize}{!}{\includegraphics[clip=]{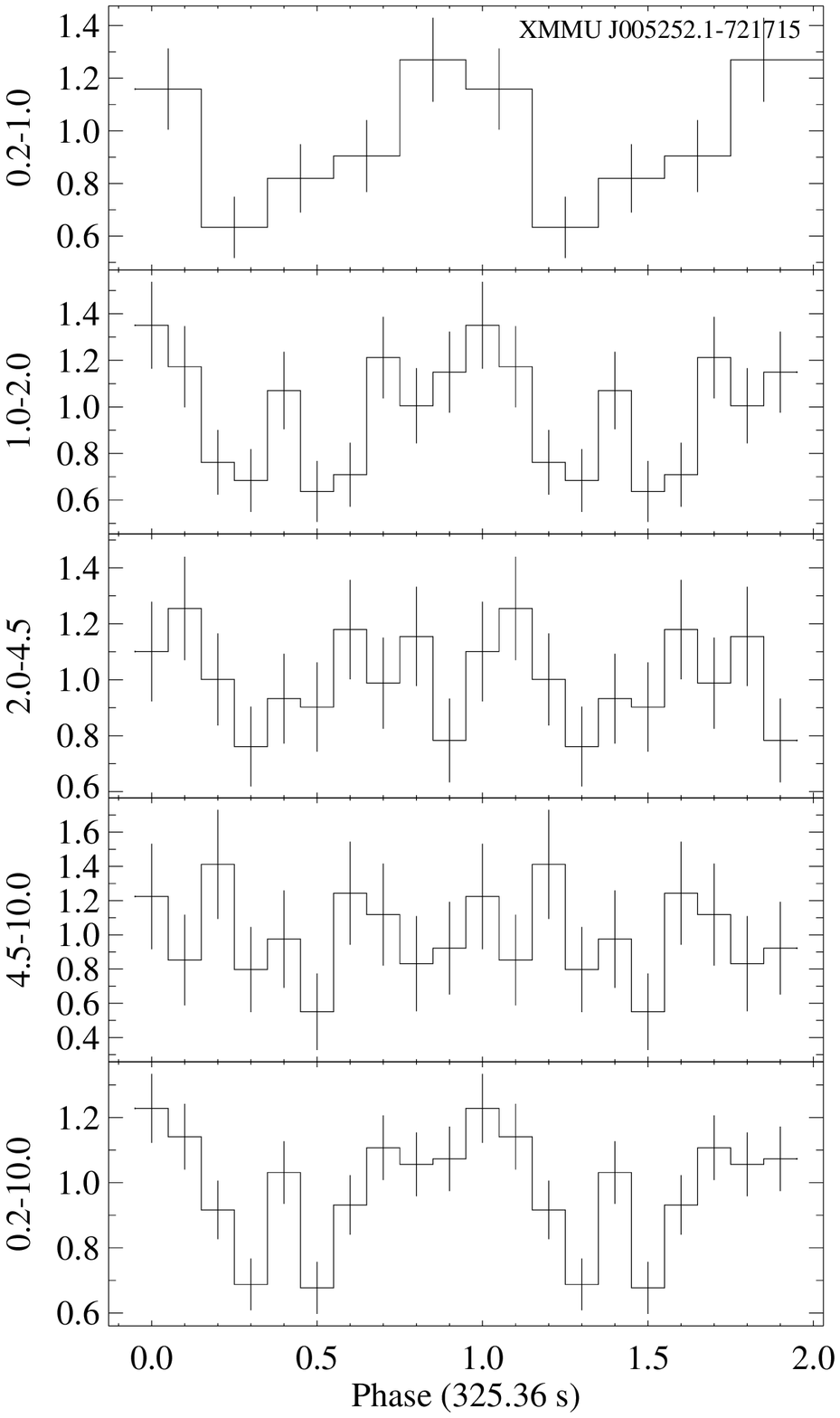}}
  \resizebox{0.33\hsize}{!}{\includegraphics[clip=]{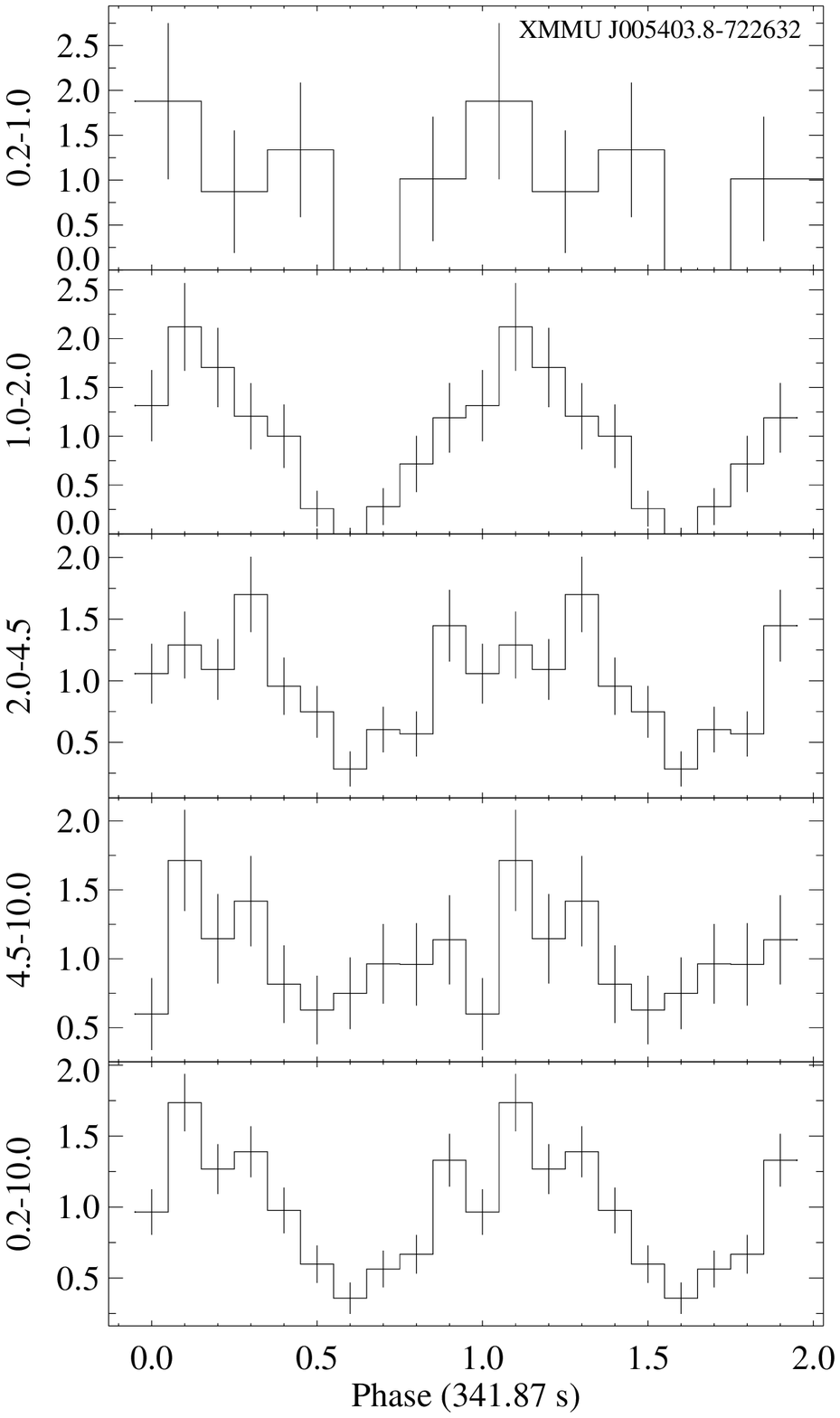}}
  \resizebox{0.33\hsize}{!}{\includegraphics[clip=]{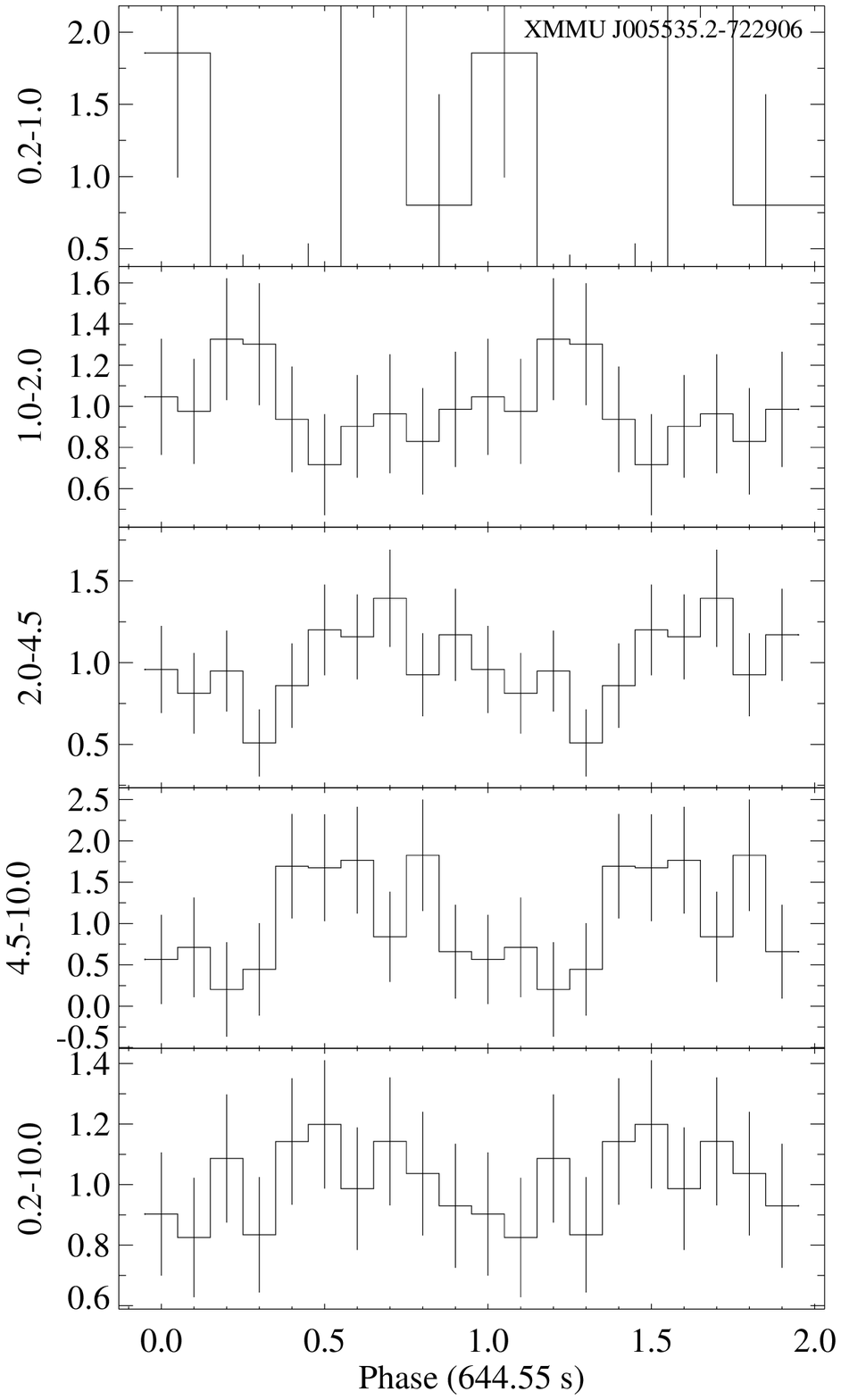}}
  }
  \caption{Folded EPIC-PN light curves of newly discovered pulsars in the standard EPIC energy bands. The panels
           show the pulse profiles for the different energies specified in keV. The intensity profiles 
           are background subtracted and normalised to the average count rate, as listed in Table~\ref{tab-rates}.}
  \label{xmmp-pulse}
\end{figure*}

This source is the third transient detected in the data from observation 0500980101.
The power spectrum shows peaks at the fundamental frequency and its first harmonic 
(Fig.~\ref{xmmp-pspec}). The pulse period is 644.55$\pm$0.72 s. The pulse profiles
show relatively shallow modulation with broad peaks (Fig.~\ref{xmmp-pulse}). The
profiles below and above 2.0 keV are nearly anti-correlated in intensity.

As optical counterpart we identify a star with V$\sim$14.6 (Table~\ref{tab-ids}) which shows 
a sudden brightness increase by $\sim$0.85 mag and $\sim$0.6 mag in the MACHO R- and B-band
light curves. A fast decline in the OGLE I-band data after the end of the MACHO observations
suggests a return to a lower brightness level (Fig.~\ref{fig-macho}). 
Such a behaviour is seen from about 15\% of the (single) Be stars in the SMC  
\citep[type-2 in][]{2002A&A...393..887M}.

\begin{table*}
\caption[]{Average EPIC-PN count rates of Be/X-ray binary pulsars.}
\begin{center}
\begin{tabular}{lrrrrr}
\hline\hline\noalign{\smallskip}
\multicolumn{1}{l}{Source} &
\multicolumn{1}{c}{0.2$-$1.0 keV}$^{(a)}$ &
\multicolumn{1}{c}{1.0$-$2.0 keV} &
\multicolumn{1}{c}{2.0$-$4.5 keV} &
\multicolumn{1}{c}{4.5$-$10.0 keV} &
\multicolumn{1}{c}{0.2$-$10.0 keV} \\

\noalign{\smallskip}\hline\noalign{\smallskip}
XMMU\,J004723.7-731226  & 3.6      & 13.5 & 15.5 & 8.6 & 41.1 \\
%
CXOU\,J005455.6-724510  & 2.1      & 2.5  & 3.3  & 1.8  & 10.0 \\
%
XMMU\,J004911.4-724939  & 1.7+6.9  & 53.7 & 79.8 & 62.9 & 206  \\
RX\,J0049.7-7323        & 8.3      & 40.8 & 51.9 & 40.0 & 142  \\
RX\,J0050.8-7316        & 4.8      & 13.2 & 19.3 & 11.0 & 67.8 \\
%
CXOU J010712.6-723533   & 5.4      & 14.8 & 23.4 & 14.6 & 58.5 \\
%
CXOU\,J010206.6-714115  & 10.3     & 12.9 & 15.1 &  8.7 & 47.0 \\
%
XMMU\,J005929.0-723703  & 3.6+43   & 138  & 140  &  93  & 419  \\
RX\,J0059.3-7223        & 2.4+15.3 & 33.5 & 27.1 & 15.8 & 94.4 \\
RX\,J0058.2-7231        &  5.4     & 18.7 & 17.7 &  9.6 & 51.1 \\
SMC\,X-3                & 8.0      & 21.5 & 26.5 & 15.7 & 73.0 \\
XMMU\,J005252.1-721715  & 12.4     & 18.0 & 17.0 &  7.3 & 54.6 \\
CXOU\,J005323.8-722715  &  8.0     & 19.5 & 20.2 & 12.4 & 61.4 \\
XMMU\,J005403.8-722632  &  1.4     &  4.8 &  8.3 &  6.4 & 21.4 \\
XMMU\,J005535.2-722906  & $-$      &  7.8 &  8.7 &  3.5 & 20.1 \\
\noalign{\smallskip}\hline
\end{tabular}
\end{center}

Count rates in \oexpo{-3} counts s$^{-1}$ from EPIC-PN. $^{(a)}$ For most sources the two energy bands
0.2$-$0.5 keV and 0.5$-$1.0 keV were combined for statistical reasons. For three cases the count rates
in the two bands are listed separately.
\label{tab-rates}
\end{table*}

\begin{figure*}
  \hbox{
  \resizebox{0.49\hsize}{!}{\includegraphics[clip=]{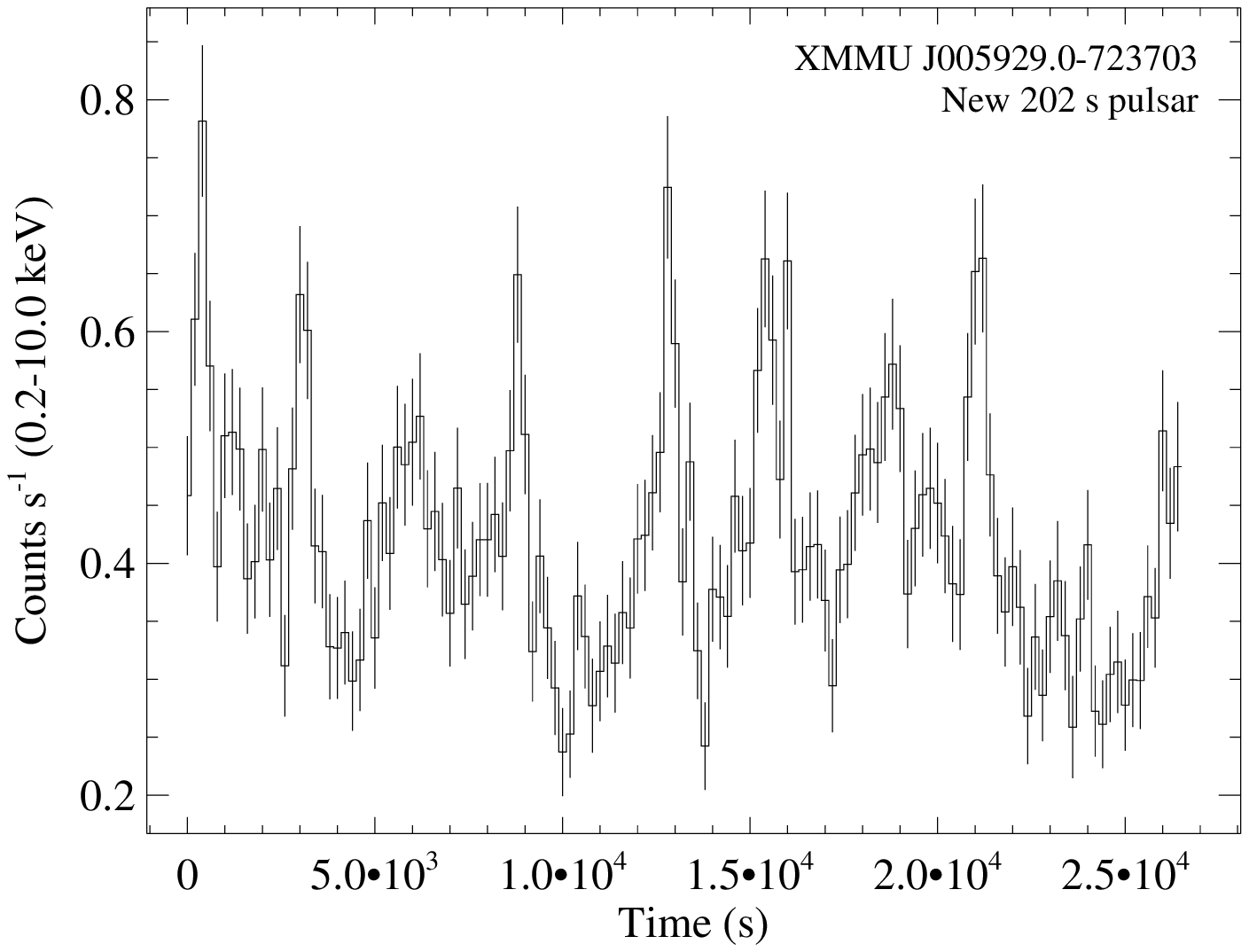}}
  \resizebox{0.49\hsize}{!}{\includegraphics[clip=]{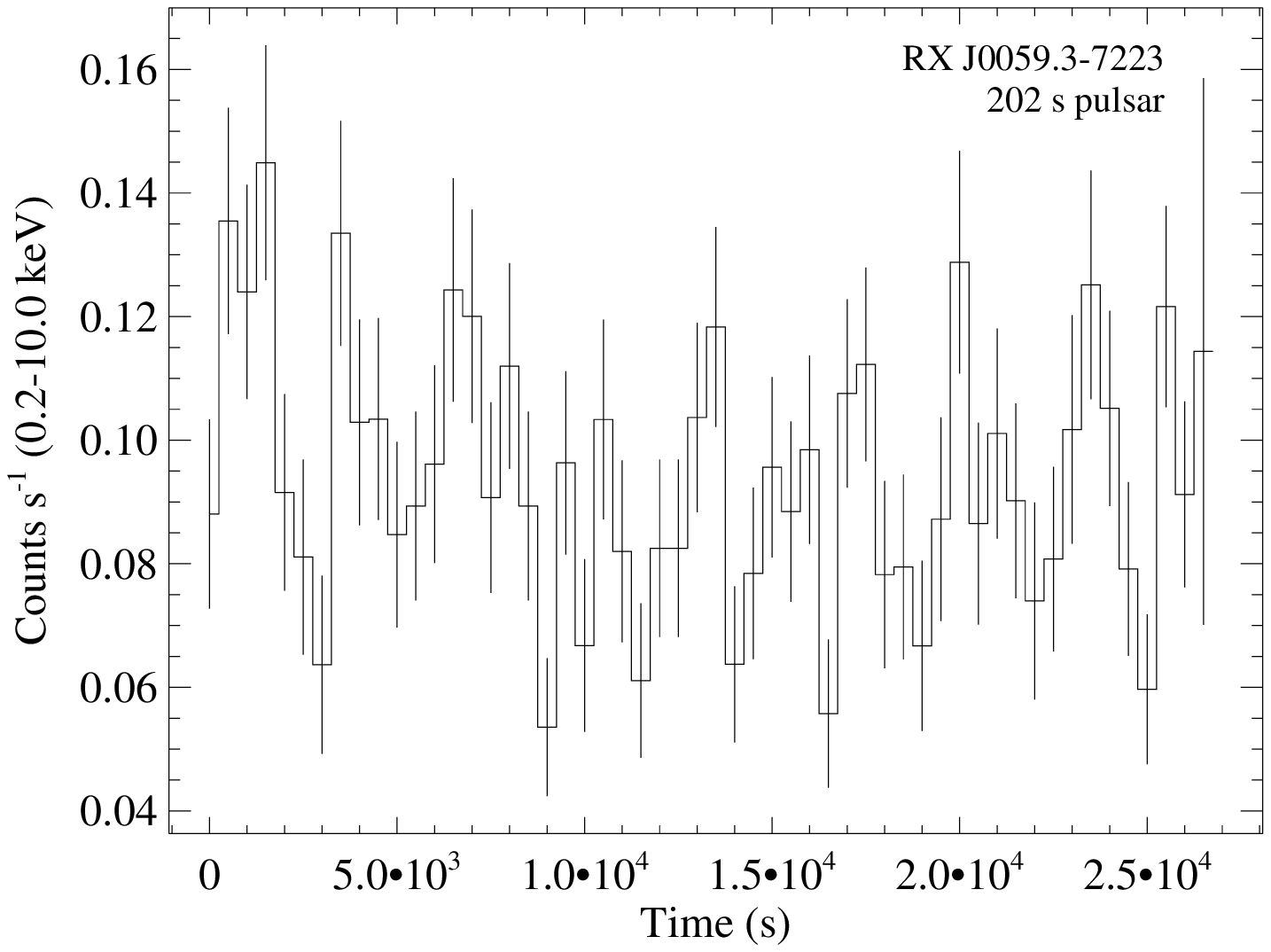}}
  }
  \caption{Broad-band (0.2$-$10.0 keV) EPIC-PN light curves of the two 202~s pulsars observed in field 0500980201. 
           The data is background subtracted and binned to 200~s and 500~s for XMMU\,J005929.0-723703
	   and RX\,J0059.3-7223, respectively. Time 0 corresponds to MJD 54257.3925.}
  \label{xmmp-lcurve}
\end{figure*}

\begin{figure*}
  \hbox{
  \resizebox{0.45\hsize}{!}{\includegraphics[angle=-90,clip=]{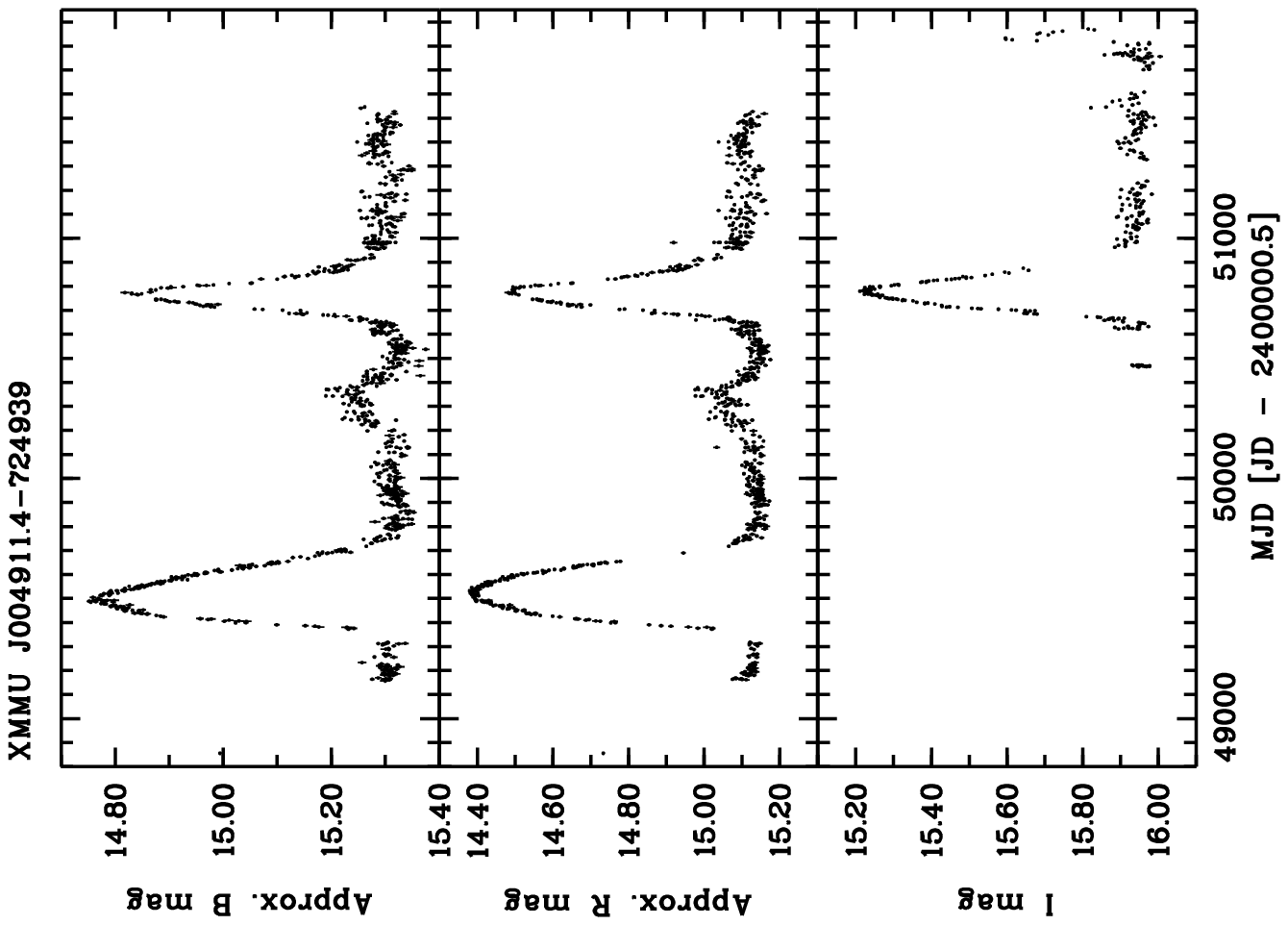}}
  \hspace{0.05\hsize}
  \resizebox{0.45\hsize}{!}{\includegraphics[angle=-90,clip=]{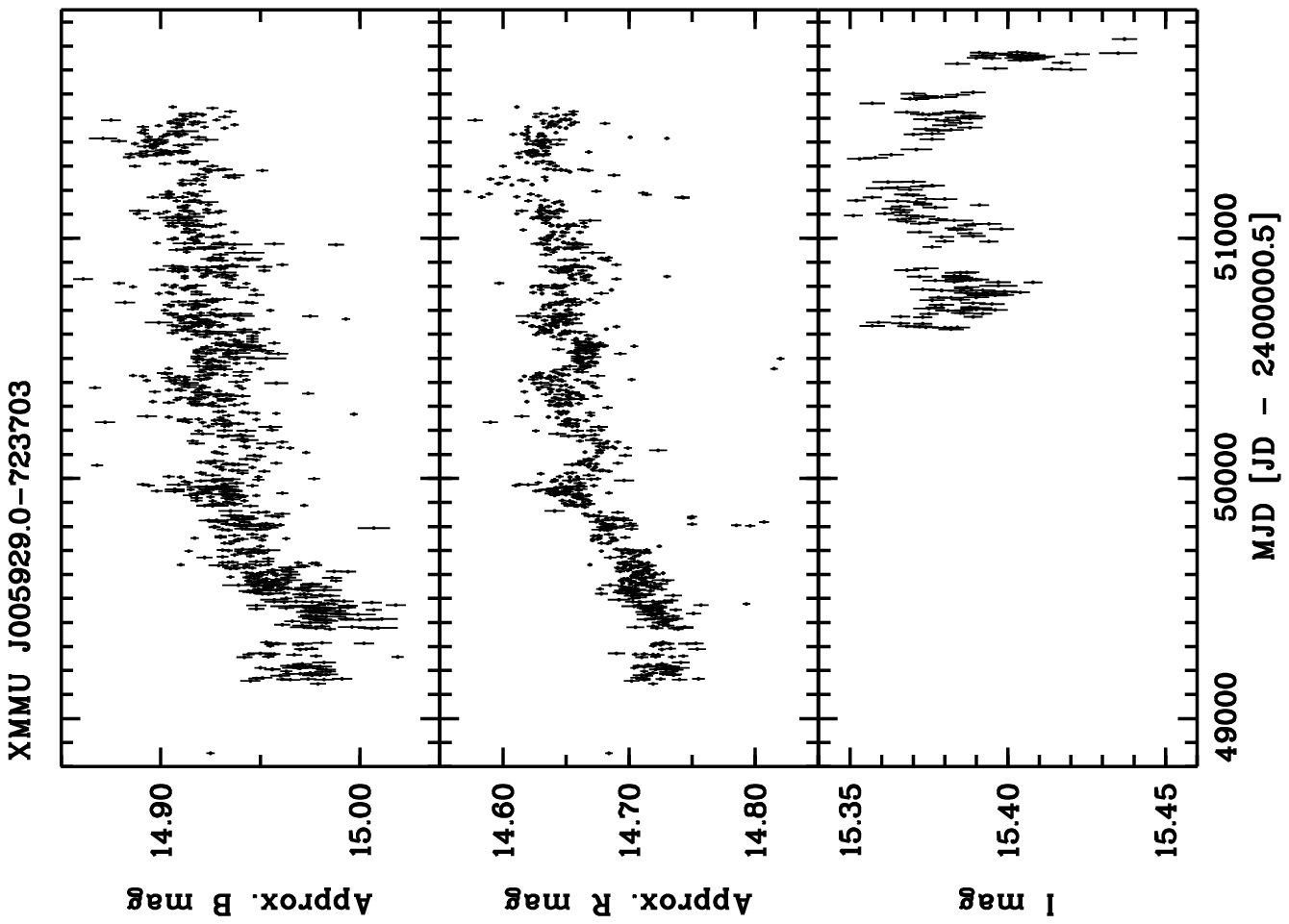}}
  }
  \hbox{
  \resizebox{0.45\hsize}{!}{\includegraphics[angle=-90,clip=]{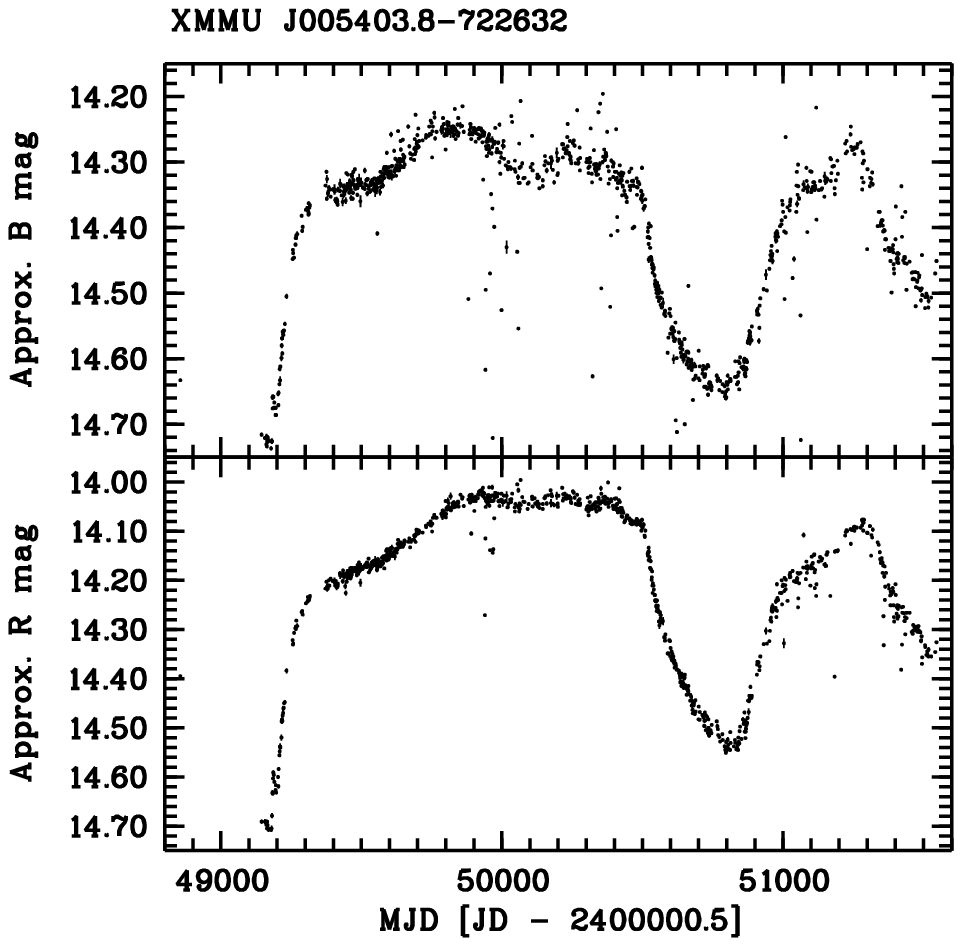}}
  \hspace{0.05\hsize}
  \resizebox{0.45\hsize}{!}{\includegraphics[angle=-90,clip=]{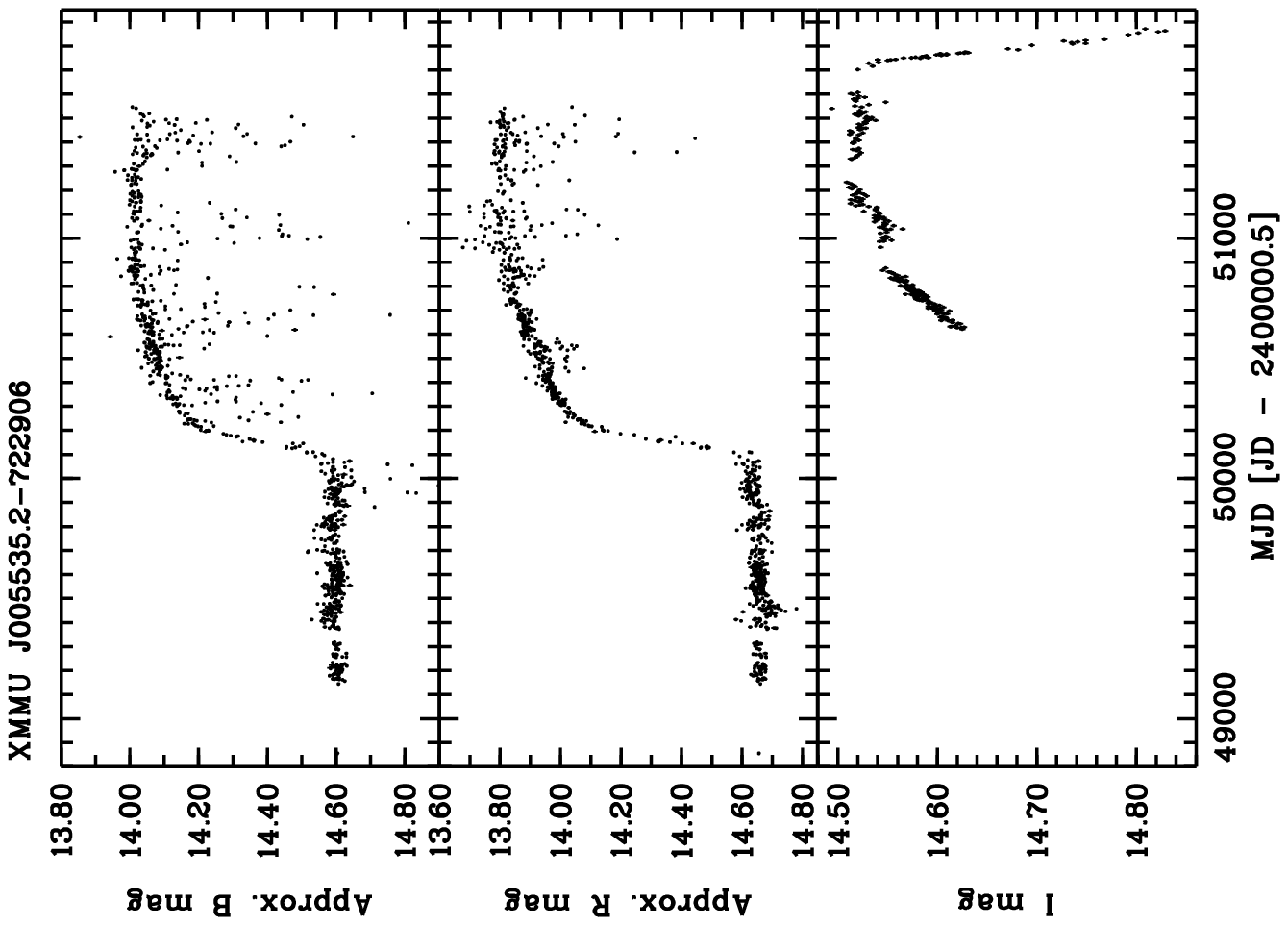}}
  }
  \caption{MACHO B- and R- and OGLE I-band (when available) light curves of the optical 
           counterparts of XMMU\,J004911.4-724939 and three new Be/X-ray binary 
	   pulsars. The optical counterparts
	   are MACHO\,208.15911.13 and OGLE\,65500 for XMMU\,J004911.4-724939,
	       MACHO\,207.16541.15 and OGLE\,139407 for XMMU\,J005929.0-723703, 
	       MACHO\,207.16202.21 for XMMU\,J005403.8-722632, and
	       MACHO\,207.16315.28 and OGLE\,137527 for XMMU\,J005535.2-722906.
	   For the MACHO data only data points with errors smaller than 
	   0.01 mag are drawn. Large excursions in the MACHO data which are not present 
	   in the OGLE data, are probably not real.}
  \label{fig-macho}
\end{figure*}

\section{Be/X-ray binaries with previously known pulse period}
\label{sect-oldpuls}

\subsection{The 263~s Be/X-ray binary pulsar XMMU\,J004723.7-731226}
\label{sect-lxharda}

\citet{2000A&A...359..573H} proposed RX\,J0047.3-7312 (= AX\,J0047.3-7312) as HMXB candidate 
due to its possible identification with the \Halp\ emission line star [MA93]\,172. 
The detection in a first \xmm\ observation on 2000 Oct. 15 supported the optical identification 
\citep[][designated the source XMMU\,J004723.7-731226]{2004A&A...414..667H}.
A finding chart for the optical counterpart of XMMU\,J004723.7-731226 was published by
\citet[][SXP264 in Fig.~1]{2005MNRAS.356..502C}, who suggested as counterpart the northern 
star in a crowded region with at least three stars. 

Our \xmm\ observations covered XMMU\,J004723.7-731226 twice.
The new, astrometrically improved X-ray positions (Table~\ref{tab-sources})
are not compatible with the proposed counterpart by \citet{2005MNRAS.356..502C}, 
but indicate the star more southwest as counterpart. This 
star also corresponds to the MCPS and OGLE stars listed in Table~\ref{tab-ids}, while the entry 
from the UBVR catalogue does not resolve the stars. 
From a previous \xmm\ observation on 2000 Oct. 15, pulsations with a period of 263.64$\pm$0.30~s 
were detected.
The new measurement of 262.23$\pm$0.66~s in Oct. 2006 indicates a slight spin-up.
In the March 2007 observation we were not able to detect any significant periodicity
from XMMU\,J004723.7-731226 (neither in the power spectra nor by folding techniques). 
This may be caused by the relative noisy structure of the pulse 
profile (Fig.~\ref{xmmo-pulse}) in combination with background flaring activity during the 
observation 0403970301. 
The X-ray luminosities during our observations on 2006 Oct. 5 and 2007 March 12 
(Table~\ref{tab-spectra}) were nearly identical, but almost a factor of 5 lower than
on 2000 Oct. 15. 
The X-ray spectrum was harder when the source was brighter (photon index of 0.76$\pm$0.04 on 
2000 Oct. 15). 
The deep dip in the broad band pulse profile present on 2000 October 15 \citep[Fig.~3 of][]{2004A&A...414..667H} 
is no longer seen anymore on 2006 Oct. 5, but this might be caused by the lower statistics in the 
new observation. Overall, the pulse profiles appear highly structured (Fig.~\ref{xmmo-pulse}).

\subsection{The 25.5~s Be/X-ray binary pulsar XMMU\,J004814.1-731003}

This pulsar was discovered in observation 0404680101 and first results were published by
\citet{2008arXiv0803.2473H}. The results from the spectral analysis are repeated in 
Table~\ref{tab-spectra}. XMMU\,J004814.1-731003 was fainter during observation 0403970301, 
about five months later (the data became public after the discovery paper). 
Scaling from the EPIC-PN 0.2$-$4.5 keV count rate 
(4.6\ct{-2} compared to 2.0\ct{-3}) with the assumption of constant spectral shape, the 
source luminosity decreased by a factor of 23. Due to the faintness of the source no pulsations 
were detected in the new observation.

\subsection{The 500~s Be/X-ray binary pulsar CXOU\,J005455.6-724510}

The pulse period of 503.5$\pm$6.7~s from this pulsar was first reported by \citet{2004ATel..217....1E} 
from a Chandra observation on 2002, July 4. This period was confirmed by \citet{2004A&A...420L..19H}
from \xmm\ data taken on 2003, Dec. 18. 
Pulse profiles and spectra were presented from the EPIC data and the pulse period during the 
\xmm\ observation was determined to 500.0$\pm$0.2 s and the source luminosity to 4\ergs{35} (0.5$-$10 keV).

During the new \xmm\ observation (0404680201) in Nov. 2006, the source luminosity was somewhat lower 
at 1.3\ergs{35}. The EPIC spectra are of low statistical quality. They can formally be fit with a 
power-law (see Table~\ref{tab-spectra}), although 
there is indication for a soft component as seen during the first \xmm\ observation.
The pulse period has further decreased to 497.5$\pm$1.0. Pulse profiles are shown in Fig.~\ref{xmmo-pulse}.

\subsection{The 74.7~s Be/X-ray binary pulsar AX\,J0049-729}

Pulsations from AX\,J0049-729 with a period of 74.7~s were first reported from RXTE data 
\citep{1998IAUC.6803....1C} and then confirmed by ASCA \citep{1998IAUC.6835....2Y}. 
The identification of the ASCA source with RX\,J0049.1-7250 \citep{1998IAUC.6840....1K} allowed 
the identification with a Be star \citep{1999MNRAS.309..421S}. 
Our improved X-ray position confirms object 1 in their finding chart as the correct counterpart.

AX\,J0049-729 was detected as faint source during observation 0403970301. The 
detected counts are insufficient for a spectral or temporal analysis.
Adopting the same X-ray spectral shape as the nearby 18.37~s pulsar XMMU\,J004911.4-724939,
scaling the EPIC-PN 0.2$-$4.5 keV count rates (8.8\ct{-1} compared to 1.8\ct{-2})
yields a luminosity of 1.1\ergs{35} (Table~\ref{tab-spectra}).

\subsection{The 18.37~s Be/X-ray binary pulsar XMMU\,J004911.4-724939}

During the XMM-Newton observation 0403970301, a bright X-ray transient, 
designated XMMU J004911.4-724939, was detected \citep{2008ATel.1453....1H},
only 83\arcsec\ away from AX\,J0049-729.
Our timing analysis revealed X-ray pulsations with a period of 18.3814$\pm$0.0001~s.
The pulse period of XMMU\,J004911.4-724939 is consistent with that of XTE\,J0055-727 
\citep[18.37$\pm$0.01;][]{2003ATel..214....1C}. This indicates, that these two sources 
are identical (the source position is 31\arcmin\ away from the RXTE pointing position, well 
within the PCA collimator response).
The pulsar XTE J0055-727 was first detected in RXTE data 
\citep{2003ATel..214....1C} when it became more active than the years before 2003
\citep{2008arXiv0802.2118G}. Although several searches for this pulsar were 
performed \citep[e.g., ][]{2004A&A...420L..19H}, it never was detected
due to the limited FOV of the imaging instruments.
The \xmm\ position and period of XMMU\,J004911.4-724939 allowed us to identify a 
Be star as optical counterpart also for XTE\,J0055-727.

We show the X-ray pulse profiles in Fig.~\ref{xmmo-pulse}. No significant pulsations 
are seen below 0.5~keV, while at higher energies a double peaked profile (0.5-1.0 keV)
develops into one with a single peak above 4.5 keV. The counting statistics
is sufficient to investigate the pulse phase dependent hardness ratios 
(Fig.~\ref{xmmp-hr}). The variations seen at the high energies indicate changes in the
intrinsic spectrum of the source.
 
The EPIC pulse-phase averaged X-ray spectra are well represented by a strongly 
absorbed hard power-law (Table~\ref{tab-spectra} and Fig.~\ref{fig-spectra}) with 
the highest luminosity in our source sample.
The X-ray position allowed us to identify the pulsar with a V = 16.0 mag Be star.
The boresight correction applied in the current work (Table~\ref{tab-sources}) 
brings the X-ray coordinates in full agreement with the position of the optical 
counterpart.
We extracted MACHO and OGLE light curves which are shown in Fig.~\ref{fig-macho}.
Two optical outbursts, lasting about 330 and 270 days with brightness increases 
by $\sim$0.55 and $\sim$0.50 mag (in MACHO approximate B magnitude), $\sim$0.8 mag 
and $\sim$0.6 (in approximate R) for the two outbursts, respectively, and 
$\sim$0.75 mag (in OGLE I) for the second outburst. 
Similar behaviour is also observed from other Be stars 
\citep{2002A&A...393..887M}. 
We conclude that XMMU J004911.4-724939 is a Be/X-ray binary pulsar in the SMC, and
identical to XTE J0055-727. \citet{2008ATel.1458....1C} reported a likely orbital period 
of 17.8 days from OGLE III data.

\subsection{The 755~s Be/X-ray binary pulsar RX\,J0049.7-7323}
\label{sect-olda}

\citet{2000A&A...359..573H} suggested RX\,J0049.7-7323 as candidate Be/X-ray binary because of
its possible counterpart, the emission line star [MA93]\,315. \Halp\ spectroscopy by 
\citet{2003MNRAS.338..428E} confirmed the Be/X-ray binary nature of RX\,J0049.7-7323.
Pulsations with a period of 755.5$\pm$0.6~s were discovered from AX\,J0049.4-7323 in a very long 
ASCA observation in April 2000 \citep{2000PASJ...52L..73Y}, with RX\,J0049.7-7323 within the error 
circle of the  ASCA source. From a first \xmm\ observation \citet{2004A&A...414..667H} confirmed
the association between the ROSAT and the ASCA source.

We detect RX\,J0049.7-7323 with a pulse period of 746.15$\pm$0.79~s (only PN data, source 
on CCD boundary in MOS) confirming the strong long-term spin-up trend of this pulsar. 
We re-determined the period from the first \xmm\ observation 
using the combined EPIC data to 752.14$\pm$0.43~s and show the spin period history in (Fig.~\ref{fig-phist}).
The period has decreased by nearly 10~s over 
7.0 years resulting in an average period derivative of 1.34~s~y$^{-1}$.
Pulse profiles are shown in Fig.~\ref{xmmo-pulse}.

\subsection{The 323~s Be/X-ray binary pulsar RX\,J0050.8-7316}
\label{sect-oldb}

RX\,J0050.8-7316 was identified with a variable Be star (V$\sim$15.4) by \citet{1997PASP..109...21C}.
Two period measurements from ASCA observations were reported from AX\,J0051-733 
\citet{1998IAUC.6853....2Y,2003PASJ...55..161Y}, which the authors identify with the ROSAT source 
by positional coincidence.
An accurate X-ray position and the detection of pulsations in \xmm\ data from 2000 Oct. 15 confirmed
this identification \citep{2004A&A...414..667H}.

During the new \xmm\ observation, RX\,J0050.8-7316 was detected with at a period of 317.44$\pm$0.26~s
continuing the spin-up seen before (Fig.~\ref{fig-phist}). The average period derivative between
the first ASCA and the last \xmm\ observation is 0.62~s~y$^{-1}$.
The RXTE period measurements show a large scatter which might partially be caused by different 
flux contributions of our newly discovered pulsar (Sect.~\ref{sect-newtra}) with similar period or by 
other nearby variable sources contributing to the X-ray flux.
In Fig.~\ref{xmmo-pulse} we present the pulse profiles which show a deep relatively
narrow feature with a energy-dependent shape.

\subsection{The 172~s Be/X-ray binary pulsar AX\,J0051.6-7311}

The 172~s pulsations of this source were first discovered in ASCA data 
\citep[AX\,J0051.6-7311 = RX\,J0051.9-7311;][]{2000IAUC.7428....3T,1997PASP..109...21C}, 
The source was right at the edge of the field of view of observation 0404680301 
(Fig.~\ref{fig-obs}), only marginally 
covered by MOS2. No sensible parameters could be derived.

\subsection{The 15.3~s Be/X-ray binary pulsar RX\,J0052.1-7319}

\citet{1999IAUC.7081....4L} detected pulsations from RX\,J0052.1-7319 in ROSAT data.
A faint source compatible with the position of this pulsar was detected in observation 
0404680301 with an EPIC-PN 0.2$-$4.5 keV count 
rate of 1.14\ct{-2} which translates into a luminosity of 4.0\ergs{34} (scaling to 
SMC\,X-3 with a count rate of 1.24\ct{-1}, i.e. assuming a power-law spectrum with $\gamma$ = 0.92 
and SMC \nh\ = 5\hcm{21}, justified due to the similar hardness ratios of the two sources).
The collected counts were also not sufficient for a period search.

\subsection{The 65.8~s Be/X-ray binary pulsar CXOU J010712.6-723533}
\label{sect-lxhardb}

Observation 0404680501 covers only one Be/X-ray binary, the pulsar CXOU\,J010712.6-723533.
It was discovered in the course of the SMC wing survey performed with Chandra on 2006 Feb. 10
with a period of 65.78$\pm$0.13~s and a luminosity of $\sim$3\ergs{36} \citep{2008MNRAS.383..330M}.
During the \xmm\ observation the source was detected as the brightest point source in 
the EPIC images with a luminosity of 6.0\ergs{35} (Table~\ref{tab-bbspectra}, see below), which is about 
four times fainter than during the Chandra observation. 
Although the absorption column density was not constrained
by the Chandra spectrum, the photon indices suggest a harder spectrum during the Chandra observation with higher
X-ray luminosity. From the EPIC data (2007 April 12) we determined the pulse period to 65.94$\pm$0.02~s, 
within the errors consistent with the Chandra measurement.
The pulse profiles (Fig.~\ref{xmmo-pulse}) are dominated by a broad peak.

The EPIC spectra of CXOU J010712.6-723533 are not well modelled by a simple power-law as the 
reduced $\chi^2$ indicates (the worst of our power-law model fits, Table~\ref{tab-spectra}). 
Systematic residuals (Fig.~\ref{fig-spectra}), similar to those observed from the 6.85~s pulsar 
XTE\,J0103-728 in the SMC \citep{2008arXiv0801.4679H} or, e.g., the Be/X-ray binary X\,Persei 
in the Milky Way \citep{2007A&A...474..137L}, indicate excess emission on top of the power-law. 
Adding a blackbody component yields similar properties as
those found for X\,Persei. We summarise characteristic parameters obtained from this model 
in Table~\ref{tab-bbspectra} and show the EPIC spectra with best fit model in Fig.~\ref{fig-bbspectra}.

\subsection{The 565~s Be/X-ray binary pulsar CXOU\,J005736.2-721934\ = XMMU\,J005735.7-721932}

This long-period pulsar was independently found as Be/X-ray binary in \xmm\ and Chandra observations.
\xmm\ observed it on 2000 October 17 and 2002 March 30 (in the same field as the 202~s pulsar 
RX\,J0059.3-7223, see next section). \citet{2003A&A...403..901S} suggested the emission line star [MA93]\,1020
as optical counterpart. The source was faint with 4.1\ergs{34} and 8.2\ergs{33}, respectively.
From a 100 ks Chandra observation on 2001 May 15, \citet{2003ApJ...584L..79M} reported the detection
of the pulse period of 564.8$\pm$0.4~s and a source luminosity of 6.2\ergs{34} (for 60 kpc).
The Chandra position was also consistent with the position of [MA93]\,1020. 
\Halp\ spectroscopy of the optical counterpart revealed a complicated profile of the 
\Halp\ line with four peaks \citep{2005MNRAS.356..502C} which the authors interpret as the 
start of a new period of mass ejection from the Be star.

During our new \xmm\ observation a source was detected right at the edge of the EPIC-PN field of view
(not covered by EPIC-MOS1) at a position compatible with that of the pulsar. There are too few counts 
for a temporal and spectral analysis.

\subsection{The 202~s Be/X-ray binary pulsar RX\,J0059.3-7223}

The ROSAT source RX\,J0059.3-7223\ was first classified as X-ray binary candidate by 
\citet{1999A&AS..136...81K} using hardness ratio criteria. It was observed twice by \xmm\ 
on 2000 October 17 and 2002 March 30 \citep{2003A&A...403..901S}. X-ray pulsations of 201.9$\pm$0.5~s 
were detected in the EPIC-MOS data of the October 2000 observation, when the source was observed at 
a large off-axis angle with the MOS detectors while it was outside the EPIC-PN field of view 
\citep{2004ApJ...609..133M}. Significant modulation in the X-ray flux was mainly detected at energies above 2 keV. 
During the 2002 March observation the source flux was 70\% of that of the earlier observation and the 
period was determined to 201.2$\pm$0.5~s \citep{2004ApJ...609..133M}.
The proposed optical counterpart (see also Table~\ref{tab-ids}) was investigated by 
\citet{2005MNRAS.356..502C}. These authors presented the MACHO and OGLE light curves in the V- and I-band,
respectively, which show very high optical variability, but no significant optical periodicity.
The X-ray and optical data clearly identify this source as Be/X-ray binary pulsar.

During our new \xmm\ observation RX\,J0059.3-7223\ was the second-brightest source in the field with about a factor 
of five lower count rate as XMMU\,J005929.0-723703, the other new 202~s pulsar. 
The light curve shows flaring variability by a factor of about 
two, but the lower statistics requires longer time binning and the flares can not be fully resolved
(Fig.~\ref{xmmp-lcurve}). Timing analysis yields a pulse 
period of 200.50$\pm$0.29~s which may indicate a slight spin-up trend over 6.7 years, but the periods
measured at the three epochs are still compatible within their 2$\sigma$ uncertainties. 
Pulse profiles and hardness ratios were derived and are shown in Figs.~\ref{xmmo-pulse} and \ref{xmmp-hr}.
The pulse profiles show little variation below $\sim$2 keV as seen before \citep{2004ApJ...609..133M}
and a broad pulse with a sharp drop 
on top of a flat light curve for energies between 2.0 and 4.5 keV. At higher energies the pulse
seems to be smeared out to a more sinusoidal profile. 
Pulse-phase averaged spectra were accumulated and are well reproduced by the power-law model with 
parameters typical for Be/X-ray binaries in the SMC (Table~\ref{tab-spectra}, Fig.~\ref{fig-spectra}).

\subsection{The 7.78~s Be/X-ray binary pulsar SMC\,X-3}

SMC\,X-3 was originally detected by \citet{1978ApJ...221L..37C} in SAS\,3 data during a strong X-ray 
outburst with 7\ergs{37}. Only recently it was identified with a previously detected 7.78~s RXTE pulsar 
by using archival Chandra data \citep{2004ATel..225....1E}. During the RXTE monitoring of the SMC
a spin-down trend in the pulse period of SMC\,X-3 is seen to around 7.79~s by the end of 2005  
\citep{2008arXiv0802.2118G}.
We detected SMC\,X-3 with a pulse period of 7.7912$\pm$0.0096~s consistent with this development.
The X-ray luminosity (Table~\ref{tab-spectra}) was more than a factor of six lower than during the
the Chandra observation on 2002 July 20 \citep{2004MNRAS.353.1286E}, when the source also showed 
a harder spectrum. The \xmm\ EPIC energy resolved pulse profiles (Fig.~\ref{xmmo-pulse}) show a relatively
smooth, broad peak at energies below 1 keV, but indicate a highly structured shape at energies 
above 1 keV.

\begin{figure*}
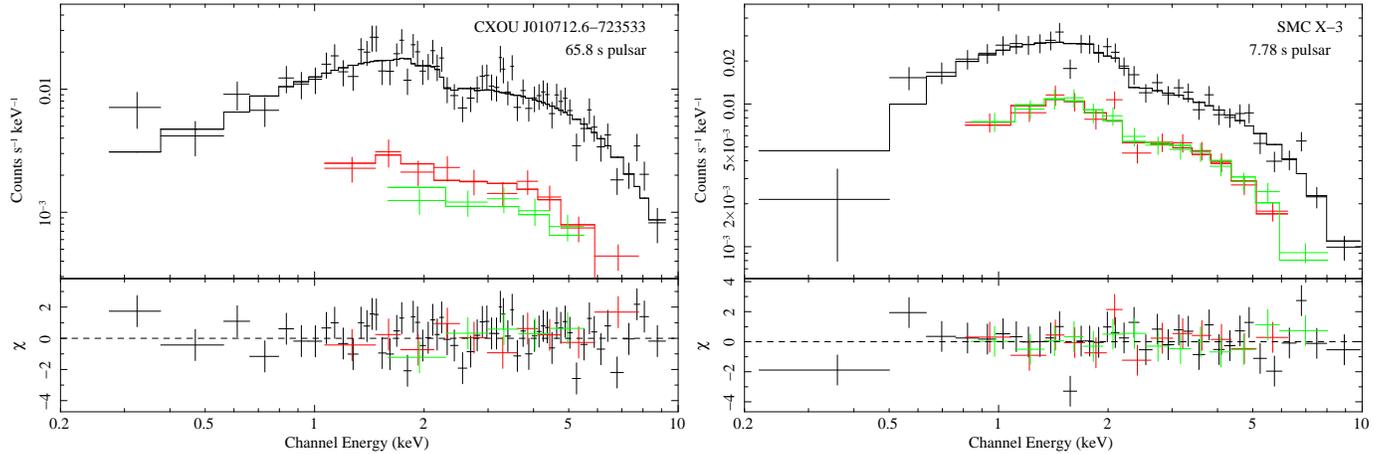

  \hbox{
  \resizebox{0.49\hsize}{!}{\includegraphics[angle=-90,clip=]{spectra_0404680501_1_bb_pow.ps}}
  \resizebox{0.49\hsize}{!}{\includegraphics[angle=-90,clip=]{spectra_0500980101_1_pl_bb.ps}}
  }
  \caption{EPIC spectra of two Be/X-ray binaries which are not well modelled by a power-law (see 
  Fig.~\ref{fig-spectra}). Including a hot blackbody component to the model yields acceptable fits.}
  \label{fig-bbspectra}
\end{figure*}

As in the case of CXOU J010712.6-723533 the power-law model fit to the EPIC spectra 
of SMC\,X-3 yields a relatively bad reduced $\chi^2$ and similarly structured residuals 
(Fig.~\ref{fig-spectra}). Adding a blackbody component improves the fit considerably
(Table~\ref{tab-bbspectra}, Fig.~\ref{fig-bbspectra}). Similar parameters for the
blackbody components from the two pulsars are inferred.

\begin{table*}
\caption[]{Spectral fit results for a power-law component with additional blackbody emission.}
\begin{center}
\begin{tabular}{lccccccc}
\hline\hline\noalign{\smallskip}
\multicolumn{1}{l}{Source} &
\multicolumn{1}{c}{SMC \nh} &
\multicolumn{1}{c}{Photon} &
\multicolumn{1}{c}{kT} &
\multicolumn{1}{c}{Radius} &
\multicolumn{1}{c}{Flux$^{(a)}$} &
\multicolumn{1}{c}{L$_{\rm x}^{(b)}$} &
\multicolumn{1}{c}{$\chi^2_{\rm r}$/dof} \\
\multicolumn{1}{c}{} &
\multicolumn{1}{c}{[\oexpo{21}cm$^{-2}$]} &
\multicolumn{1}{c}{Index} &
\multicolumn{1}{c}{keV} &
\multicolumn{1}{c}{km} &
\multicolumn{1}{c}{erg cm$^{-2}$ s$^{-1}$} &
\multicolumn{1}{c}{erg s$^{-1}$} &
\multicolumn{1}{c}{} \\

\noalign{\smallskip}\hline\noalign{\smallskip}
CXOU\,J010712.6-723533         & $<$1.5                 & 0.23$\pm$0.23  & 1.3$\pm$0.2          & 0.74 & 1.4\expo{-12}  &  6.0\expo{35} &  1.17/71   \\
\noalign{\smallskip}\hline\noalign{\smallskip}
SMC\,X-3                       & 2.0$^{+1.6}_{-1.0}$    & 0.79$\pm$0.07  & 1.1$^{+0.3}_{-0.2}$  & 0.73 & 8.5\expo{-13}  &  3.8\expo{35} &  0.98/56   \\
\noalign{\smallskip}\hline\noalign{\smallskip}
\end{tabular}
\end{center}
$^{(a)}$ Observed 0.2-10.0 keV flux.
$^{(b)}$ Source intrinsic X-ray luminosity in the 0.2-10.0 keV band (corrected for absorption)
for a distance to the SMC of 60 kpc \citep{2005MNRAS.357..304H}.
\label{tab-bbspectra}
\end{table*}

\subsection{The 138~s Be/X-ray binary pulsar CXOU\,J005323.8-722715}

This pulsar was discovered in a Chandra observation on 2002 July 20 \citep{2004MNRAS.353.1286E} 
with a period of 138.04$\pm$0.61~s (98\% confidence). We confirm this source as pulsar and determined
the period to 139.136$\pm$0.016~s from the \xmm\ observation on 2007 June 23, indicating
significant spin-down of the neutron star. 
We present the pulse profiles of CXOU\,J005323.8-722715 in Fig.~\ref{xmmo-pulse} showing a 
broad, single peak in all EPIC energy bands similar to that seen in the broad band Chandra data,
but with much increased statistical quality.

\subsection{The 59~s Be/X-ray binary pulsar XTE\,J0055-724}

The pulse period of this source was discovered in RXTE data \citep{1998IAUC.6818R...1M}
and subsequent observations with BeppoSAX refined the pulse period to 58.969$\pm$0.001~s
and the positional uncertainty to a radius of 50\arcsec\ \citep{1998IAUC.6818R...2S}.
This uncertainty was further reduced to 10\arcsec\ by the identification of XTE\,J0055-724 = 
1SAX\,J0054.9-7226 with RX\,J0054.9-7226 \citep{1998IAUC.6822....2I} which led to
the identification of the optical counterpart \citep{1999MNRAS.309..421S}.
No broad band X-ray spectrum could be obtained yet.
Also in the EPIC data, XTE\,J0055-724 was detected only as weak source with just too few counts for 
spectral fitting (spectral parameters are largely unconstrained). 
The 0.2$-$4.5 keV count rate of 2.67\ct{-2} from the EPIC-PN camera translates into a luminosity of 
9.2\ergs{34} (scaling to the nearby 138~s pulsar CXOU\,J005323.8-722715 with a count rate of 
8.70\ct{-2}, i.e. assuming a power-law spectrum with $\gamma$ = 0.97 and SMC \nh\ = 2.9\hcm{21}
which is justified by very similar hardness ratios of the two sources).
A pulse period search around the expected value yielded a period of 58.858$\pm$0.001~s, i.e. a net 
spin-up of 0.11~s over the 9.4 years between the BeppoSAX and \xmm\ observations.

\begin{figure*}
  \hbox{
  \resizebox{0.32\hsize}{!}{\includegraphics[clip=]{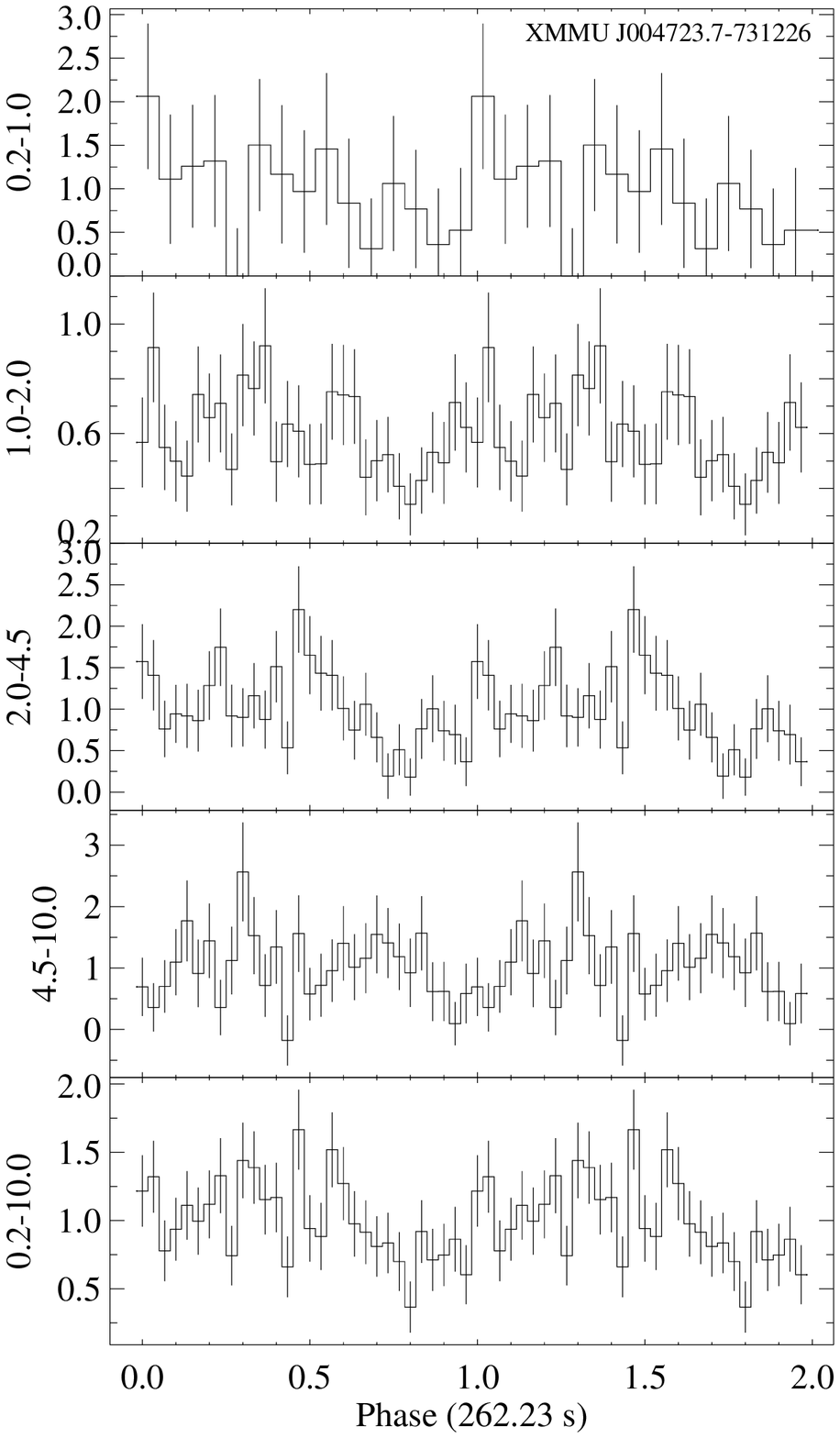}}
  \resizebox{0.32\hsize}{!}{\includegraphics[clip=]{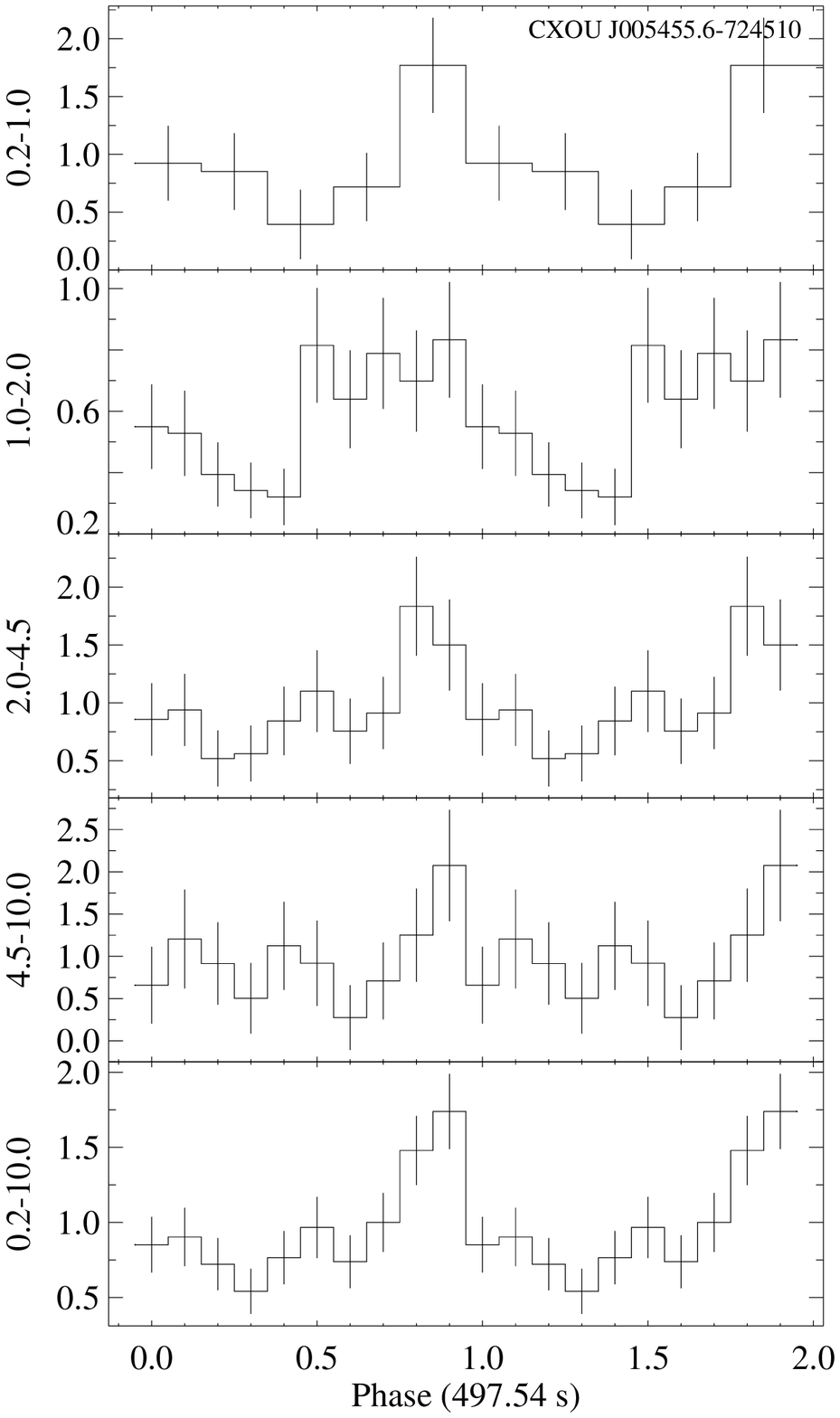}}
  \resizebox{0.32\hsize}{!}{\includegraphics[clip=]{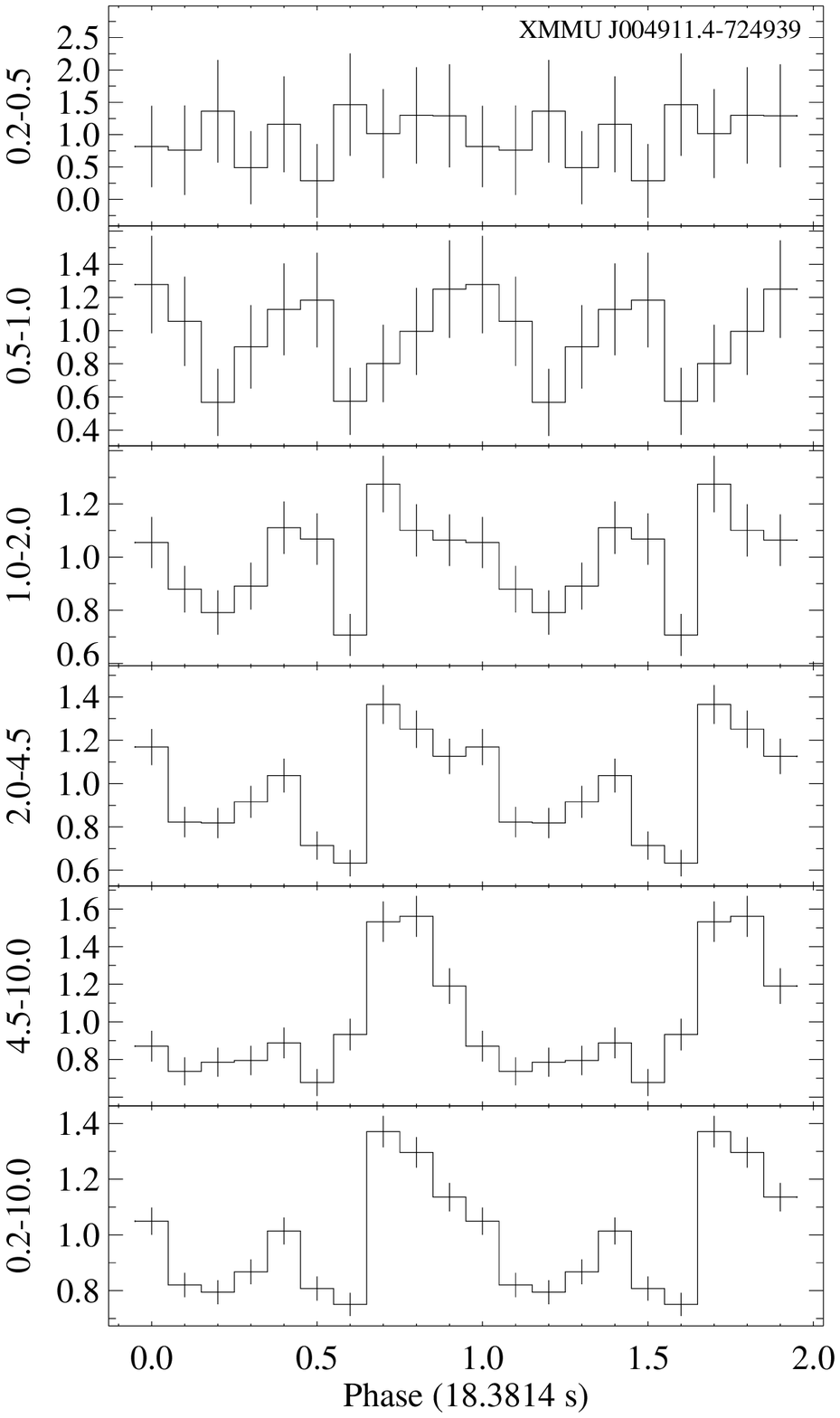}}
  }
  \hbox{
  \resizebox{0.32\hsize}{!}{\includegraphics[clip=]{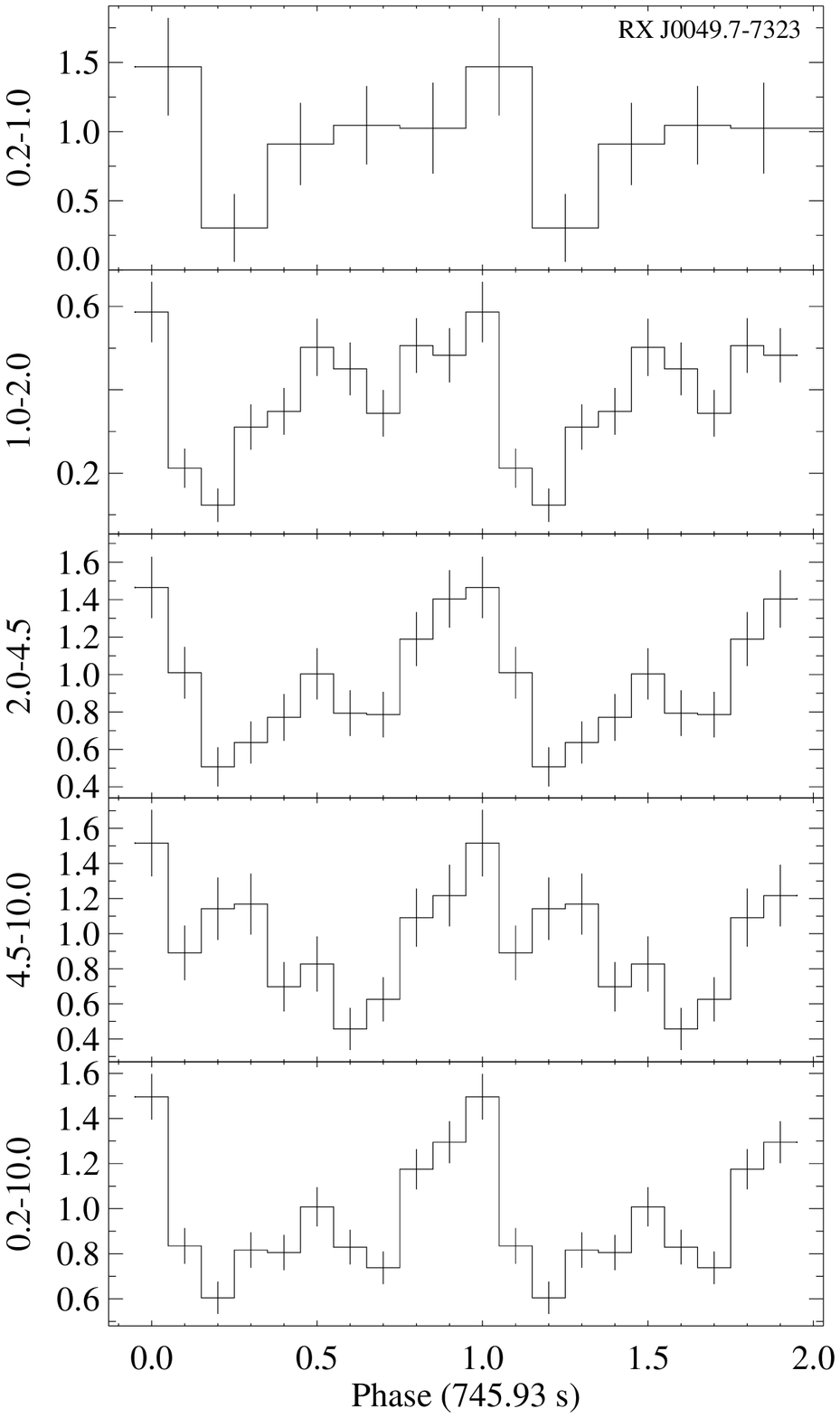}}
  \resizebox{0.32\hsize}{!}{\includegraphics[clip=]{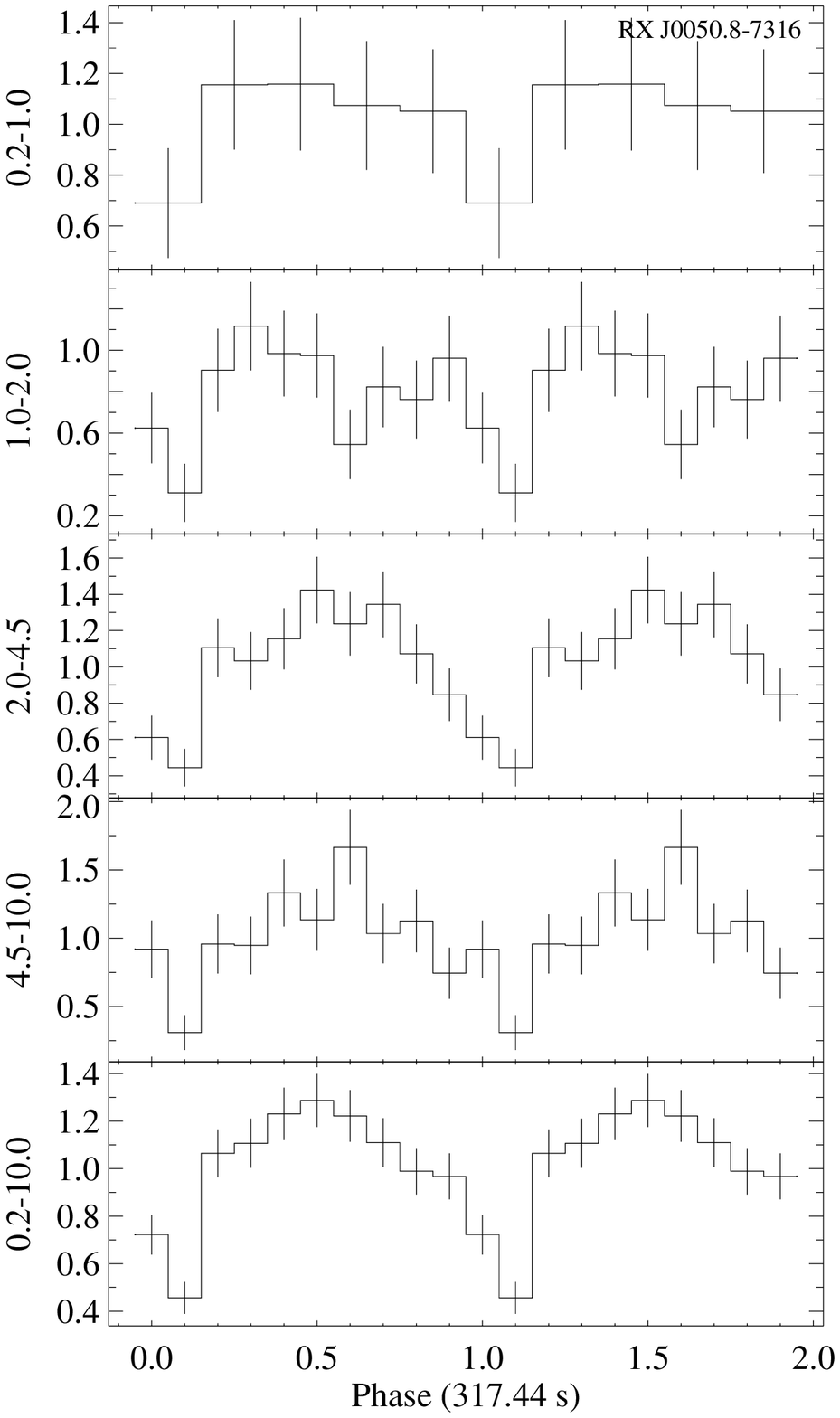}}
  \resizebox{0.32\hsize}{!}{\includegraphics[clip=]{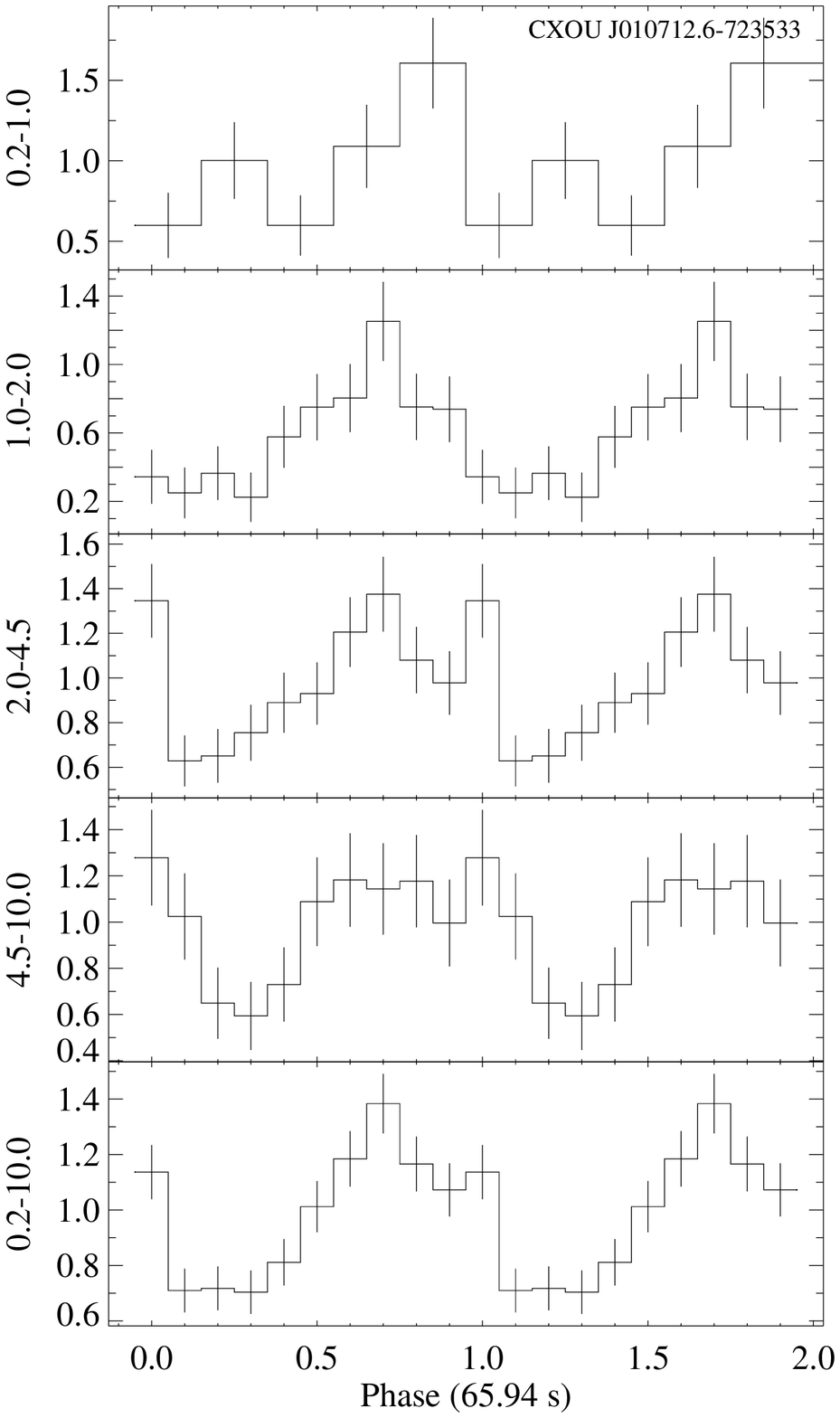}}
  }
  \caption{Folded EPIC-PN light of Be/X-ray binaries with previously identified pulse periods. 
           Presentation as in Fig.~\ref{xmmp-pulse}.
          }
  \label{xmmo-pulse}
\end{figure*}
\addtocounter{figure}{-1}
\begin{figure*}
  \begin{center}
  \hbox{
  \resizebox{0.33\hsize}{!}{\includegraphics[clip=]{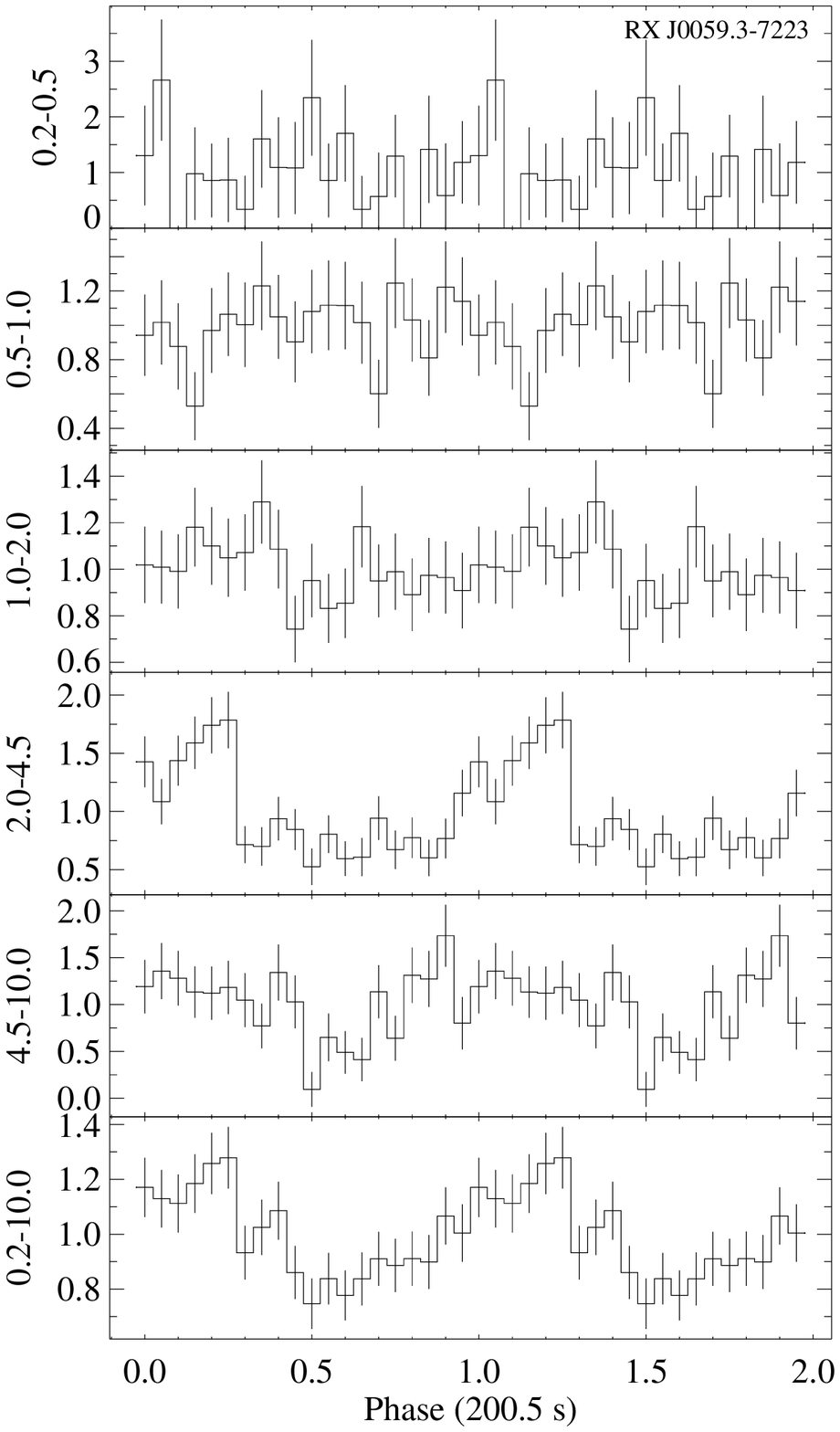}}
  \resizebox{0.33\hsize}{!}{\includegraphics[clip=]{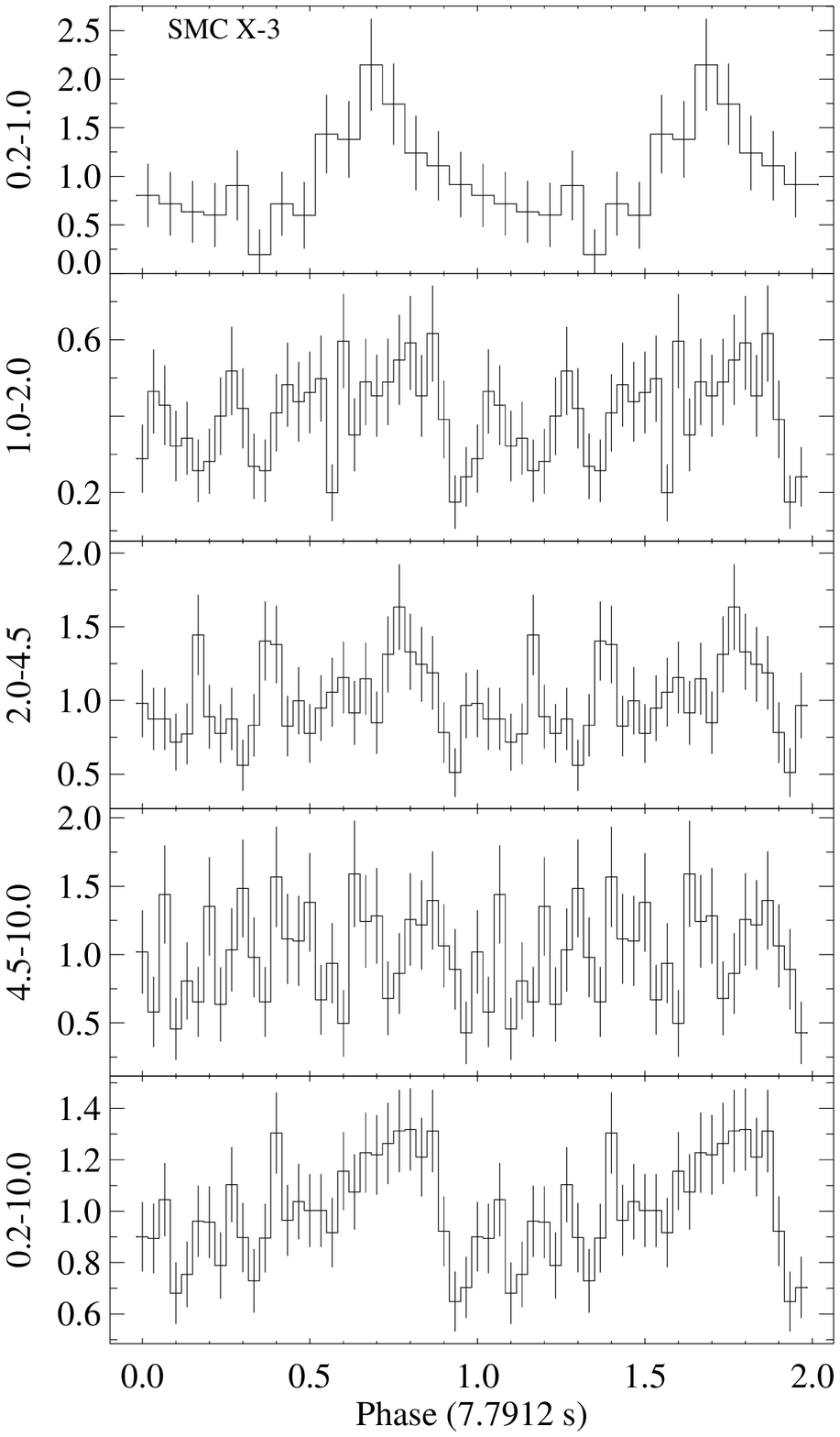}}
  \resizebox{0.33\hsize}{!}{\includegraphics[clip=]{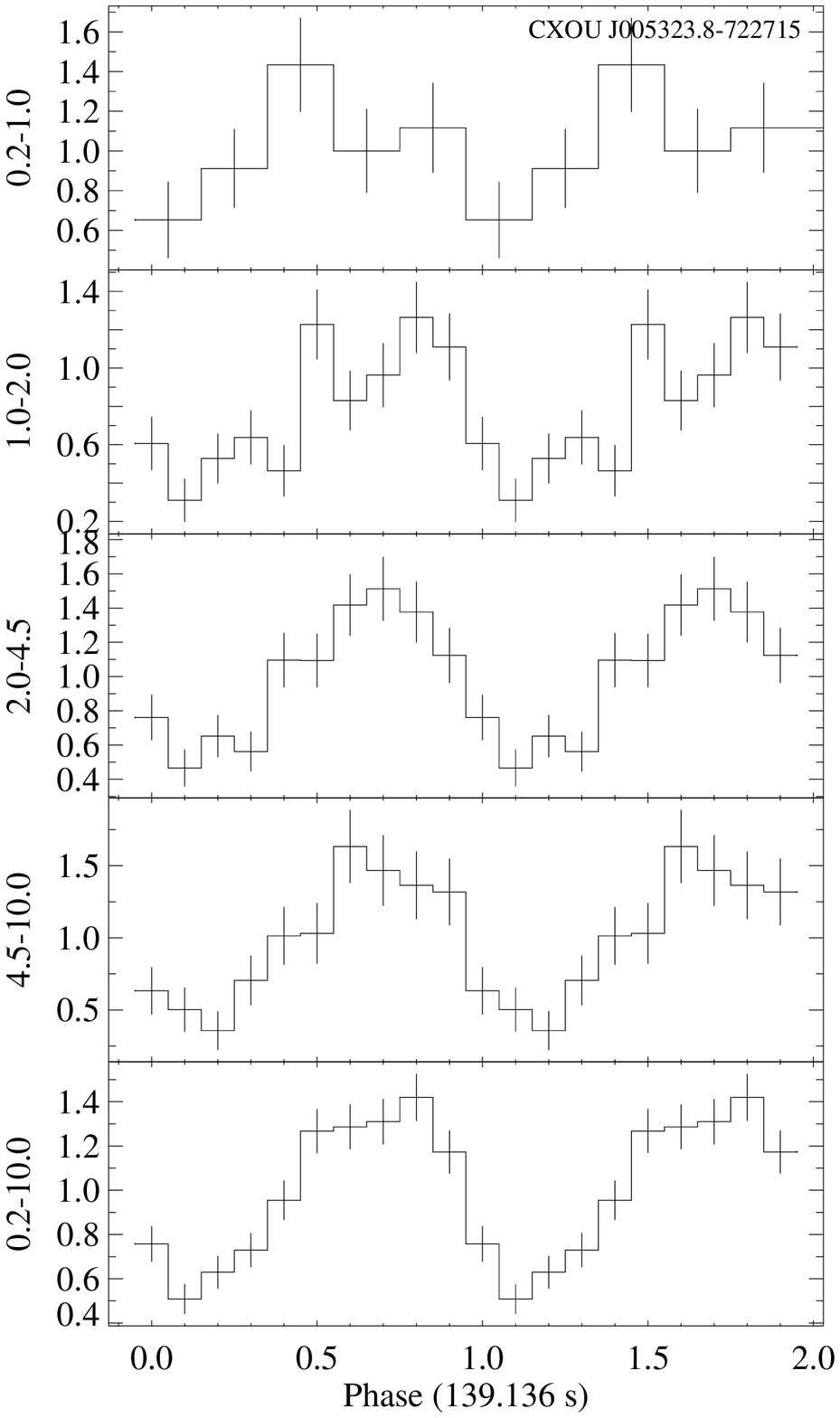}}
  }
  \end{center}
  \caption{Continued}
\end{figure*}

\begin{figure*}
  \hbox{
  \resizebox{0.33\hsize}{!}{\includegraphics[clip=]{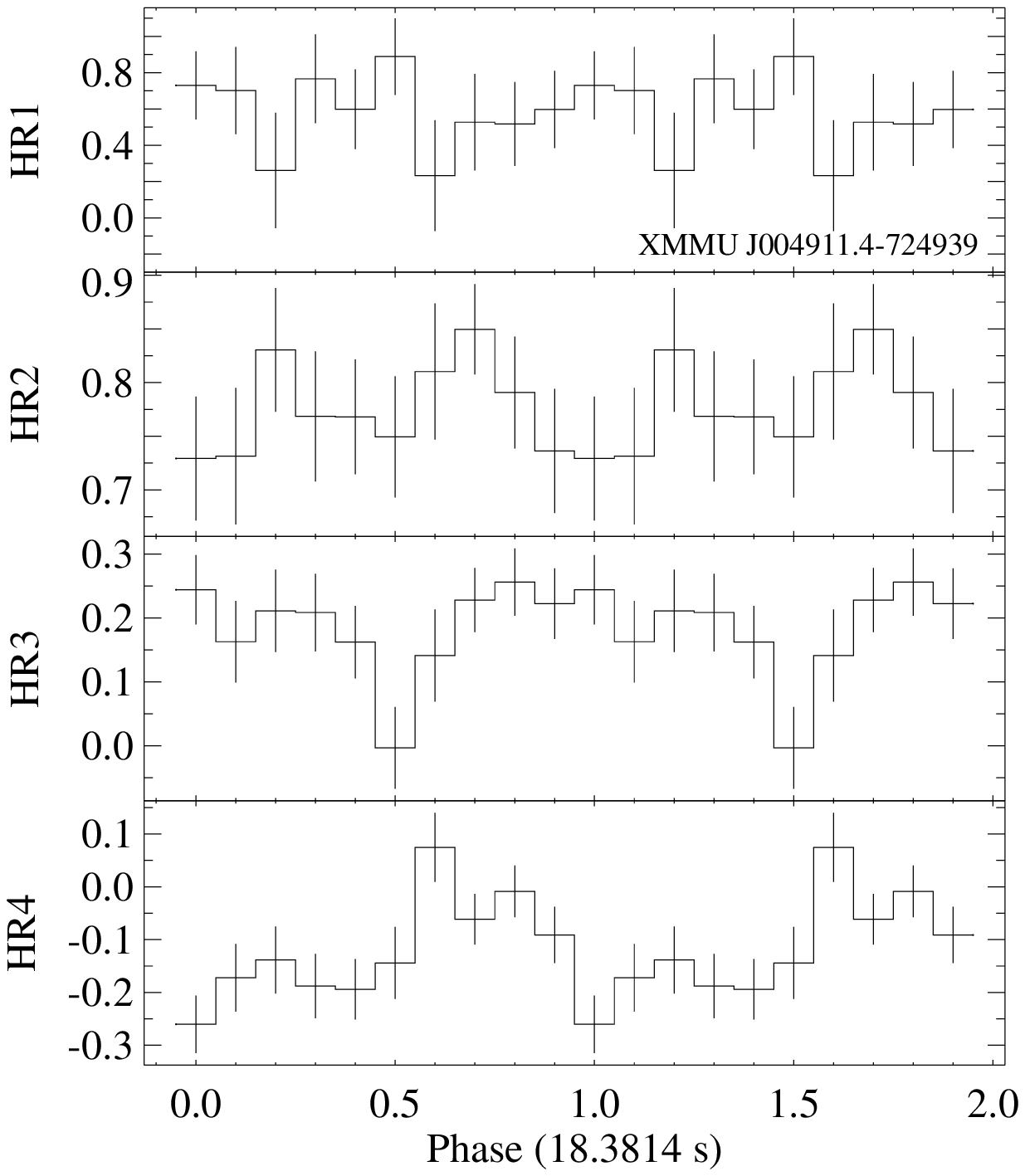}}
  \resizebox{0.33\hsize}{!}{\includegraphics[clip=]{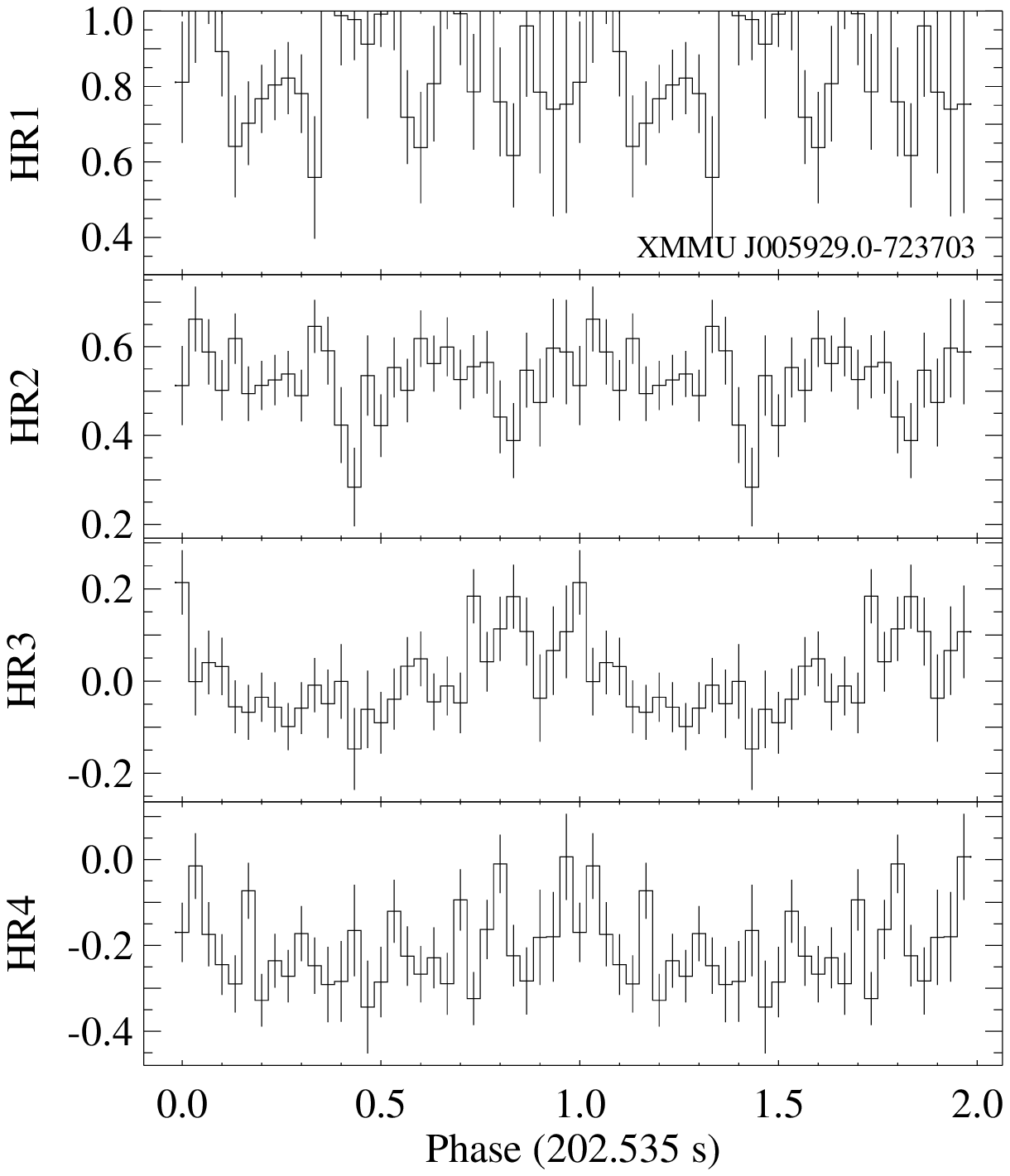}}
  \resizebox{0.33\hsize}{!}{\includegraphics[clip=]{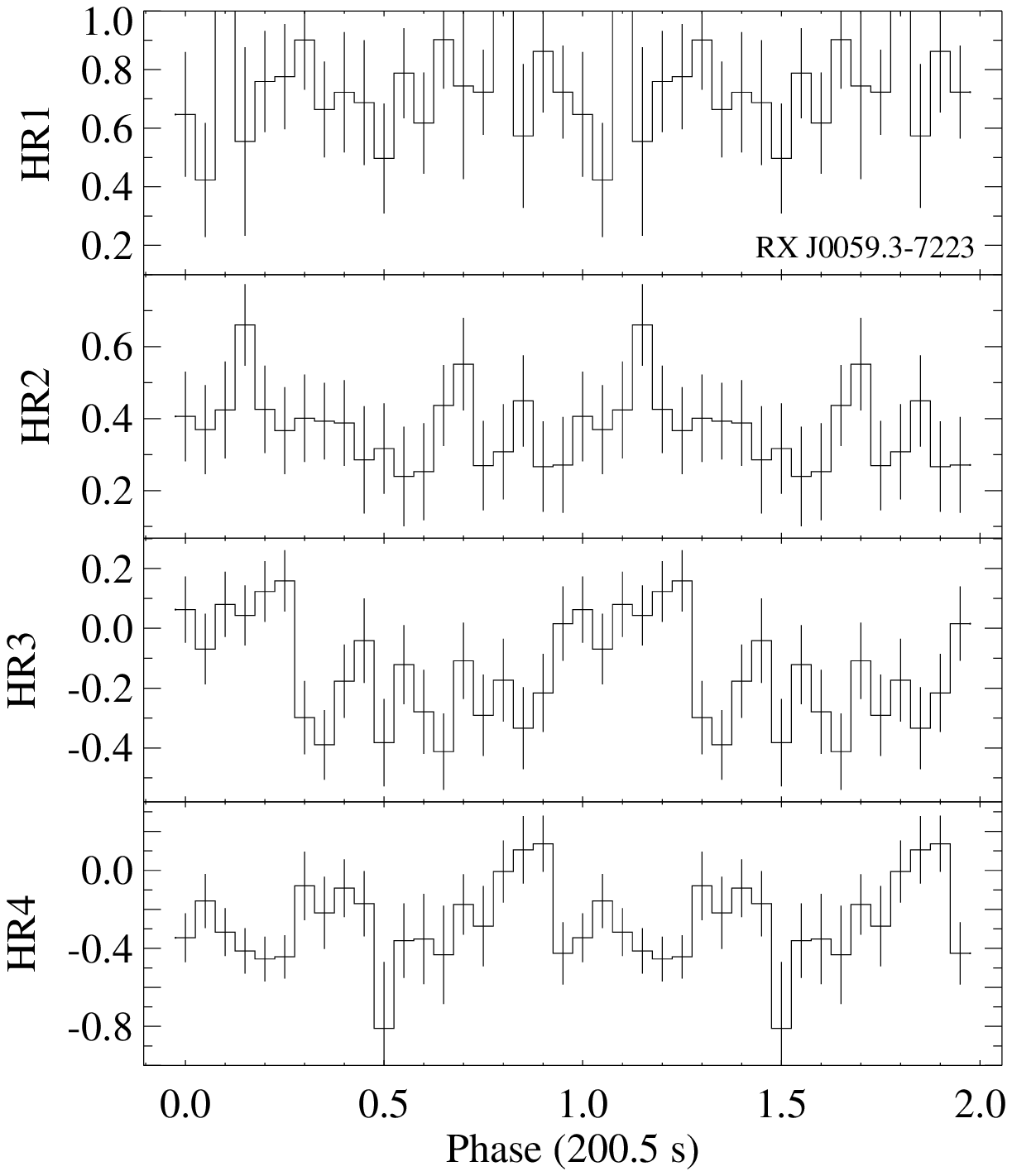}}
  }
  \caption{Hardness ratios HR1, HR2, HR3 and HR4 as function of pulse phase of three Be/X-ray pulsars, 
           derived from their folded EPIC-PN light curves shown in Figs.~\ref{xmmp-pulse} and \ref{xmmo-pulse}.
           The 18.37~s pulsar XMMU\,J004911.4-724939 was observed in field 0403970301 while
	   the two pulsars XMMU\,J005929.0-723703 and RX\,J0059.3-7223 with periods close to 202~s 
           are both located in field 0500980201.}
  \label{xmmp-hr}
\end{figure*}

\section{Be/X-ray binaries without detected X-ray pulsations}
\label{sect-nopuls}

\subsection{The Be/X-ray binary RX\,J0048.5-7302}

This ROSAT source was proposed as candidate Be/X-ray binary by \citet{2000A&A...359..573H} 
due to its identification with the \Halp\ emission line star [MA93]\,238. From a first \xmm\ 
observation \citet{2004A&A...414..667H} reported an improved X-ray position which supported
the optical identification. The first EPIC spectra were also consistent with a power-law with 
a spectral index of 0.9, typical for HMXBs. The optical counterpart was covered by the 
2dF survey of the SMC and \citet{2004MNRAS.353..601E} obtained a spectral type of B0-5(II),
strongly suggesting RX\,J0048.5-7302 as Be/X-ray binary.

Our boresight corrected X-ray position from the new \xmm\ observation on 2007 March 12 
is within 0.2\arcsec\ 
consistent with the optical counterpart, fully confirming the association. The source was detected 
with an EPIC-PN count rate of 3.1\ct{-3}, not sufficient for spectral or for timing analysis.
On 2000 Oct. 15 the count rate was 2.7\ct{-2}, a factor of 8.5 higher. Scaling the X-ray luminosity
given in \citet{2004A&A...414..667H} by this factor (adjusting the distance to 60~kpc) yields the 
value given in Table~\ref{tab-spectra}.

\subsection{The Be/X-ray binaries RX\,J0049.2-7311 and RX\,J0049.5-7310}

These two ROSAT sources with an angular distance of 80\arcsec\ were both discussed as
possible counterparts of the ASCA 9.13~s pulsar AX\,J0049-732 \citep{2003PASJ...55..161Y}.
RX\,J0049.5-7310 was proposed as candidate Be/X-ray binary by 
\citet{2000A&A...359..573H} due to its identification with the \Halp\ emission line 
star [MA93]\,300 and therefore favoured as counterpart to AX\,J0049-732 by 
\citet{2000A&A...361..823F}. Later, also for RX\,J0049.2-7311 a \Halp\ emission line star
was found as optical counterpart \citep[][Table~\ref{tab-ids}]{2005MNRAS.356..502C} 
and because the position of RX\,J0049.2-7311 is formally closer to the best ASCA position 
of AX\,J0049-732 these authors preferred this as more likely identification.
However, due to the relatively wide point spread function of the ASCA telescope
the extraction radius of 3\arcmin\ used for the analysis of AX\,J0049-732 
\citep{2000ApJS..128..491Y} encompasses both ROSAT sources and a secure identification of 
AX\,J0049-732 with either of the ROSAT sources can only be made through the detection
of pulsations.

RX\,J0049.2-7311 and RX\,J0049.5-7310 were both detected in the \xmm\ observation 
0404680101 in Oct. 2006 at a similar X-ray luminosity (Table~\ref{tab-spectra}). 
Their X-ray spectra are typical for HMXBs and given their optical counterparts, which 
are both consistent with Be stars, they are clear cases of Be/X-ray binaries in the SMC.
Both sources were also detected in a previous \xmm\ observation in Oct. 2000
\citep{2004A&A...414..667H}. While RX\,J0049.2-7311 was at the same brightness level,
RX\,J0049.5-7310 was almost a factor of 4 fainter during the earlier \xmm\ observation.
No significant pulsations were detected in the range between 0.2~s and 2000~s in 
the EPIC data of either observation. The non-detection of 9.13~s pulsations might
even suggest, that none of the two ROSAT sources is the counterpart of the ASCA source.

During the April 2007 observation RX\,J0049.5-7310 was not detected. RX\,J0049.2-7311 was 
again at a similar brightness level as seen by \xmm\ before. Scaling from the 
PN 0.2$-$4.5 keV count rate (0404680101: 5.0\ct{-2} compared to 0403970301: 6.6\ct{-2})
with the assumption of a constant spectral shape, the source luminosity of RX\,J0049.2-7311 
increased slightly by a factor of 1.3 from Oct. 2006 to March 2007 (Table~\ref{tab-spectra}).

Despite the non-detection of any significant periodicity from either of the two sources,
we confirm both as Be/X-ray binaries from their optical counterpart and their
X-ray spectral properties.

\subsection{The Be/X-ray binary RX\,J0049.5-7331}

RX\,J0049.5-7331 is another candidate Be/X-ray binary proposed by 
\citet{2000A&A...359..573H} due to its likely identification with the \Halp\ emission line 
star [MA93]\,302. The X-ray position obtained from the EPIC images (within 0.17\arcsec\ of 
the optical position from MCPS, Table~\ref{tab-ids}) further supports this identification. 
The DSS2 red image shows a pair of close bright stars with similar brightness and colours 
which is not resolved in the UBVR catalogue entry. 
The X-ray position is consistent with the northern of the two stars.
The X-ray spectrum of RX\,J0049.5-7331 (Table~\ref{tab-spectra}) indicates a relatively hard
power-law with little absorption in the SMC in addition to the Galactic foreground column density.
The EPIC X-ray spectrum and its optical counterpart clearly identify RX\,J0049.5-7331
as Be/X-ray binary in the SMC.

\subsection{The Be/X-ray binary RX\,J0101.8-7223}

This source was detected in ROSAT PSPC \citep[source 220 in the catalogue of][]{2000A&AS..142...41H}
and HRI \citep[source 97;][]{2000A&AS..147...75S} pointed observations. 
\citet{2000A&A...359..573H} suggested it as Be/X-ray binary candidate because of a possible 
identification with the \Halp\ emission line object [MA93]\,1288.
However, this needed confirmation due to the relatively large ROSAT position error of $\sim$8\arcsec.
Our improved \xmm\ position (within 1.3\arcsec\ of the optical position of [MA93]\,1288 
from MCPS, Table~\ref{tab-ids}) supports the identification with the emission line star. 
The star was included in the 2dF survey of the SMC and \citet{2004MNRAS.353..601E} infer a 
spectral type of B0-5(II). 

No X-ray pulsations have been detected so far from this Be/X-ray binary. We also did not detect 
any significant periodic signal in the EPIC data between 0.2~s and 2000~s. The EPIC spectra
are consistent with an absorbed power-law with typical values for \nh\ and photon index 
(Table~\ref{tab-spectra}) for Be/X-ray binaries. The X-ray and optical properties point 
to the identification of RX\,J0101.8-7223 as a Be/X-ray binary.

\begin{figure}
  \resizebox{0.98\hsize}{!}{\includegraphics[angle=-90,clip=]{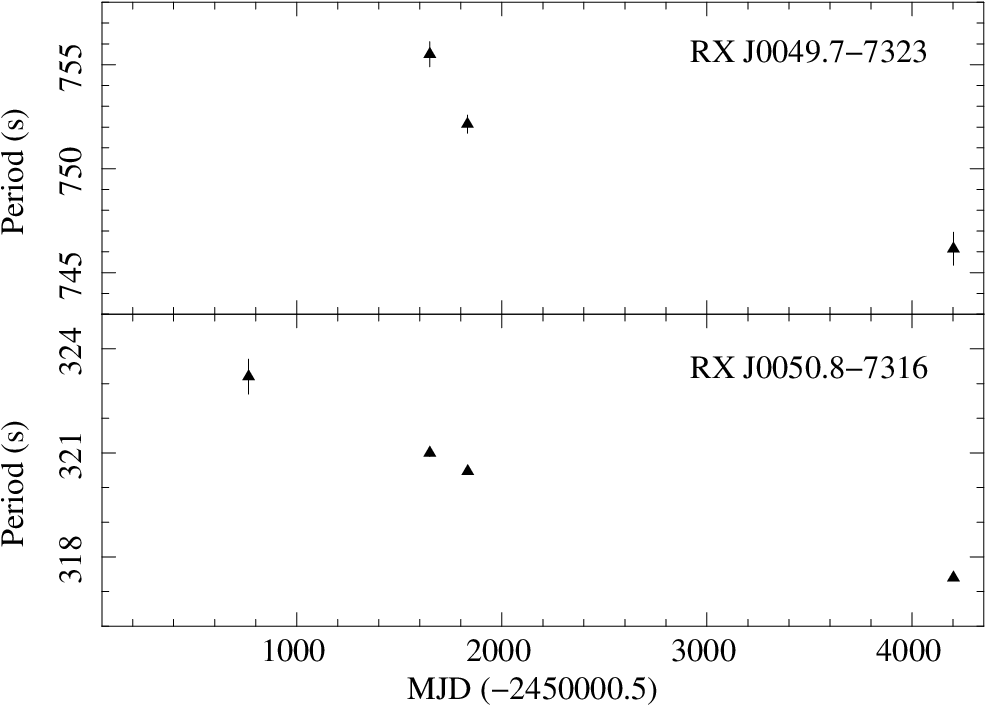}}
  \caption{Spin period history of two SMC pulsars covered by our observations. 
           The first measurement for RX\,J0049.7-7323 (=~AX\,J0049.4-7323, see Sect.~\ref{sect-olda})
	   and the first two measurements for RX\,J0050.8-7316 (=~AX\,J0051-733, see Sect.~\ref{sect-oldb})
	   were reported from ASCA data \citep[][respectively]{2000PASJ...52L..73Y,2003PASJ...55..161Y}.
	   The last two data points for each pulsar are derived from \xmm\ observations.}
  \label{fig-phist}
\end{figure}

\section{Discussion and conclusions}

We detected a large sample of twenty-six Be/X-ray binaries in eight \xmm\ observations 
of the SMC between Oct. 2006 and June 2007. After astrometric boresight corrections the
accurate X-ray positions allowed us either to confirm previously proposed optical counterparts 
or, in cases of the newly discovered systems, to identify their counterparts.
The properties of all optical counterparts (like brightness, colours and long-term 
temporal behaviour) are consistent with Be/X-ray binaries. No candidate for a supergiant 
HMXB was found. 

Twenty of the sources were observed with luminosities above $\sim$\oergs{35},
providing X-ray spectra and light curves for a detailed, systematic study.
Most of the X-ray spectra can be modelled by an absorbed power-law, typical for HMXBs.
The average power-law index is 0.93 with 90\% of the values between 0.71 and 1.27,
which is in agreement with the peak of the distribution at 1.0 found by 
\citet{2004A&A...414..667H} for a smaller sample.
\citet{2007MNRAS.376..759M} report harder spectra from four Be/X-ray binary pulsars
detected in the Chandra SMC wing survey and discuss possibilities for systematic
differences between pulsars in the wing and the bar of the SMC.
We see a general trend of harder power-law spectra when the sources are brighter
(e.g. XMMU\,J004723.7-731226, Sect.~\ref{sect-lxharda}; 
      CXOU\,J010712.6-723533, Sect.~\ref{sect-lxhardb})
which suggests that the harder spectra in the wing pulsars might at least partly 
be due to a brightness selection effect. From the short ($\sim$10~ks) Chandra 
observations, spectra could only be derived when the sources were bright.

The highest luminosities were reached by the 18.37~s pulsar XMMU\,J004911.4-724939 and 
the new 202~s pulsar XMMU\,J005929.0-723703 with 5.5\ergs{36} and 4.5\ergs{36}, respectively,
while all other sources with EPIC spectra were observed with luminosities between \oergs{35} 
and \oergs{36}. Six Be/X-ray binaries were detected only as faint sources (for four of them 
we estimate their luminosity 
from the count rates, the other two are located at the edge of the field of view) 
and about 5-10 Be/X-ray binaries (uncertain positions and unclear classifications make the 
numbers uncertain) were not detected in our eight fields.
I.e. we now know about 31-36 Be/X-ray binaries in the eight fields from which 26 (about 70-85\%) 
are detected above a luminosity of 4\ergs{34} (the weakest detected sources have count rates down to 
1\ct{-3} for EPIC-PN which converts to 4\ergs{33} for a typical HMXB power-law spectrum).

For twelve of the Be/X-ray binaries covered by our \xmm\ observations, 
orbital ephemeris obtained from RXTE X-ray data are available \citep{2008arXiv0802.2118G}. 
All our clear detections are from orbital phases near zero (which defines maximum X-ray
intensity as seen by RXTE). Consistently, three pulsars observed around phase 0.4 were
found to be either faint or not detected. 
The 74.7~s pulsar AX\,J0049-729, the 59~s pulsar XTE\,J0055-724 and the 82.4~s pulsar
XTE\,J0052-725 were relatively faint during the \xmm\ observations close to phase 0, but
the uncertainties in their ephemeris do not allow an exact determination of the orbital
phase. Therefore, it is not clear if the outbursts were missed by the \xmm\ observations
or if they did not happen. Overall, the luminosity states in which we found
the Be/X-ray binaries during the \xmm\ observations are all consistent with the 
orbital variations determined by RXTE. This also suggests, that we did not observe
any type II outburst with \xmm.

Systematic deviations from the power-law model are seen in the spectra of a few 
of our investigated Be/X-ray binaries (Fig.~\ref{fig-spectra}). In particular 
CXOU\,J010712.6-723533 and SMC\,X-3 yield bad power-law fits (in terms of reduced 
$\chi^2$ as can be seen from Table~\ref{tab-spectra}). Additional cases might be 
RX\,J0050.8-7316 and XMMU\,J005252.1-721715. However, spectra with better statistical 
quality are required to prove this. Expanding the model with a blackbody component 
yields acceptable fits for  CXOU\,J010712.6-723533 and SMC\,X-3. The inferred 
blackbody temperatures are around 1.2 keV and the sizes of the emitting areas are 
relatively small (radius $\sim$0.7 km). These spectral characteristics are 
similar to those of RX\,J0146.9+6121 and X\,Persei 
\citep{2006A&A...455..283L,2007A&A...474..137L}. These authors interprete the blackbody 
component as emission from the hot polar caps of the accreting neutron star 
\citep[for a discussion of the origin of soft emission excesses see][]{2004ApJ...614..881H}.

We discovered X-ray pulsations from four new transient Be/X-ray binaries 
\citep[for the 325~s pulsar XMMU\,J005252.1-721715 an independent discovery from 
Chandra data was reported by ][]{2008arXiv0803.3941C} and found pulsations 
from two previously known Be/X-ray binaries. The pulse periods range 
between 202~s and 967~s, the latter being the second longest known from SMC pulsars.
The high statistical quality of the EPIC data allowed us to investigate pulse
profiles in several different energy bands, revealing a large variety of pulse shapes
and different energy dependence.

All of our investigated pulsars with pulse periods longer than 150~s and known period 
history over at least a few years show secular spin-up of the neutron star. They all exhibit
more or less regular outburst activity as seen by RXTE \citep{2008arXiv0802.2118G}.
At shorter spin periods at least two of the pulsars show long-term spin-down:
the 138~s pulsar CXOU\,J005323.8-722715, which was detected only twice by RXTE 
at high brightness level \citep{2008arXiv0802.2118G} and the 7.78~s pulsar SMC~X-3.
The periods determined for the 18.37~s pulsar XMMU\,J004911.4-724939 are 
consistent with no change. This fits in the general picture of the spin evolution
of accretion powered pulsars, in which the neutron star is spun-up by accretion
until an equilibrium period is reached 
\citep[see, e.g. equations 8 and 9 in ][]{1997ApJS..113..367B}. The fact that the majority 
of long-period pulsars in the SMC show secular spin-up, suggests however a much longer 
timescale for spinning up the neutron star, as was also seen for HMXB pulsars 
in the Milky Way. High resolution BATSE measurements of the long-term spin evolution
of HMXB pulsars have shown that the small long-term spin-up rates are a consequence
of frequent transitions between spin-up and spin-down which can occur on timescales 
of less than 10 days \citep{1997ApJS..113..367B}. To confirm such a behaviour
for SMC Be/X-ray binary pulsars would require a monitoring program with \xmm\ with frequent 
observations for several weeks.

With the new discoveries, we now know of several pairs of SMC pulsars with very 
similar periods. In particular two 202~s pulsars exist with an angular distance 
of only 13.8\arcmin.
RX\,J0059.3-7223 was discovered with a period of 201.9$\pm$0.5~s in October 2000, 
which decreased to 200.5$\pm$0.3~s, and XMMU\,J005929.0-723703 was discovered 
with 202.52$\pm$0.02~s.
A similar case is XMMU\,J005403.8-722632, newly discovered in this work with a 
period of 341.87$\pm$0.15~s, and SAX\,J0103.2-7209 with the earliest measured 
period of 348.9$\pm$0.1~s \citep[in ASCA data from May 1996; ][]{1998IAUC.7009....3Y}, 
which decreased to 341.2$\pm$0.5~s in October 2000 \citep{2004A&A...414..667H}. 
These two pulsars are located 45.2\arcmin\ apart. 
Due to the overlapping ranges of spin periods which evolve with time and the 
projected proximity of many Be/X-ray binaries in the SMC some of these pulsar pairs
my not be differentiated by non-imaging instruments which needs to be considered when 
interpreting the spin period evolution.

\begin{figure}
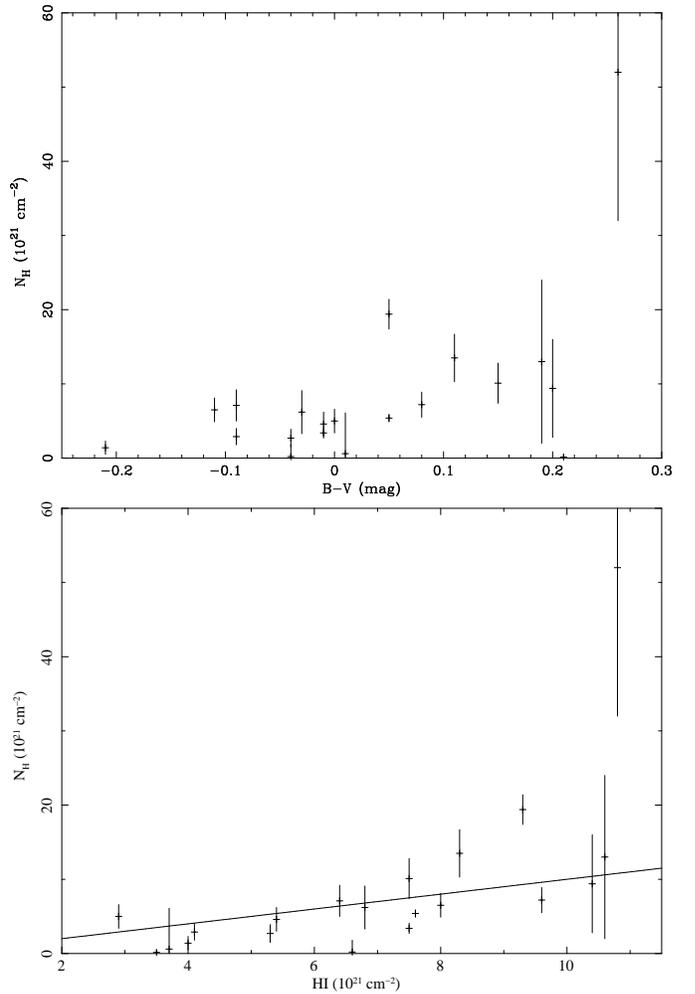

  \resizebox{0.98\hsize}{!}{\includegraphics[angle=-90,clip=]{mcps-NH.ps}}
  \resizebox{0.98\hsize}{!}{\includegraphics[angle=-90,clip=]{NH-HI.ps}}
  \caption{Equivalent hydrogen column density derived from the X-ray spectra as a function of
           the optical colour index B-V (top) and as a function of the total \Hone\ column
	   density of the SMC. The line in the bottom panel marks \nh\ = \Hone.}
  \label{fig-absorption}
\end{figure}

The SMC absorption component varies over a wide range between a few \ohcm{20}  
and several \ohcm{22}.
It is evident, that the sources in field 0500980101 show similar values of low 
absorption, while those of field 0404680101, near the emission nebula N19, 
show all very high absorption. 
For further investigation we list the total \Hone\ column density of the SMC in the direction of
each source in the last column of Table~\ref{tab-spectra}.
In Fig.~\ref{fig-absorption} the X-ray measured column density \nh\ is plotted as functions of the
B-V colour index of its optical counterpart (from MCPS in Table~\ref{tab-ids}) and the total 
SMC column density as derived from \Hone\ measurements. In both cases a correlation is 
indicated, although with considerable scatter in \nh. The scatter is expected 
as 1) the sources are located at different depth in the SMC and 2) a large part of the 
absorption can originate locally in the Be/X-ray binary systems 
\citep[suggested by strong variations of \nh\ over longterm timescales as seen, 
e.g., from RX\,J0103.6-7201; ][]{2005A&A...438..211H}. 
The general increase of the X-ray measured \nh\ with the total SMC column density suggests 
that a significant fraction of the X-ray absorption seen in the X-ray spectra of Be/X-ray binaries
arises in the interstellar medium of the SMC. 

\begin{acknowledgements}
The XMM-Newton project is supported by the Bundesministerium f\"ur Wirtschaft und 
Technologie/Deutsches Zentrum f\"ur Luft- und Raumfahrt (BMWI/DLR, FKZ 50 OX 0001)
and the Max-Planck Society. 
The ``Second Epoch Survey'' of the southern sky was produced by the
Anglo-Australian Observatory (AAO) using the UK Schmidt Telescope.
Plates from this survey have been digitised and compressed by the ST ScI.
Produced under Contract No. NAS 5-26555 with the National Aeronautics
and Space Administration.
This paper utilises public domain data obtained by the MACHO Project, jointly
funded by the US Department of Energy through the University of California,
Lawrence Livermore National Laboratory under contract No. W-7405-Eng-48, by
the National Science Foundation through the Center for Particle Astrophysics
of the University of California under cooperative agreement AST-8809616, and
by the Mount Stromlo and Siding Spring Observatory, part of the Australian
National University.
\end{acknowledgements}

\bibliographystyle{aa}
\bibliography{general,myrefereed,myunrefereed,mcs,hmxb,ism,ins,cv}

\end{document}